\documentclass[sn-mathphys-num]{sn-jnl}
\usepackage[title]{appendix}%
\usepackage{booktabs}
\usepackage{array}
\usepackage{hyperref}       
\usepackage{url}            
\usepackage{booktabs}       
\usepackage{amsfonts}       
\usepackage{graphicx}
\usepackage{natbib}
\usepackage{doi}
\usepackage{graphicx}%
\usepackage{multirow}%
\usepackage{amsmath}%
\usepackage{amsthm}%
\usepackage{mathrsfs}%
\usepackage[title]{appendix}%
\usepackage{textcomp}%
\usepackage{manyfoot}%
\usepackage{algorithm}%
\usepackage{algorithmicx}%
\usepackage{algpseudocode}%
\usepackage{listings}%
\usepackage{hyperref}
\usepackage{cleveref}
\usepackage{multirow}
\usepackage{array}
\usepackage{colortbl} 
\usepackage{hhline} 
\usepackage{booktabs} 

\usepackage[utf8]{inputenc} 
\usepackage[T1]{fontenc}    
\usepackage{hyperref}       
\usepackage{url}            
\usepackage{booktabs}       
\usepackage{amsfonts}       
\usepackage{nicefrac}       
\usepackage{microtype}      
\usepackage{lipsum}		
\usepackage{graphicx}
\usepackage{natbib}
\usepackage{doi}

\usepackage{colortbl}  

\usepackage{booktabs} 
\usepackage{threeparttable}
\usepackage{bbding}
\usepackage{ulem}
\usepackage{float}
\usepackage{subfig}
\usepackage{makecell}
\makeatletter
\newcommand*\bigcdot{\mathpalette\bigcdot@{.5}}
\newcommand*\bigcdot@[2]{\mathbin{\vcenter{\hbox{\scalebox{#2}{$\m@th#1\bullet$}}}}}
\makeatother
\usepackage{ragged2e}

\usepackage[table,xcdraw]{xcolor}
\definecolor{tablecolor1}{RGB}{216, 214, 196} 
\definecolor{tablecolor2}{RGB}{236, 234, 224} 
\usepackage{multirow}
\usepackage{adjustbox}
\usepackage{colortbl}
\usepackage{array}
\usepackage{algorithm}
\usepackage{algpseudocode}

\usepackage{etoolbox} 
\newcommand{\numberedsection}[1]{%
  \refstepcounter{section} 
  \par\vspace{18pt} 
  {\noindent\Large\bfseries\thesection\quad #1\par} 
  \vspace{12pt}
}

\newcommand{\numberedsubsection}[1]{%
  \refstepcounter{subsection}  
  \par\vspace{12pt} %
  {\noindent\large\bfseries\thesubsection\quad #1\par} %
  \vspace{6pt} %
}

\begin{document}

\title[AirMorph]{AirMorph: Topology-Preserving Deep Learning for Pulmonary Airway Analysis}




\author[1,2]{\fnm{Minghui} \sur{Zhang}}\email{minghuizhang@sjtu.edu.cn}
\author[1,2]{\fnm{Chenyu} \sur{Li}}\email{lichenyu@sjtu.edu.cn}
\author[3]{\fnm{Fangfang} \sur{Xie}}\email{xiefang314@126.com}
\author[1,2]{\fnm{Yaoyu} \sur{Liu}}\email{lyyu19@sjtu.edu.cn}
\author[1]{\fnm{Hanxiao} \sur{Zhang}}\email{hanxiao.zhang@sjtu.edu.cn}
\author[1]{\fnm{Junyang} \sur{Wu}}\email{sjtuwjy@sjtu.edu.cn}
\author[3]{\fnm{Chunxi} \sur{Zhang}}\email{zhcx1999@163.com}
\author[2]{\fnm{Jie} \sur{Yang}}\email{jieyang@sjtu.edu.cn}
\author*[3]{\fnm{Jiayuan} \sur{Sun}}\email{sunjy1976@126.com}
\author*[1]{\fnm{Guang-Zhong} \sur{Yang}}\email{gzyang@sjtu.edu.cn}
\author*[1,2]{\fnm{Yun} \sur{Gu}}\email{yungu@ieee.org}

\affil[1]{
\orgdiv{Institute of Medical Robotics}, 
\orgname{Shanghai Jiao Tong University}, 
\orgaddress{\street{800 Dongchuan RD. Minhang District}, 
\city{Shanghai}, 
\postcode{200240}, 
\country{CHINA}}}
\affil[2]{
\orgdiv{Institute of Image Processing and Pattern Recognition}, 
\orgname{Shanghai Jiao Tong University}, 
\orgaddress{\street{800 Dongchuan RD. Minhang District}, 
\city{Shanghai}, 
\postcode{200240}, 
\country{CHINA}}}
\affil[3]{
\orgdiv{Department of Respiratory Endoscopy,Department of Respiratory and Critical CareMedicine}, 
\orgname{Shanghai Chest Hospital}, 
\orgaddress{\street{}
\city{Shanghai}, 
\postcode{20030}, 
\country{CHINA}}}


\abstract{
  Accurate anatomical labeling and analysis of the pulmonary structure and its surrounding anatomy from thoracic CT is getting increasingly important for understanding the etiology of abnormalities or supporting targetted therapy and early interventions. Whilst lung and airway cell atlases have been attempted, there is a lack of fine-grained morphological atlases that are clinically deployable. In this work, we introduce AirMorph, a robust, end-to-end deep learning pipeline enabling fully automatic and comprehensive  airway anatomical labeling at lobar, segmental, and subsegmental resolutions that can be used to create digital atlases of the lung. Evaluated across large-scale multi-center datasets comprising diverse pulmonary  conditions, the AirMorph consistently outperformed existing segmentation and  labeling methods in terms of accuracy, topological consistency, and  completeness. To simplify clinical interpretation, we further introduce a compact anatomical signature quantifying critical morphological airway features—including stenosis, ectasia, tortuosity, divergence, length, and complexity. When applied to various pulmonary diseases such as pulmonary ﬁbrosis, emphysema, atelectasis, consolidation, and reticular opacities, it demonstrates strong discriminative power, revealing disease-speciﬁc morphological  patterns with high interpretability and explainability. Additionally, AirMorph supports eﬃcient  automated branching pattern analysis, potentially enhancing bronchoscopic navigation planning and procedural safety, oﬀering a valuable clinical  tool for improved diagnosis, targeted treatment, and personalized patient care.
}



\maketitle

\begin{figure}[!h]
\centering
\includegraphics[width=0.78\linewidth]{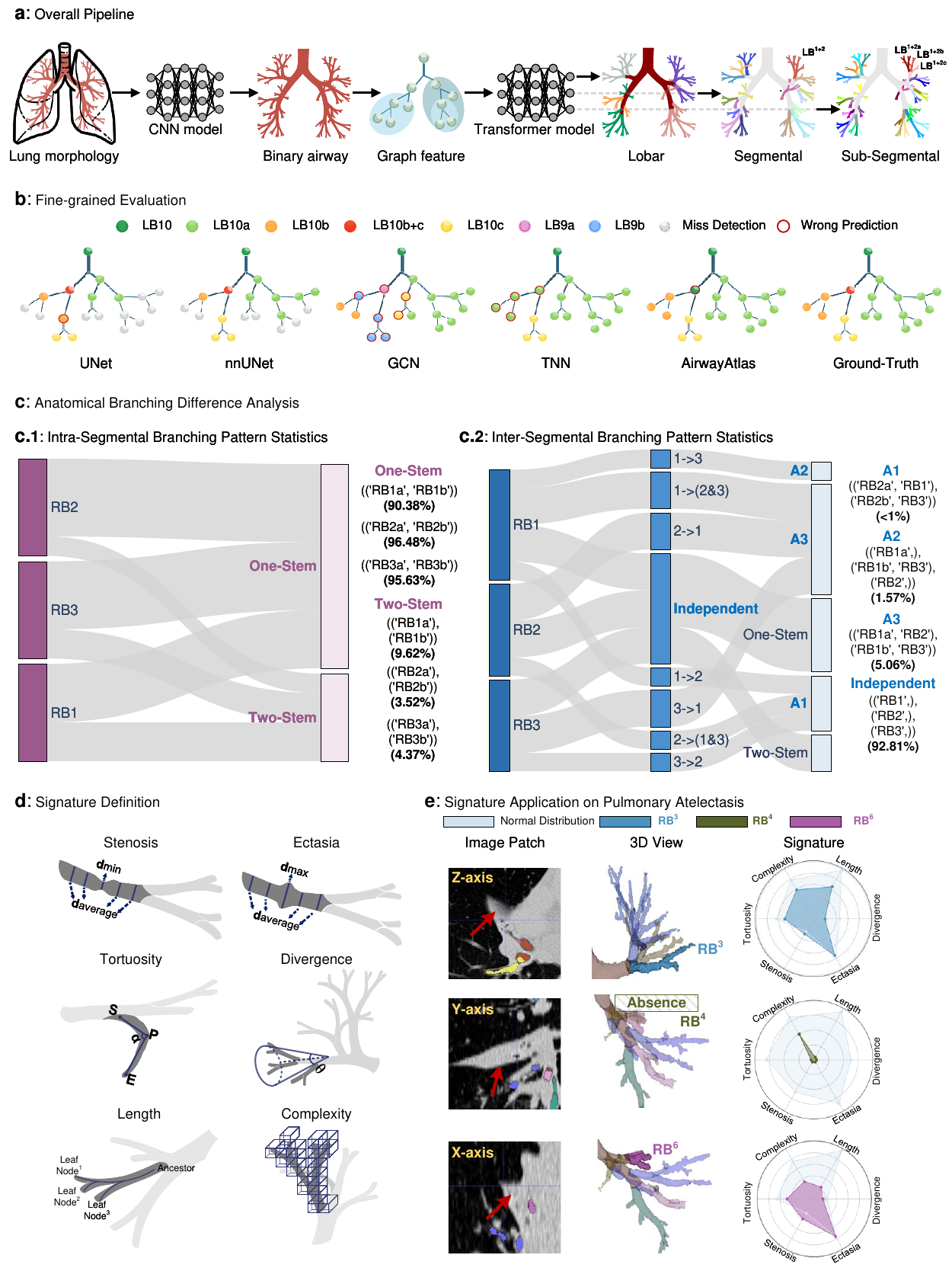}
\caption{
Overview of the AirMorph's development, fine-grained evaluation, and clinical applications. 
a) Model development. AirMorph is a fully automated framework for extracting subsegmental anatomical bronchi from thoracic CT scans. 
It comprises three stages: (1) binary airway modeling from CT scans,
(2) feature extraction from a graph-based representation of the airway tree, and (3) anatomical labeling based on branch-wise features.
b) Evaluation. AirMorph supports unified and fine-grained evaluation of both binary airway modeling and anatomical labeling via graph node-level performance metrics.
c)-d) Clinical applications. AirMorph facilitates the real-world clinical applications. 
c) It enables efficient analysis of airway branching patterns across the entire bronchial tree.
d) Fine-grained airway signatures quantify structural abnormalities between patient cohorts with lung disease and healthy controls.
}\label{fig::AirMorph_MainDigram}
\end{figure}

\numberedsection{Introduction}\label{sec::intro}

The pulmonary airway refers to the network of structures within the respiratory system that conducts air from the external environment to the lungs and facilitates gas exchange. 
The morphological anatomy of the pulmonary airway includes the detailed structural features of the airway, which are critical for diagnosing respiratory diseases. 
It is also the natural path for endobroncial biopsy and treatment. 
While lung and airway cell atlases have provided valuable insights into the cellular composition, heterogeneity, and molecular states of pulmonary structures \cite{sikkema2023integrated, kadur2022human}, 
their primary focus has been on transcriptomic and molecular profiling at the cell level. Despite their contributions to understanding cell-type diversity and interactions in both healthy and diseased tissues, 
these atlases often lack detailed morphological characterization and precise anatomical localization—features that are essential for clinical translation and procedural planning.

Previous efforts have established morphological atlases for various organs, including the brain~\cite{evans2012brain,iqbal2019developing}, heart~\cite{anderson1996anatomy,litjens2019state}, and liver~\cite{couinaud1989surgical,couinaud1999liver}. 
These morphological atlases have significantly advanced clinical diagnostics and personalized therapeutic interventions by providing high-resolution anatomical frameworks. 
However, the pulmonary airway poses unique challenges, including fine-grained branching structure complexity, high inter-individual anatomical variability, 
and subtle morphological changes associated with diverse pulmonary diseases. These intrinsic difficulties have impeded the automatic development of comprehensive and detailed morphological airway atlases.
Due to the complex structure of airway, it is however challenging to manually annotate the airway based on medical imagings, which cannot fully extract the fine-grained distal bronchi and time-consuming. 
Without large-scaled and detailed annotation of airways, it is also difficult to analyze the distribution of anatomical variations and correltion of diseases. 
Therefore, previous works only focus on the statics of certain regions, like left upper lobe \cite{he2022anatomical,deng2022anatomical}, right lower lobe \cite{zhu2023branching}, with manual check of binary airway segmentations. 
Recently, the automated methods have been proposed for binary segmentation, lobar/segmental and subsegmental labeling of airway. Previous airway segmentation benchmarks \cite{lo2012extraction,zhang2023multi,nan2024hunting} 
and methods \citep{charbonnier2017improving,xu2015hybrid,meng2017tracking,qin2019airwaynet,nadeem2020ct,zheng2021alleviating,nan2023fuzzy,zhang2023towards,wang2024accurate} lack a fine-grained evaluation of airway 
reconstruction. In many cases, assessing the overall quality of airway reconstruction may not be necessary. Instead, the reconstruction quality of local segments and subsegments surrounding lesions is of greater clinical 
importance. Therefore, branch-wise detailed evaluation holds significant relevance for clinical applications. Recent approaches to airway anatomical labeling face several limitations that prevent them from automatically 
achieving subsegmental-level labeling. Most works rely on the ground truth airway shape to predict segmental-level anatomical labeling \citep{xie2025efficient,huang2024bcnet}. While \cite{yu2022tnn} extends to subsegmental 
classification, it depends heavily on ground truth data annotated from CT images and manually refined features, rendering it incapable of operating directly from raw CT scans. This reliance diminishes its clinical applicability. 
Only a few works \citep{nadeem2020anatomical,xie2024structure, chau2024branchlabelnet} attempt to directly infer anatomical airway labeling from CT scans. However, these methods are limited to incomplete segmental-level anatomical 
labeling on relatively small-scale datasets, further highlighting the gap in achieving comprehensive, automated airway labeling.

In this work, we proposed \textit{\textbf{AirMorph}}, a digitalized atlas for pulmonary airway, a fully automated framework that jointly performs binary airway segmentation and hierarchical anatomical labeling from chest CT scans. 
As shown in Fig.\ref{fig::AirMorph_MainDigram}, the proposed method enables the automated identification and labeling of airway anatomies at the lobar, segmental, and subsegmental levels. 
To systematically assess the clinical applicability of AirMorph, we establish a fine-grained evaluation framework that considers both topological segmentation quality and anatomical labeling fidelity.
We first evaluate the framework on the Primary dataset, which contains 620 CT volumes with expert-defined ground truth annotations. AirMorph achieves a Dice Similarity Coefficient (DSC) of 84.40\%, 83.30\%, and 74.38\%, and 
a tree length detected (TLD) rate of 92.27\%, 92.68\%, and 85.56\% at the lobar, segmental, and subsegmental levels, respectively. 
The fine-grained recognition accuracy at the subsegmental level reaches 94.12\%, surpassing other state-of-the-art (SOTA) methods.
Robustness under disease-induced variations ($n=120$) was assessed using CT scans exhibiting substantial structural abnormalities. \textit{AirMorph} achieved a labeling accuracy of 83.52\% even in severely deformed cases, 
surpassing state-of-the-art (SOTA) methods by more than 15 percentage points. In parallel, the model's generalization capability was evaluated across three external,
multi-site test cohorts ($n=2403$), where \textit{AirMorph} reconstructed approximately 60\% more airway branches than competing approaches. 
These results highlight the framework’s scalability, robustness to anatomical variability, and potential utility in diverse clinical and population-scale applications.

Beyond segmentation and labeling, {AirMorph} supports efficient, large-scale analysis of bronchial branching patterns. Leveraging its graph-based airway representation, the framework automatically characterizes 
intra- and inter-segmental variations without requiring manual verification. Applied to over 3,000 CT scans from diverse cohorts, AirMorph revealed a comprehensive taxonomy of segmental and subsegmental branching types, 
including co-trunk structures and rare anatomical variants. For instance, the most prevalent $LB^{1+2}$ configuration, featuring a single-stem bifurcation into $LB^{1+2^{a+b}}$ and $LB^{1+2^{c}}$, was consistently observed across 
all cohorts, accounting for 69.95\% of cases. This automated pattern analysis significantly reduces the assessment time from hours to seconds per case, thereby facilitating scalable anatomical studies and enhancing the interpretability 
of patient-specific airway architectures in clinical planning.


To further enhance clinical interpretability, we introduce the \textbf{\textit{AirwaySignature}}, a compact and anatomically aligned representation derived from the labeled bronchial tree. 
This signature encodes six quantitative morphological descriptors, stenosis, ectasia, tortuosity, divergence, length, and complexity, capturing both local abnormalities and global structural variations across the airway. 
Leveraging these features, AirwaySignature enables efficient localization and visualization of alterations, supporting comparative morphology analysis across disease cohorts. 
In a large-scale study involving five pulmonary disease types (including fibrosis, atelectasis, consolidation, emphysema and reticular opacities), the signature revealed statistically significant morphological deviations 
compared to healthy controls, highlighting its diagnostic relevance. This abstraction facilitates rapid screening of structurally abnormal branches, provides a clinically intuitive tool for targeted navigation, and supports 
downstream radiomic integration by focusing feature extraction on anatomically relevant regions.

\begin{figure}[!t]
\centering
\includegraphics[width=0.8\linewidth]{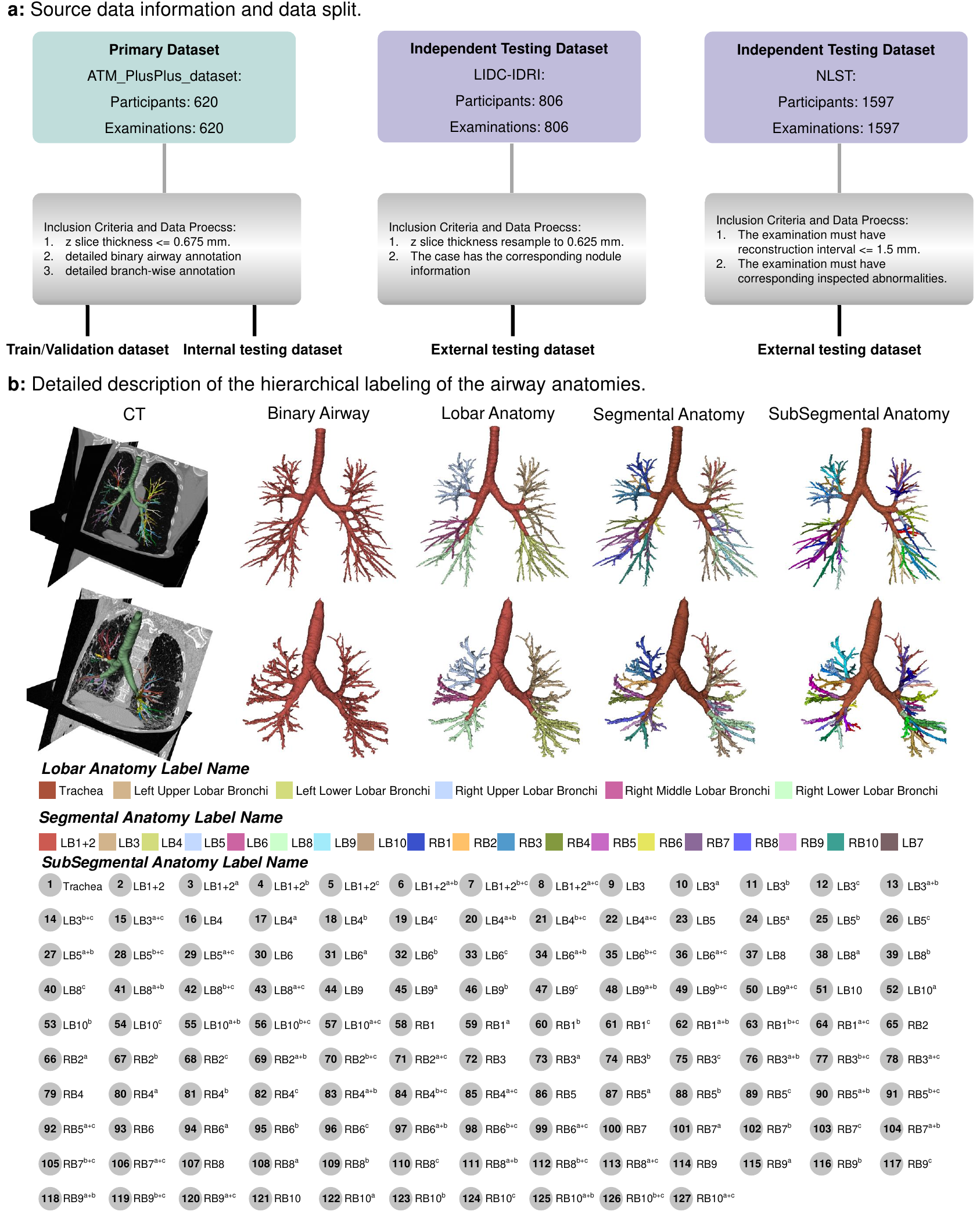}
\caption{
Detailed information of the datasets for AirMorph. a) A total of 3,023 thoracic computed tomography (CT) scans were collected from four distinct institutions and partitioned into a primary dataset and independent test datasets.
ATM++ serves as the primary dataset, comprising 620 patients with detailed annotations that include both binary airway labels and subsegmental-level anatomical labels.
An additional 806 patients from the LIDC-IDRI dataset and 1,597 patients from the NLST dataset were also included in this study.
b) The fine-grained annotation includes the binary airway from CT scans, along with five lobar anatomies, nineteen kinds of segmental anatomies, and one hundred and twenty-seven subsegmental anatomies. 
Additionally, the first row represents the airway anatomies of patients with mild conditions, wherea the last row depicts structural alterations observed in patients with advanced pulmonary fibrosis.
}\label{fig::data}
\end{figure}

\begin{figure}[!t]
\centering
\includegraphics[width=0.76\linewidth]{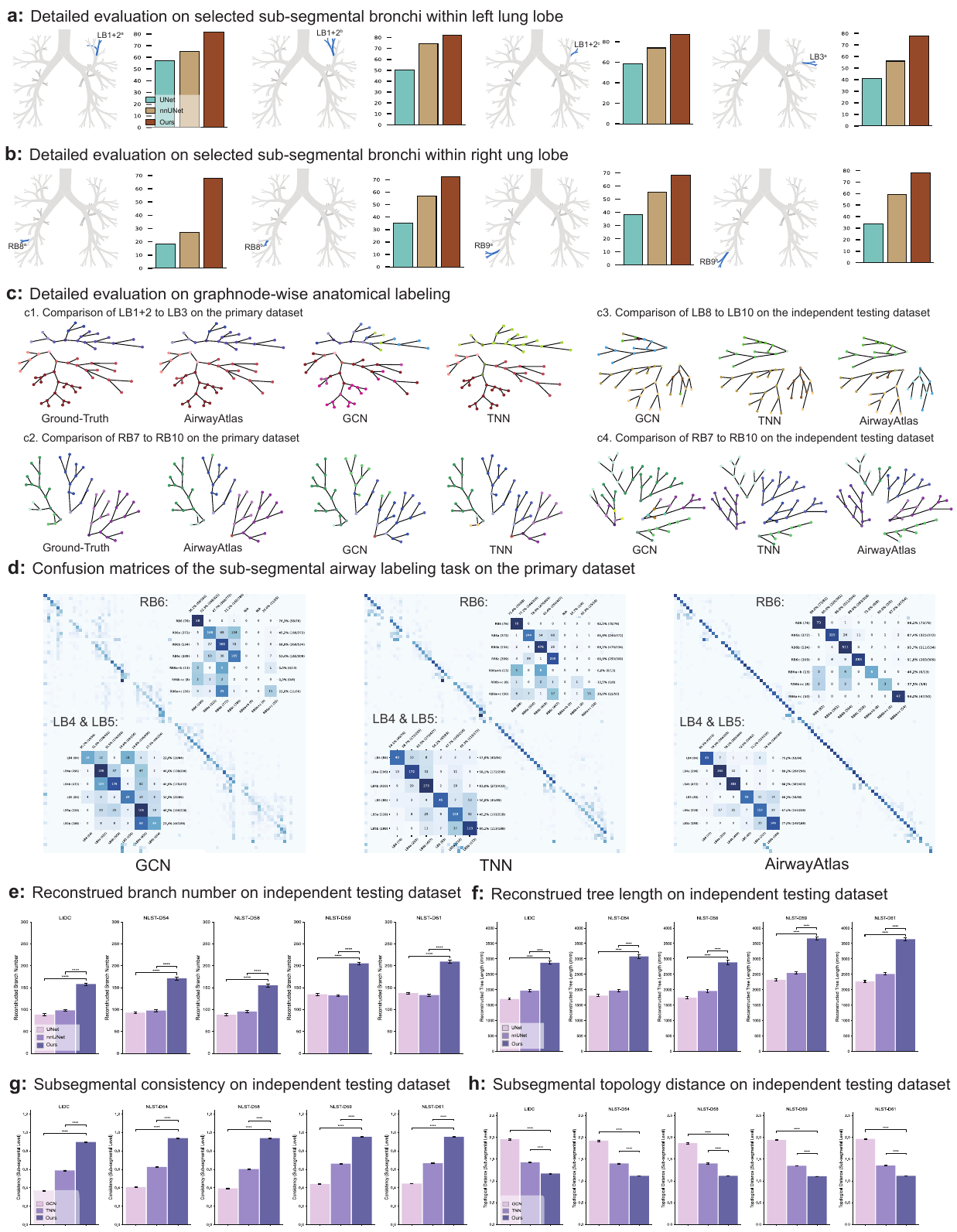}
\caption{
Fine-grained evaluation of AirMorph. 
a-b) Detailed evaluation of the tree length detection rate of AirMorph compared with UNet and nnUNet at the subsegmental branch level.
a) compares selected bronchi in the left upper lobe: $LB^{{1+2}^{a}}$, $LB^{{1+2}^{b}}$, $LB^{{1+2}^{c}}$, and $LB^{{3}^{a}}$.
b) shows corresponding comparisons for the right lower lobe: $RB^{{8}^{a}}$, $RB^{{8}^{b}}$, $RB^{{9}^{a}}$, and $RB^{{9}^{b}}$.
c) Graphnode-wise anatomical labeling performance across representative bronchial regions. Comparisons are shown for the primary dataset (c1-c2) and the external independent test set (c3-c4), involving AirMorph, GCN, and TNN.
d) Confusion matrices illustrating subsegmental labeling accuracy on the primary dataset across different models. 
e-f) Ablation Study on external independent test sets.  
e) Total number of reconstructed airway branches. 
f) Cumulative tree length of the predicted airway structures. 
g) Subsegmental labeling consistency for adjacent bronchi. 
h) Topological distance of predicted airway graphs.
P-values are specified as $\ast$p < 0.05, $\ast\ast$p < 0.01, $\ast\ast\ast$p < 0.001, $\ast\ast\ast\ast$p < 0.0001, n.s, not significant. 
}\label{fig::finegrained_evaluation_AirMorph}
\end{figure}

\makeatletter
\def\hlinew#1{%
  \noalign{\ifnum0=`}\fi\hrule \@height #1 \futurelet
  \reserved@a\@xhline}
\makeatother
\begin{table*}[!t]
\centering 
\caption{
Fine-grained evaluation of the binary airway modeling and the anatomical airway labeling.
tree length detected rate (TLD, \%), branch number detected rate (BND, \%), Dice similarity coefficient (DSC), centerline Dice (clDice), and sensitivity
are employed to evaluate the integrity of the binary airway modeling. 
For anatomical branch labeling, predicted subtree consistency (TreeCons, \%), topological distance (TopoDist), accuracy, precision, and sensitivity
are used to evaluate classification performance.
}
\label{tab::finegrained_evaluation_seg_and_labeling}
\renewcommand\arraystretch{1.4}
\resizebox{0.9\textwidth}{!}{
\begin{tabular}{>{\centering\arraybackslash}p{2cm}>{\centering\arraybackslash}p{2cm}>{\centering\arraybackslash}p{2cm}>{\centering\arraybackslash}p{2cm}>{\centering\arraybackslash}p{2cm}>{\centering\arraybackslash}p{2cm}}
\hlinew{1.0pt}
\rowcolor{tablecolor1}
\multicolumn{6}{c}{\textbf{Airway Atlas: Binary Airway Modeling}}       \\ \hlinew{0.8pt}
\rowcolor{tablecolor2}
\multicolumn{6}{l}{\textit{Lobar}}                                      \\ \hlinew{0.8pt}
\textbf{Method}      & \textbf{TLD}    & \textbf{BND}    & \textbf{DSC}   & \textbf{clDice} & \textbf{Sensitivity}  \\ \rowcolor{tablecolor2}
UNet        & 59.27 & 63.64 & 78.28 & 74.52  & 69.00   \\
nnUNet      & 73.82 & 79.33 & 82.27 & 85.62  & 73.98\\\rowcolor{tablecolor2}
AirMorph & \textbf{92.27} & \textbf{96.58} & \textbf{84.40}  & \textbf{89.35}  & \textbf{94.36} \\\hlinew{0.8pt}
\multicolumn{6}{l}{\textit{Segmental}}                          \\ \hlinew{0.8pt}\rowcolor{tablecolor2}
\textbf{Method}      & \textbf{TLD}    & \textbf{BND}    & \textbf{DSC}   & \textbf{clDice} & \textbf{Sensitivity}      \\
UNet        & 57.78 & 61.68 & 72.78 & 70.60   & 65.11 \\\rowcolor{tablecolor2}
nnUNet      & 72.91 & 78.18 & 79.52 & 83.00     & 72.38  \\
AirMorph & \textbf{92.68} & \textbf{94.94} & \textbf{83.00}    & \textbf{88.00}  & \textbf{92.41} \\\hlinew{0.8pt}
\rowcolor{tablecolor2}
\multicolumn{6}{l}{\textit{SubSegmental}}                         \\ \hlinew{0.8pt}
\textbf{Method}      & \textbf{TLD}    & \textbf{BND}    & \textbf{DSC}   & \textbf{clDice} & \textbf{Sensitivity}    \\ \rowcolor{tablecolor2}
UNet        & 60.33 & 65.55 & 66.67 & 68.10   & 61.51  \\ 
nnUNet      & 70.08 & 76.32 & 70.67 & 76.07  & 65.39\\ \rowcolor{tablecolor2}
AirMorph & \textbf{85.56} & \textbf{89.15} & \textbf{74.38} & \textbf{81.19}  & \textbf{83.87} \\\hlinew{0.8pt}
\rowcolor{tablecolor1}
\multicolumn{6}{c}{\textbf{Airway Atlas: Airway Anatomical Labeling}}   \\\hlinew{0.8pt}
\multicolumn{6}{l}{\textit{Lobar}}                                      \\\hlinew{0.8pt} \rowcolor{tablecolor2}
\textbf{Method}     & \textbf{TreeCons}    & \textbf{TopoDist}    & \textbf{Accuracy}   & \textbf{Precision}     & \textbf{Sensitivity}       \\
GCN         &  98.16     &   0.1225    & 96.73  &   96.64     &  95.77    \\\rowcolor{tablecolor2}
TNN         &  \textbf{100.0}    &   0.0349   & 98.00  &  98.02      &   96.82      \\
AirMorph &  99.99    &  \textbf{0.0260}   & \textbf{99.06}  &  \textbf{99.20}      &   \textbf{98.52}         \\\hlinew{0.8pt}
\rowcolor{tablecolor2}
\multicolumn{6}{l}{\textit{Segmental}}                                  \\\hlinew{0.8pt}
\textbf{Method}     & \textbf{TreeCons}    & \textbf{TopoDist}    & \textbf{Accuracy}   & \textbf{Precision}     & \textbf{Sensitivity}   \\\rowcolor{tablecolor2}
GCN         &  67.38     &  0.9032     &  83.09  & 75.92      &  81.51         \\
TNN         &  83.56     &  1.3455     & 80.56   &  70.94      &   78.18          \\\rowcolor{tablecolor2}
AirMorph &  \textbf{99.47}     &  \textbf{0.1024}     & \textbf{97.42}  &   \textbf{96.84}     &  \textbf{97.12}        \\\hlinew{0.8pt}

\multicolumn{6}{l}{\textit{SubSegmental}}                               \\ \hlinew{0.8pt}\rowcolor{tablecolor2}
\textbf{Method}     & \textbf{TreeCons}    & \textbf{TopoDist}    & \textbf{Accuracy}   & \textbf{Precision}     & \textbf{Sensitivity}     \\
GCN         &    65.93   &  2.2772   & 63.89 &  52.83      &  57.31        \\\rowcolor{tablecolor2}
TNN         &    86.88  & 2.1972      & 69.61 &   59.45     &   64.10         \\
AirMorph &    \textbf{98.02}   &   \textbf{0.4808}    & \textbf{94.12} &   \textbf{93.12}     &  \textbf{93.33}        \\ \hlinew{0.8pt}
\end{tabular}
}
\end{table*}

\makeatletter
\def\hlinew#1{%
\noalign{\ifnum0=`}\fi\hrule \@height #1 \futurelet
\reserved@a\@xhline}
\makeatother
\begin{table*}[th]
\renewcommand\arraystretch{1.4}
\caption{Statistical analysis of diverse airway branching patterns is presented, including both intra-segmental and intra-subsegmental levels. The analysis spans from the five lobes to segmental bronchi, and from all eighteen segmental bronchi to their corresponding subsegmental branches.
Abbreviations are used for common branching types: Mono- for monofurcation, Bi- for bifurcation, Tri- for trifurcation, Quadri- for quadrifurcation, and Quint- for quintfurcation.
}
\centering
\label{tab::statis_branch_pattern}
\resizebox{0.82\textwidth}{!}{
 &                                                                                           &                                                                    &                                                                                      \\ \hline
\end{tabular}}
\end{table*}

\begin{figure}[!t]
\centering
\includegraphics[width=0.85\linewidth]{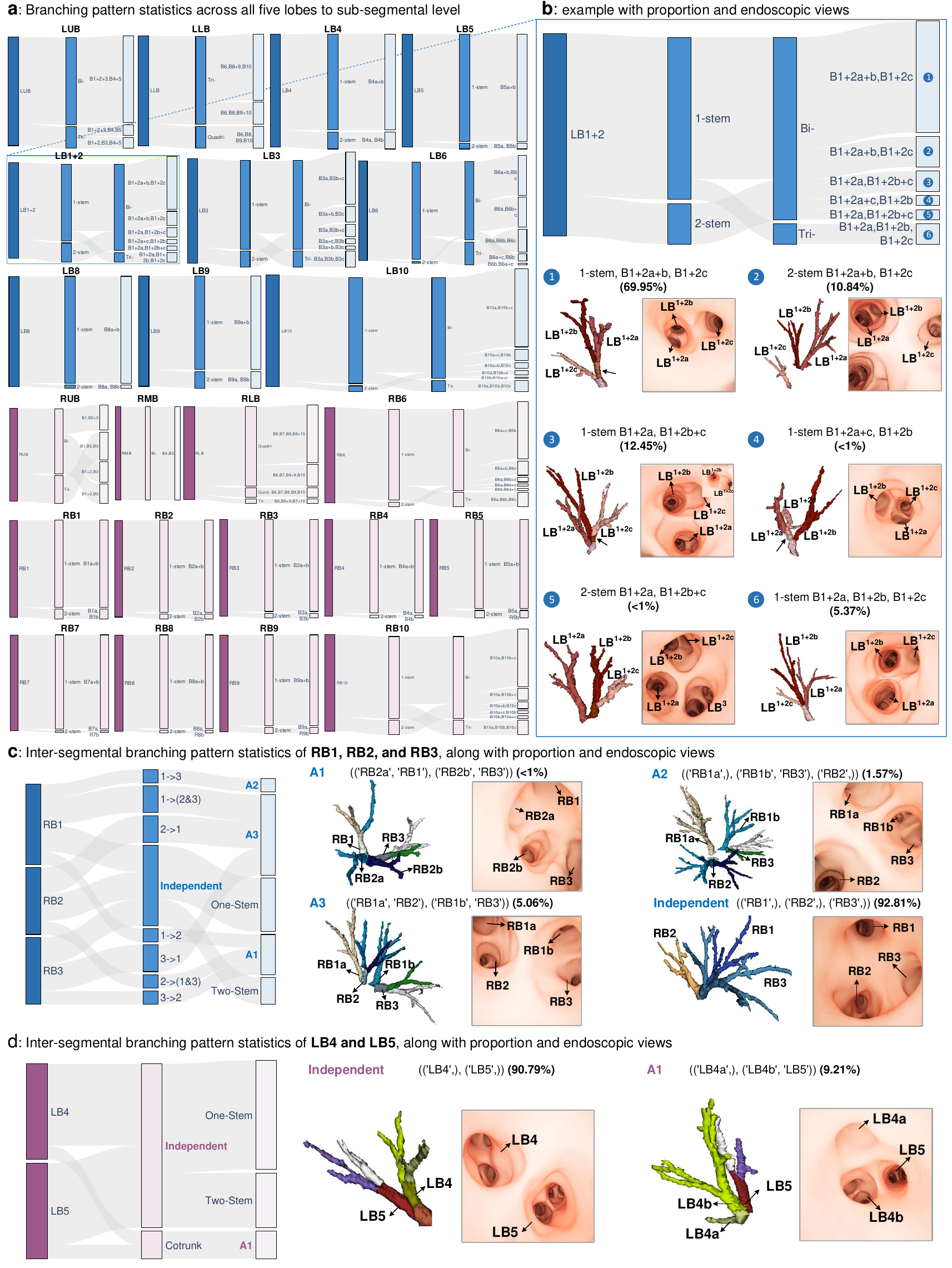}
\caption{
Comprehensive characterization of airway branching patterns across lobar, segmental, and subsegmental levels using AirMorph.
a) Distribution of branching pattern types across all five lobes, from lobar to subsegmental level, showing the prevalence of one-stem, two-stem, bifurcated, trifurcated, and co-trunk configurations.  
b) Take the $LB^{1+2}$ as a representative segmental branch, and visualize the airway branching pattern statistics along with their anatomical mapping through 3D endoscopic views.  
c) Inter-segmental branching pattern analysis among $RB^{1}$, $RB^{2}$, and $RB^{3}$, visualized with annotated 3D bronchoscopic views.
d) Inter-segmental pattern analysis of $LB^{4}$ and $LB^{4}$, showing the prevalence of independent and co-trunk configurations.
}\label{fig::branch_pattern_with_endo_view}
\end{figure}

\begin{figure}[!t]
\centering
\includegraphics[width=0.81\linewidth]{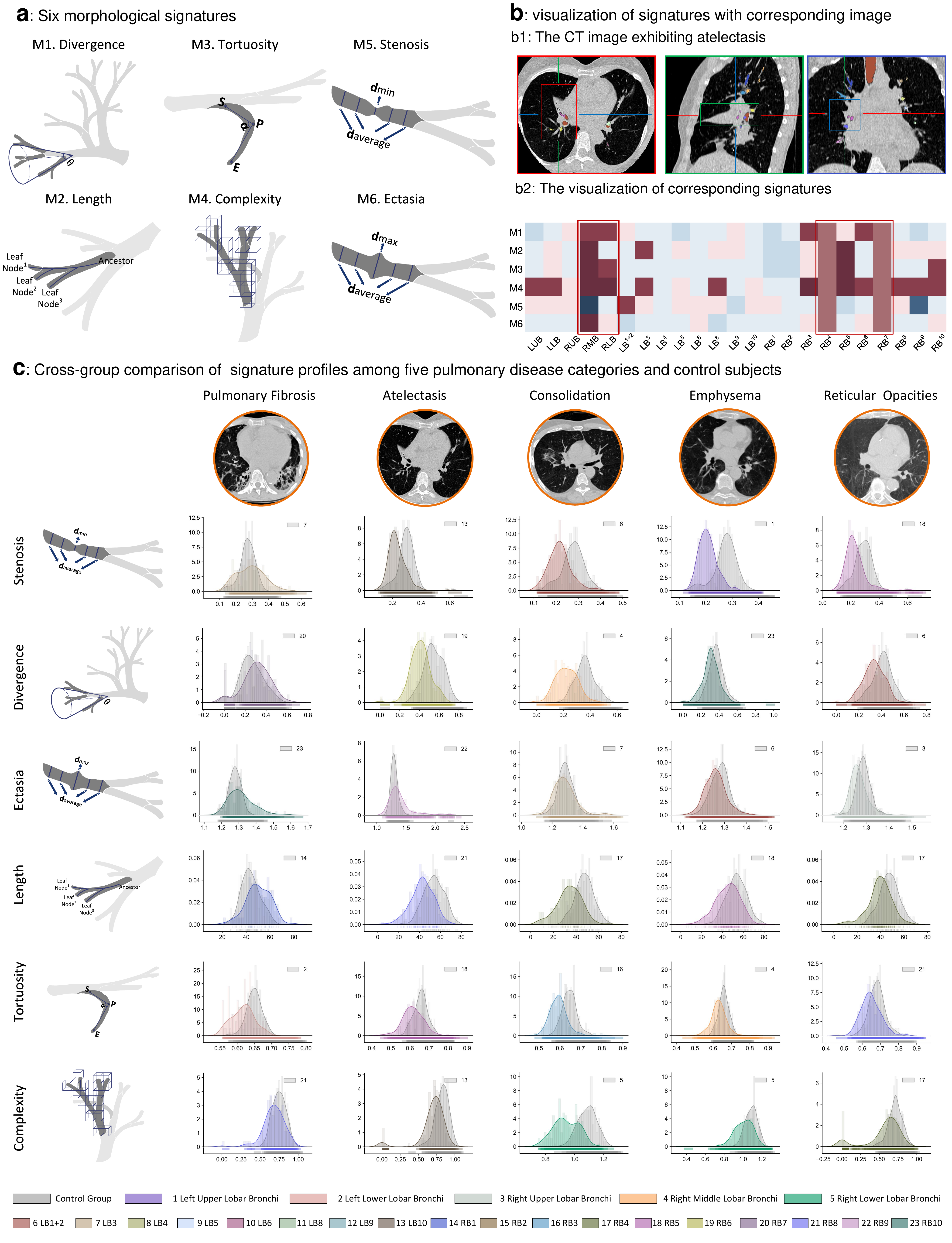}
\caption{
AirwaySignature representation and morphological comparison across disease types.
a) Definition of six airway morphological descriptors computed from anatomically labeled branches using AirMorph: stenosis, ectasia, tortuosity, divergence, length, and complexity. 
b) Visualization of the AirwaySignature matrix in a representative case with atelectasis. 
b1 shows the corresponding CT scan highlighting the lesion regions happen in the right middle and lower lobe. 
b2 displays anatomical branch-wise airway signatures distributions. Color encoding indicates statistical deviation from the healthy reference distribution. 
c) Disease-level comparison of AirwaySignature profiles across five pulmonary conditions (pulmonary fibrosis, emphysema, atelectasis, consolidation, and reticular opacities) and healthy controls. 
Distinct morphological trends are visible among subgroups, supporting the discriminative power of AirwaySignature.
}\label{fig::AirMorph_morpho_distribution_presentation}
\end{figure}

\numberedsection{Results}\label{sec::results}

\numberedsubsection{Dataset Characteristic}
\noindent \textbf{Dataset summary}. In total, 3,023 thoracic computed tomography scans were retrospectively collected from four different sites for digital analysis of the pulmonary airway (Fig.\ref{fig::data}.a)). 
The primary dataset extends previously established binary airway annotations from ATM'22 \cite{zhang2023multi} and AIIB'23 \cite{nan2024hunting} by introducing fine-grained anatomical airway labeling. 
This primary dataset comprises scans from 620 patients and is partitioned into training/validation and internal testing subsets.
The external test dataset consists of scans derived from the two largest publicly available lung-related cohorts: the Lung Image Database Consortium and Image Database Resource Initiative (LIDC-IDRI) \cite{armato2011lung, armato2015data}, 
and the National Lung Screening Trial (NLST) \citep{national2011reduced, nlst2013data}. A total of 806 patients from LIDC-IDRI and 1,597 patients from NLST were included in this study.
Moreover, this research involves five distinct lung diseases. Specifically, patients from the AIIB'23 dataset were diagnosed with pulmonary fibrosis, while patients from the NLST dataset exhibited four representative pulmonary conditions: 
atelectasis (NLST-D54, n=252), pulmonary consolidation (NLST-D58, n=175), emphysema (NLST-D59, n=686), and pulmonary reticular opacities (NLST-D61, n=484).

\noindent \textbf{Reference standards of the dataset}. 
The annotation protocol for the AirMorph dataset was developed based on characteristic imaging features and informed by clinical anatomical priors (Fig.\ref{fig::data}.b)). 
The annotation follows a hierarchical approach, beginning with binary airway tree delineation from CT images and subsequently assigning anatomical branch labels guided by clinical anatomical conventions. 
Specifically, the binary airway tree is annotated in a branch-by-branch manner using a backtracking strategy initiated from the tracheal root in the CT scan. 
In accordance with clinical anatomical standards \cite{netter2014atlas}, the binary airway tree is further partitioned into five lobar-level anatomical regions—including the left upper, left lower, right upper, right middle, 
and right lower lobar bronchi—along with 19 segmental labels (9 in the left lung and 10 in the right lung) and 127 subsegmental anatomical labels. 
Our AirMorph provides large-scale anatomical annotations extending to subsegmental airway branches across a wide spectrum of lung diseases. A predefined labeling dictionary 
comprising 127 categories was established to account for all plausible subsegmental branches and their common trunking configurations. For instance, the segmental branch $LB^{1+2}$ can 
be further categorized into sub-branches such as $LB^{1+2^{a}}$, $LB^{1+2^{b}}$, $LB^{1+2^{c}}$, $LB^{1+2^{a+b}}$, $LB^{1+2^{a+c}}$, $LB^{1+2^{b+c}}$, and $LB^{1+2^{a+b+c}}$.
To ensure cross-cohort applicability, the labeling protocol was standardized across all datasets regardless of imaging source or disease condition. 
All annotations were initially delineated and independently verified by two radiologists, each with over five years of professional experience. 
Subsequently, all labels were reviewed and refined by a senior radiologist with more than ten years of clinical experience to ensure consistency and anatomical accuracy.

\numberedsubsection{AirMorph: Automated Airway Labeling Model}
We present AirMorph, a deep learning model designed to fully extract fine-grained anatomical bronchi from thoracic CT scans. It efficiently and accurately produces comprehensive anatomical 
representations across the entire bronchial tree (Fig.\ref{fig::AirMorph_MainDigram}). 
The AirMorph model was trained and internally evaluated on a primary dataset with complete subsegmental annotations, and subsequently validated on two external datasets to demonstrate generalizability (Methods). 
The first stage of AirMorph reconstructs the binary airway tree from CT scans, leveraging our prior work, ATM'22 \cite{zhang2023multi}. Two encoder-decoder architectures were adopted for binary airway modeling, 
optimized using topology-preserved objective functions and intra-class discrimination sample solution \cite{zhang2023towards, zheng2021alleviating, yu2022break} (Methods).
In the second stage, AirMorph constructs a branch-wise graph representation to support subsequent anatomical labeling. Specifically, a minimum path-cost tree extraction method (MPC-Skel) is employed to derive a robust skeleton 
from the binary airway segmentation. MPC-Skel effectively suppresses spurious branches generated by conventional methods and alleviates the burden of manual correction \cite{lee1994building, shit2021cldice, menten2023skeletonization}, 
as demonstrated in Fig. \ref{fig::comparison_graph_building} and Table \ref{tab::AirMorph_cls_generalization_on_pulmonary_fibrosis}. Bifurcation and trifurcation points are then identified by analyzing the 26-connected local 
neighborhood of each skeleton voxel, voxels with more than three connected neighbors are classified as branching points. This enables the binary airway tree to be partitioned into discrete structural branches yet not anatomically assigned, 
which serve as the basis for subsequent anatomical labeling. On average, the binary airway tree is partitioned into 266 discrete branches ($266.4 \pm 73.5$) across all datasets.
After obtaining the partitioned structural branches, the binary airway is abstracted into a graph representation, where each graph node corresponds to an individual branch.
To assign anatomical labels to these branches, eleven features are computed for each node (Table \ref{tab::AirMorph_manual_feature_definition}), comprising one structural feature, three positional features, 
and seven morphological features (feature definition in Table \ref{tab::AirMorph_manual_feature_definition}). The structural feature encodes the generation of the branch, with the trachea defined as generation one and the generation number increasing with each bifurcation.
The three positional features capture the average relative location of each branch along the x-, y-, and z-axes with respect to the trachea.
The seven morphological features describe the geometric properties of each branch, including both length and orientation. Specifically, the angle features quantify the intersection angles between the branch vector and the x-, y-, and z-axes.
The length features include the geodesic distance between the branch endpoints as well as the projected lengths along each axis, calculated based on the branch orientation.
The aforementioned graph node-wise features are treated as tokens and fed into Transformer blocks to model cross-attention among branches, ultimately enabling anatomical label assignment for each token.
To further improve labeling consistency, hierarchical anatomical relationships are incorporated into a U-shaped Transformer-based framework, which assigns labels in a sequential manner—progressing from lobar 
to segmental and finally to subsegmental levels. Moreover, to mitigate the substantial inter-individual variability, a soft-subtree consistency module is introduced to encapsulate clinically meaningful anatomical information 
by organizing segmental branches into subtrees, enhancing the structural consistency across varying cases.

AirMorph is the first robust framework to generate hierarchical anatomical airway structures from CT scans in a fully automated, end-to-end manner. Implementation details are provided in the Methods section.

\numberedsubsection{AirMorph: Unified and Fine-grained Evaluation}
A robust evaluation of pulmonary airway modeling must address both anatomical completeness and clinical interpretability, especially in the subsegmental regions most relevant to diagnosis and intervention. 
To this end, we establish a unified evaluation framework for AirMorph that systematically quantifies performance of both the topological fidelity of the binary airway modeling and the accuracy of the fine-grained airway anatomical labeling. 

\noindent \textbf{Internal unified and fine-grained evaluation.}
We first evaluate AirMorph on the internal primary dataset with complete subsegmental-level annotations 
(Fig. \ref{fig::finegrained_evaluation_AirMorph}, Table \ref{tab::finegrained_evaluation_seg_and_labeling}, 
Fig. \ref{fig::finegrained_evaluation_seg}, Fig. \ref{fig::comparison_airway_seg_quantative}, 
Fig. \ref{fig::comparison_graph_building}, Fig. \ref{fig::comparison_airway_cls_quantative}, Table \ref{tab::AirMorph_seg_generalization_on_pulmonary_fibrosis}, 
Table \ref{tab::AirMorph_cls_generalization_on_pulmonary_fibrosis}, Table \ref{tab::AirMorph_graph_building}
). Performance is assessed at multiple anatomical levels for both binary airway modeling and anatomical branch labeling.
The evaluation incorporates a combination of topology-aware metrics and commonly used accuracy measures.
Specifically, the tree length detected rate (TLD, \%) and branch number detected rate (BND, \%) are employed to evaluate the integrity of the reconstructed airway 
topology \cite{zhang2023multi}, while the predicted subtree consistency (TreeCons, \%) and topological distance (TopoDist) quantify the topological accuracy of anatomical branching.
For binary airway modeling, additional metrics include the Dice similarity coefficient (DSC), centerline Dice (clDice), and sensitivity.
For anatomical branch labeling, accuracy, precision, and sensitivity are used to evaluate classification performance. 
Across all anatomical levels, AirMorph consistently outperforms established comparative methods, including UNet \cite{ronneberger2015u}, nnUNet \cite{isensee2021nnu}, GCN \cite{kipf2017semi}, and TNN \cite{yu2022tnn}.  
At the lobar level, AirMorph achieves near-saturation performance, with a tree length detected rate (TLD) of 92.27\%, branch number detected rate (BND) of 96.58\%, and clDice of 89.35\% (Table~\ref{tab::finegrained_evaluation_seg_and_labeling}).  
At the segmental level, performance remains high---exceeding 92\% TLD, 94\% BND, and 88\% clDice---while UNet drops to below 60\% TLD, 62\% BND, and 71\% clDice. 
Compared with nnUNet, AirMorph shows a substantial margin of improvement, surpassing it by over 20\% in TLD, 15\% in BND, and 5\% in clDice.
Importantly, at the subsegmental level---where airways are most numerous, thinnest, and clinically relevant---AirMorph still maintains strong performance, 
achieving 85.56\% TLD, 89.15\% BND, and 81.19\% clDice, demonstrating its unique capability to preserve fine terminal branches with high topological fidelity.
Anatomical labeling performance demonstrates comparably strong results. 
At the subsegmental level, AirMorph achieves a classification accuracy of 94.12\% and a TreeCons of 98.02\%, 
indicating its ability to assign correct anatomical labels while preserving bronchial topology. 
In contrast, TNN and GCN achieve substantially lower accuracies of 69.61\% and 63.89\%, respectively. 
TopoDist, which reflects misalignment between semantic labels and structural topology, 
is also the lowest for AirMorph (0.48 vs.\ > 2.2 for comparative methods), confirming its ability to achieve fine-grained anatomical localization 
without compromising topological consistency. 

A vivid qualitative illustration of the unified evaluation framework is presented in Fig. \ref{fig::AirMorph_MainDigram}.b. 
When evaluated using the semantic graph node representation, branch-level detection failures in UNet and nnUNet underscore their limited ability 
to preserve the topological integrity of the airway tree, in contrast to the superior performance of AirMorph.
Meanwhile, misclassifications in anatomical labeling by GCN and TNN are also evident in the graph-based representation. 
Compared with these methods, AirMorph achieves both the most complete topological reconstruction and the most accurate anatomical classification.
AirMorph also demonstrates strong performance in fine-grained evaluation (Fig. \ref{fig::finegrained_evaluation_AirMorph}.a--d and Fig. \ref{fig::finegrained_evaluation_seg}). 
Fig. \ref{fig::finegrained_evaluation_seg} supplements the TLD evaluation across subsegmental bronchi, where AirMorph consistently outperforms both UNet and nnUNet on every individual branch. 
In particularly challenging regions, the improvement is substantial. For example, AirMorph achieves over 30\% higher TLD on $RB^{6^{a}}$ and over 40\% on $RB^{8^{a}}$ compared with UNet and nnUNet 
(TLD on $RB^{6^{a}}$: UNet: 30.0\%, nnUNet: 49.6\%, AirMorph: 77.7\%, $p < 0.01$; TLD on $RB^{8^{a}}$: UNet: 18.5\%, nnUNet: 27.0\%, AirMorph: 67.7\%, $p < 0.01$).
Fine-grained evaluation is further demonstrated through graphnode-wise anatomical labeling (Fig. \ref{fig::finegrained_evaluation_AirMorph}.c--d). 
Fig. \ref{fig::finegrained_evaluation_AirMorph}.c visualizes labeling accuracy using an abstract graph representation, where each node corresponds to an individual airway branch 
and edges indicate anatomical connectivity. Node color reflects the ground-truth category. It is clearly observed that AirMorph produces the most accurate graph-level labeling compared to GCN and TNN.
In contrast, GCN and TNN tend to misclassify a substantial number of nodes, including erroneous assignments within the same anatomical category or complete mislabeling of subtrees. 
For example, TNN incorrectly predicts the entire $LB^{3}$ region (Fig. \ref{fig::finegrained_evaluation_AirMorph}.c1). 
The fine-grained evaluation is further illustrated by the confusion matrix in Fig. \ref{fig::finegrained_evaluation_AirMorph}.e. 
Misclassifications predominantly occur within intra-segmental classes or between anatomically adjacent subsegments. 
Overall, AirMorph exhibits the lowest risk of semantic-structural misalignment among all comparative methods.
For instance, in the $RB^6$ region, misclassification is particularly severe for GCN and TNN. 
The classification accuracy of $RB^{6^{c}}$ is only 21.2\% for GCN and 61.4\% for TNN, whereas AirMorph achieves 89.0\%. 
Additionally, GCN fails to detect the $RB^{6^{b+c}}$ co-trunk branch, and TNN predicts it correctly in only 16.7\% of cases. 
In contrast, AirMorph achieves a precision of 60.0\% on this difficult co-trunk branch.
As another example of intra-segmental misclassification, for the subsegmental branch $LB^{5^{a}}$, GCN incorrectly assigns 47 instances to $LB^{4^{a}}$ and 62 to $LB^{4^{b}}$, 
resulting in a precision of only 16.8\%. In comparison, AirMorph attains a precision of 71.2\% for $LB^{5^{a}}$, with only 4 and 5 false classifications to $LB^{4^{a}}$ and $LB^{4^{b}}$, respectively.

\noindent \textbf{Robustness under disease-induced variations.}
Further, AirMorph demonstrates robustness in modeling airway anatomical structures in the presence of pulmonary diseases, which often introduce imaging artifacts and 
morphological alterations to the airway tree \cite{walsh2018deep, walsh2015relationship}. 
To assess its performance under such challenging conditions, a cohort of patients with pulmonary fibrosis ($n = 120$) was included for internal testing (Fig.~\ref{fig::data}). 
Patients with pulmonary fibrosis often exhibit a characteristic honeycombing pattern, characterized by the presence of small cystic airspaces 
in the peripheral regions of the lung. These structures may appear visually similar to bronchial airways on CT scans, thereby increasing the complexity of airway modeling. 
In addition, fibrotic lungs frequently present with bronchiectasis, which introduces significant morphological alterations to the airway tree and further complicates the task of accurate anatomical labeling.
UNet and nnUNet suffer a substantial performance decline on patients with pulmonary fibrosis, detecting less than 50\% of the airway structure in both tree length detected (TLD) and branch number detected (BND) metrics 
across all anatomical levels (Table \ref{tab::AirMorph_seg_generalization_on_pulmonary_fibrosis}). In contrast, AirMorph demonstrates strong generalization performance, 
achieving 78.49\% TLD and 81.49\% BND at the lobar level, 73.48\% TLD and 76.80\% BND at the segmental level, and 66.46\% TLD and 70.88\% BND at the subsegmental level. 
Furthermore, for voxel-wise evaluation, AirMorph outperforms both UNet and nnUNet by over 10 percentage points on the most fine-grained branches. 
At the subsegmental level, AirMorph achieves a DSC of 57.63\% compared to 42.43\% for UNet and 44.15\% for nnUNet ($p < 0.01$). 
Similarly, AirMorph obtains a clDice score of 61.94\%, surpassing UNet (43.70\%) and nnUNet (46.41\%) with statistical significance ($p < 0.01$).
These complementary advantages highlight the robustness and superior generalization ability of AirMorph under disease-induced imaging and structural variations.
Similar statistical trends are observed in the anatomical labeling task on patients with pulmonary fibrosis (Table \ref{tab::AirMorph_seg_generalization_on_pulmonary_fibrosis}). 
AirMorph maintains high labeling precision across all anatomical levels, achieving 97.04\%, 88.69\%, and 80.59\% at the lobar, segmental, and subsegmental levels, respectively.
In contrast, GCN and TNN experience severe performance degradation, particularly at finer levels of the airway hierarchy. 
At the segmental level, GCN and TNN achieve 60.25\% and 69.35\% precision, respectively, while at the subsegmental level, their performance drops to 33.85\% and 53.39\%.
AirMorph not only achieves high classification accuracy, but also preserves topological consistency under disease-induced variations. 
At the segmental level, it achieves a TreeCons of 97.26\%, significantly outperforming TNN (81.34\%) and GCN (55.33\%) ($p < 0.01$). 
TopoDist further confirms the advantage, with AirMorph scoring 0.6198 compared to 1.1834 for TNN and substantially higher values for GCN ($p < 0.01$).
The qualitative results further demonstrate the effectiveness of AirMorph (Fig. \ref{fig::comparison_airway_seg_quantative}, Fig. \ref{fig::comparison_airway_cls_quantative}). 
AirMorph preserves a more complete airway structure without compromising voxel-level segmentation accuracy (Fig. \ref{fig::comparison_airway_seg_quantative}).
Notably, in regions with pronounced morphological alterations—such as the left lower lobe in Case 3 and the right lower lobe in Case 4, AirMorph successfully reconstructs 
the full airway architecture, whereas comparative methods fail to capture these challenging subtree structures.

\noindent \textbf{Generalization to external multi-site test cohorts.} AirMorph is further evaluated on large external test cohorts from LIDC-IDRI and five sub-cohorts of NLST (D54, D58, D59, D61), which include over 2,400 CT scans 
covering various disease types. Despite considerable variations in scan resolution, disease-induced distortion, and inter-institutional heterogeneity, AirMorph demonstrates strong generalization capacity across all cohorts 
(Fig. \ref{fig::finegrained_evaluation_AirMorph}.e-h, Fig. \ref{fig::comparison_airway_seg_qualitative}, 
Fig. \ref{fig::comparison_airway_cls_qualitative}, Table \ref{tab::AirMorph_qualitative_seg}, Table \ref{tab::AirMorph_qualitative_cls}). 
In binary airway modeling, AirMorph consistently outperforms UNet and nnUNet across all anatomical levels, from lobar to subsegmental branches (Table~\ref{tab::AirMorph_qualitative_seg}). 
We adopted two qualitative metrics to assess performance: the total reconstructed tree length ($mm$) and the number of reconstructed airway branches.
AirMorph achieves airway reconstructions of 2876.13 $mm$, 3078.86 $mm$, 2890.10 $mm$, 3668.33 $mm$, and 3643.27 $mm$ on the LIDC-IDRI and NLST (D54, D58, D59, D61) datasets, respectively. 
These values consistently exceed those of UNet and nnUNet by over 1000~mm in total tree length.
In addition, AirMorph reconstructs approximately 60\% more airway branches compared to the comparative methods. 
This improvement reflects the model's ability to detect a greater number of distinct airway branches, rather than merely overextending existing ones, 
demonstrating its effectiveness in capturing the full extent of the bronchial structure. 
Fig. \ref{fig::comparison_airway_seg_qualitative} further confirms the effectiveness of the AirMorph, where surrounding the lesion regions, AirMorph reconstructs more complete and clinical valuable branches than other methods.
For the subsequent airway anatomical labeling, we propose the use of TreeCons and TopoDist as automated metrics to evaluate topological consistency and semantic alignment.
Fig ~\ref{fig::comparison_airway_cls_qualitative} presents a qualitative comparison of anatomical labeling results. 
AirMorph produces highly consistent predictions within segmental and subsegmental classes. 
Following a clinical reader study, the majority of these high-consistency results generated by AirMorph were confirmed to be anatomically correct, highlighting the reliability of the model in complex anatomical regions.
The statistical analysis (Fig. \ref{fig::finegrained_evaluation_AirMorph}.e-h) on the external multi-site test cohorts demonstrates the significant advantages of AirMorph in both anatomical labeling accuracy and topological consistency.

\numberedsubsection{AirMorph: Efficient Branching Pattern Analysis}
\noindent \textbf{Efficient automated pattern analysis}.
AirMorph enables fully automated and efficient analysis of bronchial branching patterns across lobar, segmental, and subsegmental levels, 
validated on a multi-center cohort of over 3000 CT scans. Unlike prior studies limited to manual annotation or partial airway regions \cite{nagashima2015analysis, maki2022pulmonary, he2022anatomical, wang2018variations}, 
Co-trunk relationships between segments were determined based on the topological structure of the airway graph. Segments that satisfied the co-trunk condition were merged using union-find, yielding the final segment-level branching patterns.
AirMorph systematically characterizes both intra- and inter-segmental branching variations through its detailed anatomical graph representation (Fig. \ref{fig::branch_pattern_with_endo_view}, Methods).
Notably, AirMorph efficiently integrates new cases without requiring manual verification of the binary airway tree, thereby enhancing the efficiency and scalability of large-scale anatomical studies of the bronchial tree.
The average branching patterns observed across the multi-center cohorts are summarized in Table \ref{tab::statis_branch_pattern}. While the statistical distributions of specific configurations exhibit some variations between cohorts, 
a consistent overarching trend emerges. For instance, the $LB^{1+2}$ branch branching pattern of the 1-stem, bifurcation, a+b co-trunk type (($LB^{1+2^{a+b}}$, $LB^{1+2^{c}}$)) occupies 69.95\% in average. Among the separate cohorts, this branching pattern proportion 
is 72.45\% in ATM'22, 65.00\% in AIIB'23, 67.89\% in LIDC-IDRI, 66.52\% in NLST-D54, 76.82\% in NLST-D58, 72.25\% in NLST-D59, 68.71\% in NLST-D61 (Table \ref{tab::statis_branch_pattern_ATM22} - \ref{tab::statis_branch_pattern_NLST-D61}). 

\noindent \textbf{Large cohorts evaluation and clinical findings}.
The standard bronchial nomenclature system \cite{netter2014atlas} fails to fully capture the inherent diversity of subsegmental bronchial branching patterns. 
In contrast, AirMorph enumerates all anatomically plausible subsegmental configurations and branching types, providing a comprehensive, predefined taxonomy. 
Following rigorous validation across large-scale cohorts, the most complete statistical representation of these branching patterns is summarized in Table \ref{tab::statis_branch_pattern}. 
Consistent with the classical nomenclature, no instances of $LB^{4^{c}}$ or $LB^{5^{c}}$ branches were observed in our dataset. Furthermore, AirMorph confirms the non-existence of certain 
hypothetical branching variants: specifically, $LB^{1+2}$ never exhibits a two-stem bifurcation with an a+c co-trunk configuration, and $LB^{10}$ does not display the two-stem, a+b co-trunk bifurcation pattern.

Compared to prior works \cite{nagashima2015analysis, maki2022pulmonary, he2022anatomical, wang2018variations}, 
AirMorph achieves consistent branching pattern results with previous studies that focused on selected airway regions. 
Moreover, AirMorph provides a comprehensive analysis across the entire airway. At the segmental level, we evaluated branching within each lobe by categorizing patterns 
based on the number of branches (e.g., bifurcation, trifurcation, etc.). At the subsegmental level, we further classified branching within each segment by first distinguishing 
between single-stem and dual-stem structures, and then by subsequent bifurcation, trifurcation, or higher-order divisions. This dual-level approach facilitates a more precise and consistent evaluation of bronchial branching.
Fig. \ref{fig::branch_pattern_with_endo_view}.b illustrates the intra-subsegmental branching pattern of LB1+2, where the most common configuration (69.95\% of cases) features a single stem that bifurcates into $LB^{1+2^{a+b}}$ and $LB^{1+2^{c}}$. 
AirMorph eliminates the need for manual verification in branching pattern statistics, reducing the analysis time from hours to seconds per case. 
This high degree of automation demonstrates strong scalability for real-time analysis in newly incoming patients.
Furthermore, AirMorph enables rule-based branching pattern analysis, extending beyond intra-segmental and intra-subsegmental configurations to include inter-segmental co-trunk structures. 
Representative inter-segmental branching variants were successfully identified and visualized (Fig. \ref{fig::branch_pattern_with_endo_view}.c,d; see Methods).
Within the inter-segmental group, the independent pattern is the most common, characterized by the absence of subsegmental arising from a common stem shared with another segment.  
Conversely, other inter-segmental variants demonstrate more complex branching. For instance, in 9.21\% of the cases, the lingular division bronchus first gives off B4a before bifurcating into $LB^{4^{b}}$ and $LB^{5}$.
Additionally, a variant pattern was observed in 10.84\% of cases within the right upper lobe, where the superior division bronchus divides into $RB^{1^{a}}+RB^{2}$ and $RB^{2^{a}}+RB^{3}$.

\noindent \textbf{Prospective clinical utility of branching pattern analysis.}
The systematic characterization of airway branching patterns is essential - not only for precise bronchial identification and reliable automated annotation \cite{mori2005method}, 
but also to support critical clinical decisions in bronchoscopy and pulmonary segmental resection \cite{kanzaki2013complete, wu2016thoracoscopic}. 
As for the endoscopic surgical planning, the endoscopic structures are complex and share high self-similarity, the detailed knowledge of patient-specific bronchial anatomy can enhance procedural precision and safety. 
Furthermore, the integration of detailed endoscopic views and local airway patches (Fig. \ref{fig::branch_pattern_with_endo_view}.b, Fig. \ref{fig::branching_pattern_with_endo_view_supplementary_part1} - \ref{fig::branching_pattern_with_endo_view_supplementary_part3}) 
provides crucial visual context, supporting more effective patient-wise navigation during bronchoscopic procedures. These capabilities underscore the prospective clinical utility of AirMorph, potentially improving diagnostic accuracy and therapeutic outcomes through enhanced anatomical understanding and visualization.

\numberedsubsection{AirMorph: Anatomical Signatures with Clinical Relevance}
\noindent \textbf{Establishment of the AirwaySignature.} 
AirMorph enables the extraction of hierarchical airway anatomical structures aligned with volumetric CT images, facilitating detailed analysis of airway morphology and image-derived structural variations. 
Such anatomical parsing is particularly valuable for identifying disease-induced alterations that often manifest in localized regions of the airway. Traditional binary airway modeling, however, lacks the resolution to reveal which specific 
branches are most affected, limiting its clinical interpretability and diagnostic utility.
To address this limitation, AirMorph provides precise anatomical labeling down to the subsegmental level, covering up various kinds of distinct bronchial types per patient. While this high-resolution 
labeling greatly enriches anatomical understanding, it also introduces a substantial volume of information. In clinical practice, this can increase the cognitive and computational burden on radiologists, 
who must manually rotate 3D reconstructions and scroll through stacks of axial slices to identify suspicious branches—often without an intuitive way to compare multiple abnormal sites simultaneously.
To overcome these challenges, we introduce \textbf{AirwaySignature}, a compact, anatomically aligned representation that encodes multi-level morphological characteristics of each bronchial branch 
(Fig.~\ref{fig::AirMorph_morpho_distribution_presentation}.a, Methods). This signature provides a quantifiable and interpretable abstraction of the airway tree, 
enabling efficient localization of structural abnormalities and facilitating downstream analysis, visualization, and disease interpretation. 
Specifically, AirwaySignature defines six morphological descriptors that capture both local and global anatomical features of individual bronchi, based on the fine-grained labeling results provided by AirMorph. 
These descriptors are designed to quantify pathological and structural variations. The six descriptors can be broadly categorized into two groups. 
The first group consists of Stenosis ($\mathcal{S}$), Ectasia ($\mathcal{E}$), and Tortuosity ($\mathcal{T}$), 
which are computed at the level of individual semantic branches. $\mathcal{S}$ quantifies the maximal degree of narrowing along a bronchial branch, measured as the greatest percentage reduction in radius relative to its proximal reference.
$\mathcal{E}$ serves as a complementary descriptor, capturing the maximal dilation ratio compared to the expected normal radius, thus reflecting airway enlargement or ectatic changes.
$\mathcal{T}$ measures the curvature-induced deformation of a bronchial branch based on its 3D trajectory, providing a geometric estimate of local bending and undulation.
The second group consists of Length ($\mathcal{L}$), Divergence ($\mathcal{D}$), and Complexity ($\mathcal{C}$), which aggregate information across semantically adjacent branches under a shared anatomical context. 
$\mathcal{L}$ represents the geodesic length of airway subtrees rooted at a common ancestor, capturing elongation or contraction across multiple connected branches.
$\mathcal{D}$ quantifies the spatial spread of descendant branches by computing the minimal enclosing cone angle, thereby characterizing the divergence pattern of airway branching.
$\mathcal{C}$ leverages fractal dimension analysis to assess the spatial irregularity and local branching density, offering a compact estimate of geometric complexity.
Together, these six morphological signatures provide a comprehensive representation of airway structure that extends beyond the capabilities of binary airway segmentation.
Moreover, based on the hierarchical anatomical labeling provided by AirMorph, each descriptor can be efficiently grouped and analyzed at the lobar, segmental, or subsegmental levels, enabling both localized and global morphological profiling.

\noindent \textbf{Evaluation of AirwaySignature on pulmonary diseases.}
To evaluate the effectiveness of AirwaySignature in capturing disease-induced airway alterations, we performed a large-scale analysis across five distinct pulmonary disease types: 
pulmonary fibrosis, emphysema, atelectasis, consolidation, and reticular opacities.The results demonstrate that AirwaySignature provides robust and interpretable indicators of pathological airway changes associated with these diseases.
(Fig. \ref{fig::AirMorph_morpho_distribution_presentation}, Fig. \ref{fig::AirMorph_morpho_distribution_within_controlgroup}, Fig. \ref{fig::AirMorph_morpho_distribution_of_D54Stenosis}, 
Fig. \ref{fig::AirMorph_morpho_heatmapwithimge_all_five_diseases}). 
As a prerequisite, we first assessed the intra-group stability of AirwaySignature in the healthy population. The control cohort was randomly divided into subgroups, 
and the six morphological descriptors were statistically analyzed across all lobar and segmental levels. The resulting distributions showed no significant intra-group differences (Fig. \ref{fig::AirMorph_morpho_distribution_within_controlgroup}), 
thereby establishing a consistent reference profile for normal airway morphology. These reference distributions were subsequently used to quantify deviations in disease cohorts and to detect statistically significant 
abnormalities associated with specific pathological conditions (Table \ref{tab::AIIB23_Morpho_Quantative_Result} - \ref{tab::NLSTD61_Morpho_Quantative_Result}). 
Marked differences in airway morphological distributions can be observed when comparing patients with pulmonary diseases to healthy controls (Fig. \ref{fig::AirMorph_morpho_distribution_presentation}.c).
For instance, in cases of pulmonary atelectasis, Stenosis for segmental branch $LB^{10}$ shows a significantly lower value ($0.23 \pm 0.06$) compared to the normal reference distribution 
($0.30 \pm 0.06$, $p < 0.01$), indicating pronounced bronchial narrowing (Fig. \ref{fig::AirMorph_morpho_distribution_of_D54Stenosis}).
As for the patients with pulmonary fibrosis, Stenosi of the detailed branch $LB^{3}$ is $0.27 \pm 0.16$, compared with the normal reference distribution ($0.27 \pm 0.06$, $p < 0.01$) shows a large variation, 
similar findings exist in the Ectasia of the lobar branch $LUB$, which is $1.28 \pm 0.25$ (normal: $1.29\pm 0.03$, $p < 0.01$, Table \ref{tab::AIIB23_Morpho_Quantative_Result}). 
In patients with pulmonary fibrosis, the stenosis descriptor for branch $LB^{3}$ exhibits substantial variability, with a distribution of $0.27 \pm 0.16$ compared to the reference value of $0.27 \pm 0.06$ ($p < 0.01$). 
Although the mean values are comparable, the increased standard deviation reflects heterogeneous airway narrowing, likely due to fibrotic traction and architectural distortion.
Similarly, the ectasia signature for the lobar branch $LUB$ shows a distribution of $1.28 \pm 0.25$ in the fibrosis cohort, compared to $1.29 \pm 0.03$ in healthy controls ($p < 0.01$), 
indicating localized airway dilation with greater inter-patient variability (Table~\ref{tab::AIIB23_Morpho_Quantative_Result}). 
The Length and Complexity descriptors do not show completely statistically significant differences in patients with pulmonary fibrosis, likely due to compensatory traction effects on the bronchial tree.
For instance, the airway length of lobar branches in the fibrosis cohort remains comparable to that in healthy controls across all five lobes: $LUB$: $47.55 \pm 9.98$ mm (vs. $45.33 \pm 5.87$ mm, $p = 0.13$);
$LLB$: $49.56 \pm 9.42$ mm (vs. $49.70 \pm 7.19$ mm, $p = 0.68$); $RUB$: $50.82 \pm 8.00$ mm (vs. $45.26 \pm 6.67$ mm, $p = 0.63$); $RMB$: $53.31 \pm 9.11$ mm (vs. $50.91 \pm 7.58$ mm, $p = 0.81$);
$RLB$: $46.29 \pm 7.93$ mm (vs. $50.10 \pm 6.24$ mm, $p = 0.23$).
These results suggest that, despite architectural distortion, airway elongation remains largely preserved in pulmonary fibrosis. 
In contrast, the length signature shows a statistically significant reduction in patients with pulmonary emphysema (Table \ref{tab::NLSTD59_Morpho_Quantative_Result}).
For example, the branch $RB^{5}$ exhibits a markedly shorter airway length in emphysema patients ($45.00 \pm 11.71$ mm) compared to healthy controls ($53.31 \pm 9.37$ mm), with a significance level of $p < 0.01$. This finding likely reflects distal airway destruction and alveolar collapse.
Beyond morphological quantification, AirwaySignature also serves as an effective auxiliary tool for automated anomaly detection. The spatial distribution of abnormal signature values aligns well with lesion regions, 
enabling the identification of locally affected airway branches and providing branch-level morphological descriptors that are both interpretable and quantitative.
As illustrated in Fig.\ref{fig::AirMorph_morpho_heatmapwithimge_all_five_diseases}.c, AirwaySignature highlights pronounced abnormalities in the right middle lobe associated with pulmonary consolidation. 
Similarly, Fig.\ref{fig::AirMorph_morpho_heatmapwithimge_all_five_diseases}.e reveals abnormalities localized to the lower lobes of both lungs in a case with reticular opacities. 
These automated and anatomically aligned visual cues, derived from AirMorph, assist clinicians in efficiently identifying regions of interest, thereby enhancing diagnostic interpretability and clinical utility.

\noindent \textbf{Prospective clinical utility of AirwaySignature.}
AirwaySignature offers a clinically meaningful abstraction of complex airway morphology by condensing high-resolution anatomical labeling into a compact, structured representation. 
This branch-aligned signature enables rapid identification and localization of disease-associated abnormalities, thereby supporting efficient region-of-interest selection in diagnostic workflows. 
Unlike voxel-level segmentation maps that require full 3D observation, AirwaySignature directly highlights structurally abnormal branches, facilitating fast screening of pathological regions such as distal bronchial narrowing or lobe-specific remodeling.
Moreover, by aggregating multi-dimensional geometric features—such as stenosis, ectasia, and tortuosity—AirwaySignature enables systematic characterization of disease-specific airway phenotypes. 
This facilitates comparative morphology studies across patient cohorts, offering quantitative insight into how different pulmonary diseases differentially affect the airway tree (Fig. \ref{fig::AirMorph_MainDigram}.e).
Finally, AirwaySignature can serve as a selective filter for radiomic analysis. By identifying branches with significant structural deviations, radiomic features can be extracted from anatomically and pathophysiologically relevant regions (Fig. \ref{fig::AirMorph_full_features_with_radiomics}, Methods). 
This targeted integration not only improves interpretability but also enhances the robustness and discriminative power of radiomics-based predictive models.
  
\numberedsection{Discussion}
In this study, we introduced AirMorph, an end-to-end deep learning framework designed for automatic and comprehensive anatomical labeling of pulmonary airway structures from thoracic CT scans, achieving unprecedented granularity at lobar, segmental, and subsegmental levels. The developed AirwaySign signature encapsulates diverse airway morphological features, demonstrating considerable clinical potential for distinguishing and analyzing various pulmonary diseases.

One of the major achievements of AirMorph is its ability to consistently outperform existing segmentation methods across all anatomical levels. This robust performance, validated across large-scale multi-center datasets, highlights its capability to preserve topological integrity even at the subsegmental level, an area historically challenging due to the complexity and variability of bronchial anatomy. Compared to previous models such as UNet and nnUNet, AirMorph exhibits significant improvements in tree length detection, branch number accuracy, and anatomical labeling precision, indicating a substantial advancement in automated airway modeling. The robust graph representation constructed by AirMorph is essential for reliable extraction of structural, positional, and morphological features (Table \ref{tab::AirMorph_graph_building}, Fig. \ref{fig::comparison_graph_building}). Compared to skeletonization \cite{lee1994building}, soft skeleton \cite{shit2021cldice}, and gradient-based soft skeletonization \cite{menten2023skeletonization}, our MPC-Skel method produces a topologically faithful airway skeleton with $\beta_{0} = 1$ and $\beta_{1} = 0$. While conventional skeletonization can also yield $\beta_{0} = 1$, it introduces spurious spikes and detrimental loops (Fig. \ref{fig::comparison_graph_building}), resulting in an average $\beta_{1}$ of 16.55. Soft-skeleton methods further degrade topological integrity, generating large volumes of false-positive branches and failing to preserve the correct airway structure.
These inaccurate skeleton representations hinder the reliable identification of bifurcation and trifurcation points, which are necessary for partitioning the airway tree into discrete structural branches used in the graph representation. 
In contrast, AirMorph overcomes these challenges by leveraging MPC-Skel to accurately partition the binary airway into clean, independent structural branches, forming a robust foundation for subsequent anatomical labeling (Table \ref{tab::AirMorph_graph_building_ablation_study_on_ATM22}, Table \ref{tab::AirMorph_graph_building_ablation_study_on_AIIB23}). 

The clinical utility of AirMorph extends beyond accurate anatomical reconstruction. The automated branching pattern analysis it facilitates enables rapid, scalable statistical assessments of bronchial structures, eliminating extensive manual annotation requirements. This functionality is essential not only for detailed anatomical studies but also significantly benefits clinical procedures like bronchoscopic navigation and segmental resection planning. Moreover, by systematically characterizing intra- and inter-segmental branching variations, AirMorph provides valuable anatomical insights that could enhance procedural safety and precision.

AirwaySignature, as an anatomical signature derived from the detailed airway modeling provided by AirMorph, represents a major innovation with substantial clinical implications. By quantifying six morphological descriptors—stenosis, ectasia, tortuosity, divergence, length, and complexity—AirwaySignature facilitates rapid and intuitive identification of pathological airway changes. Our analysis demonstrates distinct morphological profiles across various pulmonary conditions, including pulmonary fibrosis, emphysema, atelectasis, consolidation, and reticular opacities. Such quantifiable morphological markers can significantly enhance diagnostic accuracy, enabling clinicians to quickly identify affected airway regions and potentially improving patient outcomes through more targeted therapeutic interventions.

By complementing cellular atlases, the morphological AirMorph has the potential to bridge the gap between high-resolution anatomical morphology and cellular-level heterogeneity, thereby enabling a more comprehensive and integrative understanding of pulmonary diseases. Such an integrated approach not only enhances diagnostic accuracy and clinical interpretability but also supports targeted interventions and personalized therapeutic strategies.

Despite these promising results, there remain several limitations to our study. While AirMorph demonstrates robust performance across diverse conditions and datasets, variations in image quality, disease severity, and anatomical anomalies could still impact model performance. Future studies should focus on further improving the adaptability of AirMorph, particularly in cases with severe disease-induced morphological distortions. Additionally, prospective clinical validation involving real-time bronchoscopic navigation and therapeutic planning will be essential to fully assess the practical utility and integration potential of AirMorph and AirwaySignature into clinical workflows.

In conclusion, AirMorph represents a significant step forward in automated pulmonary airway analysis, providing comprehensive and clinically relevant anatomical labeling. Coupled with the AirwaySignature, it offers a powerful tool for the precise characterization and interpretation of airway pathology, paving the way for improved clinical diagnostics and targeted therapeutic interventions in pulmonary medicine.

\numberedsection{Methods}
\numberedsubsection{Detailed Dataset Information and Data Processing}
\noindent \textbf{Primary Dataset.} 
The primary dataset extends both the ATM'22 dataset \cite{zhang2023multi} and the AIIB'23 dataset \cite{nan2024hunting} by enriching the binary airway annotations with multi-level semantic labels, 
covering lobar, segmental, and subsegmental anatomy.
In ATM'22, each chest CT scan has an axial resolution of 512 $\times$ 512 pixels and a spatial resolution ranging from 0.500 to 0.919 mm, with a slice thickness between 0.450 and 1.000 mm. 
Similarly, the AIIB'23 dataset comprises cases with more than 120 slices per scan, each exceeding 512 $\times$ 512 pixels in-plane resolution, and voxel spacing between 0.417 and 0.926 mm. 
The number of slices ranges from 146 to 947, with slice thicknesses between 0.400 and 2.000 mm.

\noindent \textbf{LIDC-IDRI.} 
The Lung Image Database Consortium and Image Database Resource Initiative (LIDC-IDRI) \cite{armato2011lung, armato2015data} is a widely used public chest CT dataset for lung image analysis. 
It contains 1,018 thoracic CT scans, each accompanied by an XML file documenting detailed nodule annotations. Slice thicknesses range from 0.6 to 3.0 mm.
ATM'22 previously included 140 scans from LIDC-IDRI. For this study, we selected an additional 800 high-quality scans from the remaining LIDC-IDRI cases. 
All selected scans were resampled to a uniform slice thickness of 0.625 mm for consistency.

\noindent \textbf{NLST Trial.} 
National Lung Screening Trial(NLST) \citep{nlst2013data} is a project conducted to determine the relationships between low-dose CT screening and lung cancer. 
4 types of pulmonary abnormalities are included in this research.
These types are as follows: 1) NLST-D54: Atelectasis, segmental or greater. 2) NLST-D58: Consolidation. 3) NLST-D59: Emphysema. 4) NLST-D61: Reticular/reticulonodular opacities.
\noindent \textbf{NLST-D54.} Atelectasis is a partial or complete collapse of the lung \citep{bankier2024fleischner}. D54 refers to atelectasis that occurs at the level of an anatomical segment or lobe. 
Atelectasis is often associated with abnormal bronchial and/or trachea displacement \citep{woodring1996types}. For a study case diagnosed with D54 in NLST dataset, it is highly possible that its anatomical structure shifts away from healthy cases. 
\noindent \textbf{NLST-D54.} Consolidation is a descriptor referring to replacement of air in one or more acini by fluid or other solid material \citep{rubens1990diseases, bankier2024fleischner}. 
One of the common causes of consolidation is pneumonia \citep{mojoli2019lung}, which potentially leads to structural changes on the airways \citep{peng2019neonatal, hsieh2023airway}.
\noindent \textbf{NLST-D59.} Emphysema is characterized by irreversible enlarged airspaces \citep{bankier2024fleischner}. It destructs the pulmonary lobules and can diffuse across the lung parenchyma. 
Severe emphysema can compress the airway, reduce total bronchial areas, potentially leading to Chronic Obstructive Pulmonary Disease (COPD)\citep{diaz2013effect, gallardo2016normalizing, bankier2024fleischner}. 
\noindent \textbf{NLST-D61.} D61 includes a variety of abnormalities. Fibrosis refers to a repair mechanism of the lung in which parenchyma is irreversibly replaced by connective tissue \citep{bankier2024fleischner}. 
Fibrosis can cause airway remodeling, architectural distortion, and volume loss inside the lung. Honeycombing represents the destruction of the lung parenchyma with loss of architecture and well-defined adjacent cyctic structures. 
It is often a CT sign of severe fibrosis \citep{bankier2024fleischner}. Reticular and reticulonodular opacities are lung CT patterns. Their presence often suggests existence of fibrosis \citep{martini2021applicability}. 
Besides abnomality types, it is also required that the included CT images meet certain criteria. In the original NLST dataset, multiple screenings are present for one participant in each study year. 
These CT scans differ in various aspects. For better generalization across different participants, at most one eligible CT scan is selected from one participant in each study year.
Below are selection criteria for these CT images: 1) Z-Axis Spacing: For better imaging quality and resample results, CT scans with z-axis spacing $\leq 2.0$ mm are included. 
2) Construction Kernel: One of the distinguishing features between CT scans for one participant in a single study year is the kernel for CT reconstruction. Based on a list of ranked kernels for different CT scanner manufacturers \citep{ardila2019end}, we select the highest ranked kernel, if available, in candidate CT scans. 
3) Airway Integrity: For full analysis of entire airway structure, a thorough visual examination is taken for selected scans. CT images that have too few slices are discarded.


\begin{figure}[t]
\centering
\includegraphics[width=0.9\linewidth]{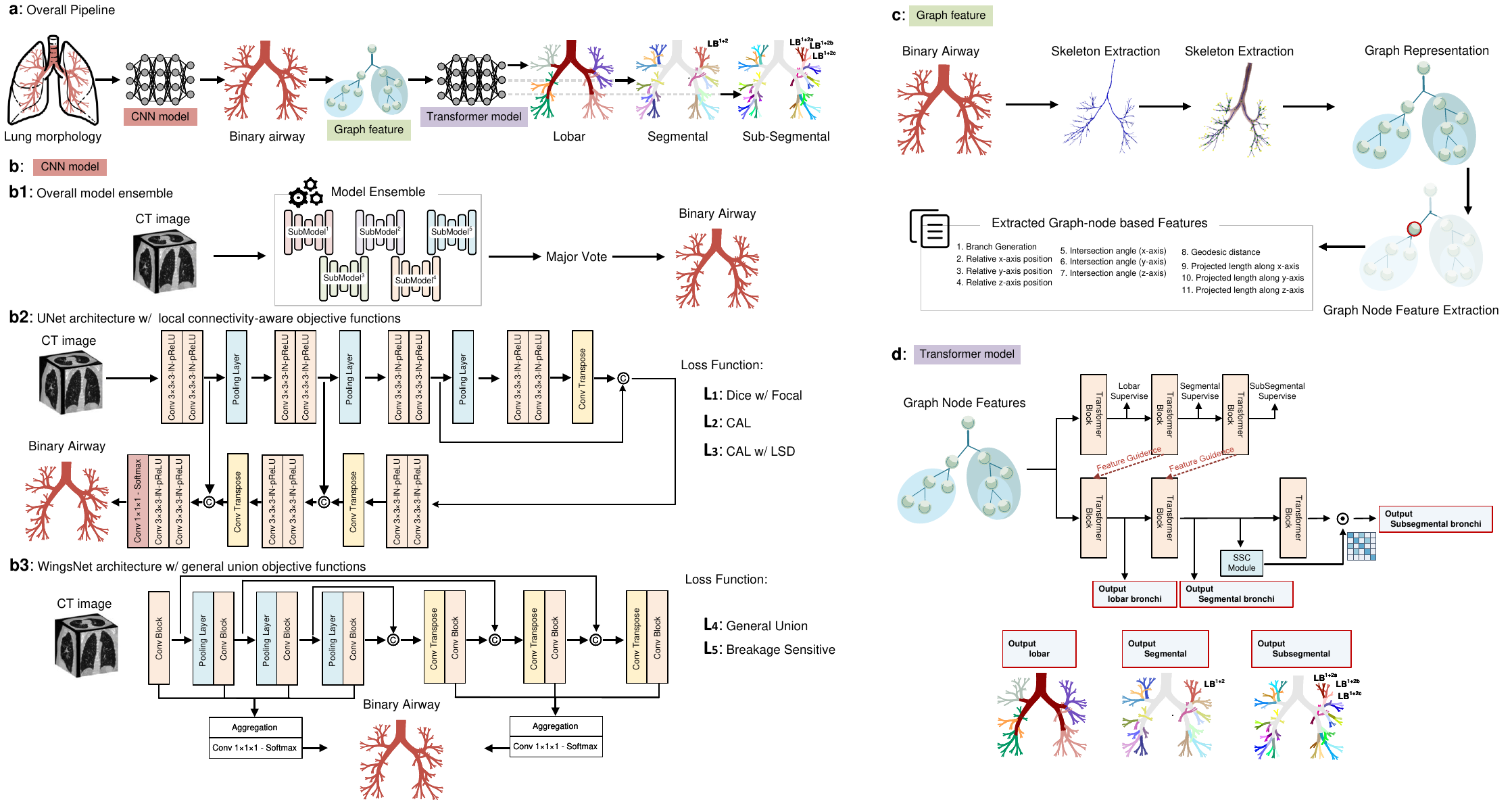}
\caption{
Detailed framework of the AirMorph. 
a) Overview of the end-to-end airway anatomical labeling pipeline.  
b) Encoder-decoder architecture used for binary airway segmentation.  
c) Graph-node based feature extraction from the binary airway tree.  
d) Hierarchical anatomical labeling of airway branches using a transformer-based module, incorporating a soft subtree consistency constraint to enhance topological coherence.
}\label{fig::AirMorph_detailed_method}
\end{figure}

\makeatletter
\def\hlinew#1{%
\noalign{\ifnum0=`}\fi\hrule \@height #1 \futurelet
\reserved@a\@xhline}
\makeatother
\begin{table*}[!t]
\renewcommand\arraystretch{1.3}
\centering
\caption{The features extracted from the graph representation of the binary airway, and then used for the anatomical branches labeling for AirMorph.}
\label{tab::AirMorph_manual_feature_definition}
\resizebox{1.0\textwidth}{!}{
\begin{tabular}{|c|c|c|c|}
\hline
\rowcolor{tablecolor1}
\textbf{Symbol}                 & \textbf{Description}           & \textbf{Symbol}                & \textbf{Description}           \\ \hline
$\mathcal{G}$          & Generation of branch                      &      $\theta_{z}$                   &  Intersection angle between branch vector and z-axis                      \\ \hline
$\mathnormal{RP}_{x}$  & Relative x-axis position of branch to treacha &     $\mathcal{L}$                    &  Geodesic distance between the branch endpoints                     \\ \hline
$\mathnormal{RP}_{y}$  & Relative y-axis position of branch to treacha   &    $\mathnormal{PL}_{x}$                      &   Projected length along x-axis of branch orientation                   \\ \hline
$\mathnormal{RP}_{z}$  & Relative z-axis position of branch to treacha    &   $\mathnormal{PL}_{y}$                          &    Projected length along y-axis of branch orientation                     \\ \hline
$\theta_{x}$           & Intersection angle between branch vector and x-axis                    &    $\mathnormal{PL}_{z}$                         &         Projected length along z-axis of branch orientation                \\ \hline
$\theta_{y}$           & Intersection angle between branch vector and y-axis                       &                       &                       \\ \hline
\end{tabular}
}
\end{table*}

\numberedsubsection{AirMorph: Model Development}
\noindent \textbf{Model design}.
The AirMorph is a fully automated framework for extracting hierarchical airway anatomical structures (Fig. \ref{fig::AirMorph_detailed_method}). It consists of three sequential stages. 
The first stage is binary airway modeling. Two encoder-decoder architectures, U-Net \cite{cciccek20163d} and WingsNet \cite{zheng2021alleviating}, are adopted for segmentation, 
and optimized using a set of advanced loss functions, including Connectivity-Aware Loss (CAL) \cite{zhang2023towards}, General Union Loss (GUL) \cite{zheng2021alleviating}, 
and Breakage-Sensitive Loss (BS) \cite{yu2022break}. These objectives prioritize topological integrity and are designed to mitigate airway leakage and breakage.
To enhance robustness, an ensemble strategy based on majority voting is employed to integrate predictions from five trained models. Details of these ensembled models are described below.
As illustrated in Fig. \ref{fig::AirMorph_detailed_method}.b, three models are based on the U-Net architectures with different loss functions. The first model is optimized by the basic segmentation loss function, Dice with Focal loss:
\[
L_{\text{Dice w/ Focal}} = -\frac{2 \sum_{\forall x} p_x g_x}{\sum_{\forall x} (p_x + g_x)} - \frac{1}{|X|} \left( \sum_{\forall x} (1 - p_x)^2 \log(p_x) \right),
\]
where the $p_x$ and $g_x$ denote the prediction and ground-truth of the voxel $x$. The second model adopts the Connectivity-Aware loss function:
\[
\mathcal{L}_{{\text{CAL}}} = \left\{ 1 - \frac{ \sum_{x=1}^{N} p_x g_x }{ \alpha_t \sum_{x} p_x + \beta_t \sum_{x} g_x } \right\} + \left\{ \sum_{x=1}^{N} \alpha_x \, \text{CE}(p_x, g_x) \right\},
\]
where $\alpha_t = 0.1$, and $\alpha_t = 0.9$. $\alpha_x$ defines the weight of each airway voxel based on the Euclidean distance to the centerline. 
Further, based on the second model, the third model introduces the local-sensitive distance objective functions:
\[
\mathcal{L}_{{\text{CAL w/ LSD}}} = \mathcal{L}_{{\text{CAL}}} + \left\| \mathrm{Dist}(g_x) - \mathrm{Dist}(p_x) \right\|_2. 
\]
Detailed calculation of the differentiable distance transform $\mathrm{Dist}$ can be referred to \cite{zhang2023towards}.
The forth model and fifth model are relied on the WingsNet framework, using the following General Union and Breakage-Sensitive Loss:
\[
\mathcal{L}_{{\text{GU}}}  = 1 - \frac{ \sum_{x=1}^{N} w_x \, p_x^{r_l} \, g_x }{ \sum_{x=1}^{N} \alpha_x  \left( \alpha \, p_x + \beta \, g_x \right)} ({r_l} = 0.7, \alpha=0.2, \beta=0.7),
\]
\[
\mathcal{L}_{{\text{BS}}} = 1 - \frac{ \sum_{i=1}^{N} p_x c_i }{ \sum_{i=1}^{N} c_i + \varepsilon}.
\]
where $c_i$ denotes the i-th voxel value on the centerline map \cite{yu2022break}. Additional supervision is introduced through label-informed sampling to improve intra-class discrimination. In particular, peripheral airway branches are emphasized via over-sampling strategies 
guided by the spatial distribution of centerline points. Furthermore, a multi-stage training strategy refines learning across iterations:  (1) lung region segmentation is used as a hard attention mechanism to suppress irrelevant areas and reduce false positives; (2) early-stage model predictions are used to identify and resample hard patches in later epochs, enabling more targeted and efficient learning.
After acquiring the binary airway segmentation, AirMorph constructs a branch-wise graph representation to facilitate subsequent anatomical labeling (Fig.~\ref{fig::AirMorph_detailed_method}.c) in the second stage. 
The binary airway tree is partitioned into individual branches based on centerline topology, and eleven graph-node level features are extracted for each branch 
(Table \ref{tab::AirMorph_manual_feature_definition}). These features characterize the structural, positional, and morphological properties of each airway segment,  serving as input for downstream anatomical classification.
In the third stage, AirMorph utilizes the graph node–level features as input tokens and feeds them into a Transformer architecture to model cross-attention among branches, enabling anatomical label assignment for each token (Fig.~\ref{fig::AirMorph_detailed_method}d).
To mitigate inter-individual anatomical variability and enhance prediction stability, a soft subtree consistency module is introduced. 
This module dynamically encodes hierarchical subtree representations by modulating the attention scores during subsegmental-level classification, thereby enforcing local structural consistency. The modified attention computation is defined as:
\[
A_{\mathrm{sub}} = \frac{(X_{\mathrm{sub}} Q_{\mathrm{sub}})(X_{\mathrm{sub}} K_{\mathrm{sub}})^T}{\sqrt{d}} \odot \hat{M}_t,
\]
where \( \hat{M}_t \) denotes the soft subtree mask, as introduced in~\cite{li2024airway}.

\noindent \textbf{Training details}.
For the binary airway modeling, We adopted a large input volume size of \(128 \times 224 \times 304\) for the CT scans. During preprocessing, dense cropping was applied around airway-centered regions to focus on anatomically relevant structures. 
A batch size of 1 was used during training due to memory constraints. Real-time data augmentation included random horizontal flipping and random rotations within the range of \([-10^\circ, 10^\circ]\). Model optimization was performed using the Adam optimizer, with an initial learning rate set to 0.002.
For the ariway anatomical labeling, We employed the Adam optimizer for training with a learning rate of 5e-4 over 600 epochs. 
For lobar and segmental airway labeling, two Transformer layers were stacked. At the subsegmental level, an additional two Transformer layers with SSC modules were introduced. Each Transformer block used 32 heads with a hidden dimension of 128.

\noindent \textbf{Evaluation metrics}. 
The evaluation metrics are designed to assess both topological preservation and accuracy in binary airway modeling and anatomical airway labeling. 
For binary airway modeling, two topology-aware metrics are used: tree length detected rate (TLD, \%) and branch number detected rate (BND, \%). 
TLD is defined as the fraction of the ground-truth airway tree length that is successfully detected, while BND measures the percentage of airway branches correctly reconstructed 
relative to the total number of ground-truth branches:
\begin{equation}
\text{TLD} = \frac{T_{\text{det}}}{T_{\text{ref}}} \times 100\%, \quad 
\text{BND} = \frac{B_{\text{det}}}{B_{\text{ref}}} \times 100\%,
\end{equation}
where \(T_{\text{det}}\) and \(B_{\text{det}}\) are the detected tree length and branch number, and \(T_{\text{ref}}\) and \(B_{\text{ref}}\) are the corresponding reference values from the ground truth.
In addition, voxel-wise overlap metrics are reported, including Dice Similarity Coefficient (DSC), centerline Dice (clDice), and Sensitivity.
For anatomical airway labeling, two graph-level metrics are proposed to evaluate topological consistency: predicted subtree consistency (TreeCons, \%) and topological distance (TopoDist). 
TreeCons measures the percentage of subtrees that are classified with consistent labels, while TopoDist quantifies the average graph distance between predicted and ground-truth matched nodes:
\begin{equation}
\text{TreeCons} = \frac{N_{\text{cs}}}{N_s} \times 100\%, \quad 
\text{TopoDist} = \frac{1}{N} \sum_{i=1}^{N} \min_{j \in \{ j \mid y_j = \hat{y}_i \}} d_{i,j},
\end{equation}
where \(N_s\) is the total number of anatomical subtrees and \(N_{\text{cs}}\) is the number of consistently labeled subtrees. 
\(N\) is the total number of nodes in the airway graph, \(d_{i,j}\) denotes the shortest path length between node \(v_i\) and node \(v_j\), 
and \(y_j\), \(\hat{y}_i\) represent the ground-truth and predicted labels, respectively.
Additionally, classification metrics such as Accuracy, Precision, and Sensitivity are used to evaluate the anatomical branch classification performance.


\numberedsubsection{Implementation of Branching Pattern Analysis}
We designed a structured pipeline to characterize bronchial branching patterns at three levels: intra-segment, intra-subsegment, and inter-subsegment. 
The corresponding implementations are outlined in Algorithm \ref{alog::intra_seg_pattern_analysis}, Algorithm \ref{alog::intra_subseg_pattern_analysis}, Algorithm \ref{alog::inter_subseg_pattern_analysis}.
In the intra-segment analysis (Algorithm \ref{alog::intra_seg_pattern_analysis} ), only lobes with all expected segmental branches present were included to reduce the impact of prediction errors. 
Co-trunk relationships between segments were determined based on the topological structure of the airway graph. Segments that satisfied the co-trunk condition were merged using union-find, 
yielding the final segment-level branching patterns.
The intra-subsegment characterization (Algorithm \ref{alog::intra_subseg_pattern_analysis}) focused on segments with all expected basic subsegments (a, b, and optionally c) to ensure structural completeness. 
A default subsegment was defined as the bronchial branch preceding the division into basic subsegments (e.g., LB6 before LB6a/b/c), while co-trunk subsegments (e.g., LB6a+b, LB6a+c) were 
defined as shared branches between two basic subsegments. Based on the presence of default and co-trunk subsegments, each segment was assigned a stem number and co-trunking type.
For the inter-subsegment pattern (Algorithm \ref{alog::inter_subseg_pattern_analysis}), co-trunk relationships between subsegments and  external segments (i.e., segments other than the subsegment belong to) were 
inferred from the airway graph topology. As in the intra-subsegment analysis, only subsegments from structurally complete segments were considered. Final co-trunk clusters were 
identified using union-find to establish standardized inter-subsegment branching patterns.

\numberedsubsection{Implementation of AirwaySignature}
\noindent \textbf{Implementation details of the morphological features.}
We desiged \textbf{six} fine-grained morphological signatures, which can be efficiently calculated based on AirMorph. These six morphological signatures 
depict local variations from different aspects, which cannot be observed from the binary modeling results. Specifically, according to the fine-grained 
anatomical labeling results, we define the group-wise morphological signatures, the basic group can be the segmental- or lobar- wise semantic branches. 
The morphological features involve \textbf{Stenosis ($\mathcal{S}$), Ectasia ($\mathcal{E}$), Tortuosity ($\mathcal{T}$), Divergence ($\mathcal{D}$), Length ($\mathcal{L}$), Complexity ($\mathcal{C}$)}, 
which can be seen in Fig.\ref{fig::AirMorph_morpho_distribution_presentation}. The detailed description are demonstrated as follow:

\noindent \textbf{Stenosis ($\mathcal{S}$).} 
To evaluate airway narrowing, stenosis is alternatively defined based on the local airway radius extracted along the centerline. 
Specifically, the stenosis rate is computed as
\[
\text{Stenosis} = \left(1 - \frac{R_{\text{narrow}}}{R_{\text{ref}}} \right) \times 100\%.
\]
The Stenosis ranges from [0\%, 100\%]. The higher value represents a more severe stenosis condition. \(R_{\text{narrow}}\) denotes the minimum radius at the site of maximal constriction, 
and \(R_{\text{ref}}\) is the mean radius of a proximal, healthy airway segment. 
This radius-based definition enables efficient and consistent stenosis quantification in centerline-based geometric models. 
To quantify the degree of bronchial narrowing, we compute a stenosis score for each bronchial segment and lobe based on the radius statistics extracted from the airway centerline. 
First, we apply a 3D Euclidean distance transform to the binary airway mask, yielding a voxel-wise approximation of the local airway radius. 
The resulting distance field is then masked by the airway skeleton to retain only the centerline voxels.
Subsequently, we parse the skeleton into individual segments using a 26-connected neighborhood and morphological junction removal, 
ensuring each branch is spatially isolated. For each parsed centerline branch, we compute its minimum and mean radius values. 
These two statistics are then used to define a local stenosis ratio, which reflects the degree of luminal constriction within the branch.
To obtain anatomically interpretable scores, we aggregate the stenosis ratios of all branches belonging to the same anatomical segment or lobe, 
as determined by previously established anatomical labels. The final segmental and lobar stenosis scores are reported as the average values across all corresponding branches.

\noindent \textbf{Ectasia ($\mathcal{E}$).} 
To complement the stenosis quantification, we additionally compute the Ectasia for each bronchial segment and lobe. The range of Ectasia is $[100\%, +\infty]$. A higher value indicates a greater degree of Ectasia.
While stenosis captures lumen narrowing, ectasia reflects potential overexpansion of the airway lumen. It is defined as the ratio between the maximum and mean radius along each parsed skeleton branch:
\[
\text{Ectasia} = \frac{R_{\max}}{R_{\text{ref}}} \times 100\%. 
\]
The final segmental and lobar ectasia scores are obtained by averaging this ratio across all centerline segments associated with the corresponding anatomical label. 

\noindent \textbf{Tortuosity ($\mathcal{T}$).}
Tortuosity is defined to characterize the local curvature of a bronchial branch based on its volumetric geometry. 
As illustrated in Fig. \ref{fig::AirMorph_morpho_distribution_presentation}.a, 
For each branch region $\Omega_l$, we first extract its physical-space voxel coordinates and apply principal component analysis (PCA) to determine the two endpoints $S$ and $E$ along the first principal axis.
Then, we identify the voxel point $P \in \Omega_l$ that exhibits the largest perpendicular distance to the line segment $\overline{SE}$:
\[
P = \arg\max_{x_i \in \Omega_l} \; \text{dist}(x_i, \overline{SE})
\]
Finally, tortuosity is computed as the twist angle formed by vectors from $P$ to the endpoints:
\[
\text{Tortuosity} = \alpha = \angle (\overrightarrow{PS}, \overrightarrow{PE})
\]
This angular measure reflects the degree of bending or deflection within the airway branch. The range first lies in $[0, \pi]$, and then normalized by $arccos$ to $[0, 1]$. 
A value close to 1 indicates a high degree of tortuosity. The final segmental and lobar tortuosity are obtained by averaging across all centerline segments associated with the corresponding anatomical label.

\noindent \textbf{Length  ($\mathcal{L}$).} 
To quantify the geometric elongation of airway segments and lobes, we define a geodesic length based on branchwise skeleton path accumulation. 
For each anatomical class $K$ (segmental or lobar bronchi), we first identify all leaf nodes labeled as $K$ in the centerline tree. The lowest common ancestor (LCA) of these leaf nodes is then determined via generation-aware traversal.
If the LCA itself belongs to class $K$, the geodesic length is computed as the average of path lengths from the LCA to each leaf node:
\[
\mathcal{L}_K = \frac{1}{|\mathcal{T}_K|} \sum_{v \in \mathcal{T}_K} \sum_{n \in \text{Path}(\text{LCA}, v)} \ell(n)
\]
where $\mathcal{T}_K$ denotes the set of leaf nodes labeled as $K$, and $\ell(n)$ is the normalized length of node $n$.
If the LCA is not labeled as $K$ (e.g., trachea), each path is re-rooted at the nearest downstream node with label $K$ to ensure anatomical relevance. 
The range is $[0, +\infty]$, and the measuring unit is $mm$.

\noindent \textbf{Divergence ($\mathcal{D}$).} 
To quantify the spatial dispersion of airway branches within a segmental or lobar class, usually measure the regions where lesions locate at, 
we define a divergence angle based on the minimal enclosing cone. For each class $K$, we first identify the set of terminal (leaf) nodes labeled as $K$, and determine their lowest common ancestor (LCA), 
which serves as the apex of the cone.
$\mathbf{v}_i$ denotes the normalized direction vector from the apex to each leaf node. We then solve for the optimal unit vector $\mathbf{u}$ on the unit sphere that maximizes the minimal cosine similarity:
\[
\mathbf{u}^* = \arg\max_{\|\mathbf{u}\|=1} \min_i \langle \mathbf{u}, \mathbf{v}_i \rangle
\]
The divergence angle is defined as twice the maximal angular deviation from $\mathbf{u}^*$:
\[
\theta = 2 \cdot \arccos\left( \min_i \langle \mathbf{u}^*, \mathbf{v}_i \rangle \right)
\]
The range first lies in $[0, \pi]$, and then normalized to $[0, 1]$. A smaller angle $\theta$ indicates lower branch separation and angular dispersion within the anatomical class, which may be affected by the pulmonary diseases.

\noindent \textbf{Complexity ($\mathcal{C}$).}
To assess the geometric complexity of airway branches within each anatomical segment or lobe, we adopt a box-counting based fractal dimension. This metric quantifies the spatial irregularity and branching density of a binarized skeleton structure.
For each class $K$, the corresponding airway skeleton region is first cropped and zero-padded to a fixed 3D volume. The volume is then covered with a set of cubic boxes of varying sizes $s \in \{2^1, 2^2, \dots, 2^k\}$. L
et $N(s)$ denote the number of non-empty boxes of size $s$ required to cover the foreground skeleton. The complexity of the branch $\mathcal{C} $ is estimated via linear regression in the log-log domain:
\[
\mathcal{C} = \lim_{s \to 0} \frac{\log N(s)}{\log (1/s)}
\]
Higher values of $\mathcal{C}$ indicate greater geometric complexity and branching richness in the corresponding airway segment.  The range of  $\mathcal{C}$ is $[0, +\infty]$, and the larger $\mathcal{C}$ denotes a more complex branch.

\noindent \textbf{Complementary information of the radiomics features.}
After the acquisition of AirMorph, radiomic signatures are computed using open-source PyRadiomics tools\citep{pyradiomics}. 19 first-order statistics, 16 gray level run length matrix statistics, 16 gray level size zone matrix statistics, 
5 neighboring gray tone difference matrix statistics and 14 gray level dependence matrix statistics are selected from 1 original image, 3 Laplacian-of-Gaussian filtered images, 8 wavelet filtered images. These amounts to 
$(19 + 16 + 16 + 5 + 14)\times(1 + 5 + 8)=840$ radiomic signatures.
The radiomic feature selection pipeline is shown in Fig.\ref{fig::radiomic_feature_selection}. From AirMorph we obtain airway anatomical labels at different levels, from which we select 23 lobar and segmental-level components.
By comparing airway morphological features between experimental group cases and control group cases on these components, we can easily select for each case in the experimental group the significant anatomical component. 
This is achieved by modeling the distribution of each morphological feature in control group as a normal distribution. If an anatomical component has $\geq 3$ out of 6 morphological features that lies outside the 2-$\sigma$ interval 
of the control group distribution, it is labeled significant. Otherwise, it is labeled insignificant.
By dividing the anatomical components into significant set and insignificant set for each case, we can collect radiomic features for both sets across all cases in the experimental group. 
T-test one is taken between significant-set radiomic features and control group radiomic features to find significant radiomic features. T-test two is taken between insignificant-set radiomic features and control group radiomic features to 
find insignificant radiomic features. The intersection of two t-test results is thus acquired and ranked, in which top-20 radiomic features are selected as airway radiomic signatures.
In order to better illustrate Airway Radiomic Signature, for dataset AIIB'23 and the four subset selected from NLST, we choose one typical case as example.
\noindent \textbf{AIIB'23}: For AIIB'23 dataset, we choose case 30 for analysis. The heatmap is shown in Fig. \ref{fig::AirMorph_full_features_with_radiomics}.a. Among the 20 top-ranked radiomic features, 3 features match well with the morphological features, 
namely: wavelet-HHH\_glszm\_SmallAreaEmphasis, wavelet-HLH\_gldm\_DependenceVariance and wavelet-LLL\_gldm\_DependenceNonUniformityNormalized. Small Area Emphasis measures the distribution of small size gray level zones, 
with greater value indicating smaller size zones. In the context of Dataset AIIB'23, this refers to the gray-level zones between fibrotic tissue on CT scans. Dependence variance measures the variance in dependence size in the CT image. 
Dependence non-uniformity measures the similarity of dependence throughout the ROI, with greater values indicating less homogeneity among dependencies. Larger variance and non-uniformity indicates complex gray-level dependencies in the ROI, 
which is a sign of fibrosis.
\noindent \textbf{NLST D58}: For D58 subset of NLST dataset, we choose CT scan of patient with ID 106018 in the second study year for analysis. The heatmap is shown in Fig. \ref{fig::AirMorph_full_features_with_radiomics}.b. 
Among the 20 top-ranked features, 4 features match well with the morphological features, namely: wavelet-LHL\_ngtdm\_Strength, log-sigma-3-0-mm-3D\_glrlm\_RunEntropy, original\_glrlm\_GrayLevelNonUniformityNormalized and 
wavelet-LHH\_ngtdm\_Strength. Ngtdm strength is high when an image with slow change in intensity but more large coarse differences in gray level intensities. High strength, together with large run entropy and high gray level 
variance and non-uniformity, indicates the ground-glass opacities in the consolidation regions inside CT scan.

\section*{\large\bfseries Data availability}
Example data with annotations will be publicly available under \url{https://github.com/EndoluminalSurgicalVision-IMR/AirMorph}. 
The authors acknowledge the National Cancer Institute for providing access to data from the National Lung Screening Trial (approved Project ID: NLST-704) and extend their gratitude to the patients who participated in the study.
The remaining datasets used in this study can be obtained through reasonable requests to corresponding authors and will be available for data sharing upon request and be reviewed and approved by an independent review panel on the basis of scientific merit.

\section*{\large\bfseries Code availability}
The code used for the implementation of AirMorph will be publicly available under \url{https://github.com/EndoluminalSurgicalVision-IMR/AirMorph}.

\bibliography{references}

\setcounter{figure}{0}
\setcounter{table}{0}
\setcounter{algorithm}{0}

\renewcommand{\thefigure}{S\arabic{figure}}
\renewcommand{\thetable}{S\arabic{table}}
\renewcommand{\thealgorithm}{S\arabic{algorithm}}

\clearpage

\section*{Supplementary Material}

The content includes the supplementary figures, tables, and algorithms. 

\tableofcontents

\clearpage

\begin{figure}[h]
\centering
\includegraphics[width=1.0\linewidth]{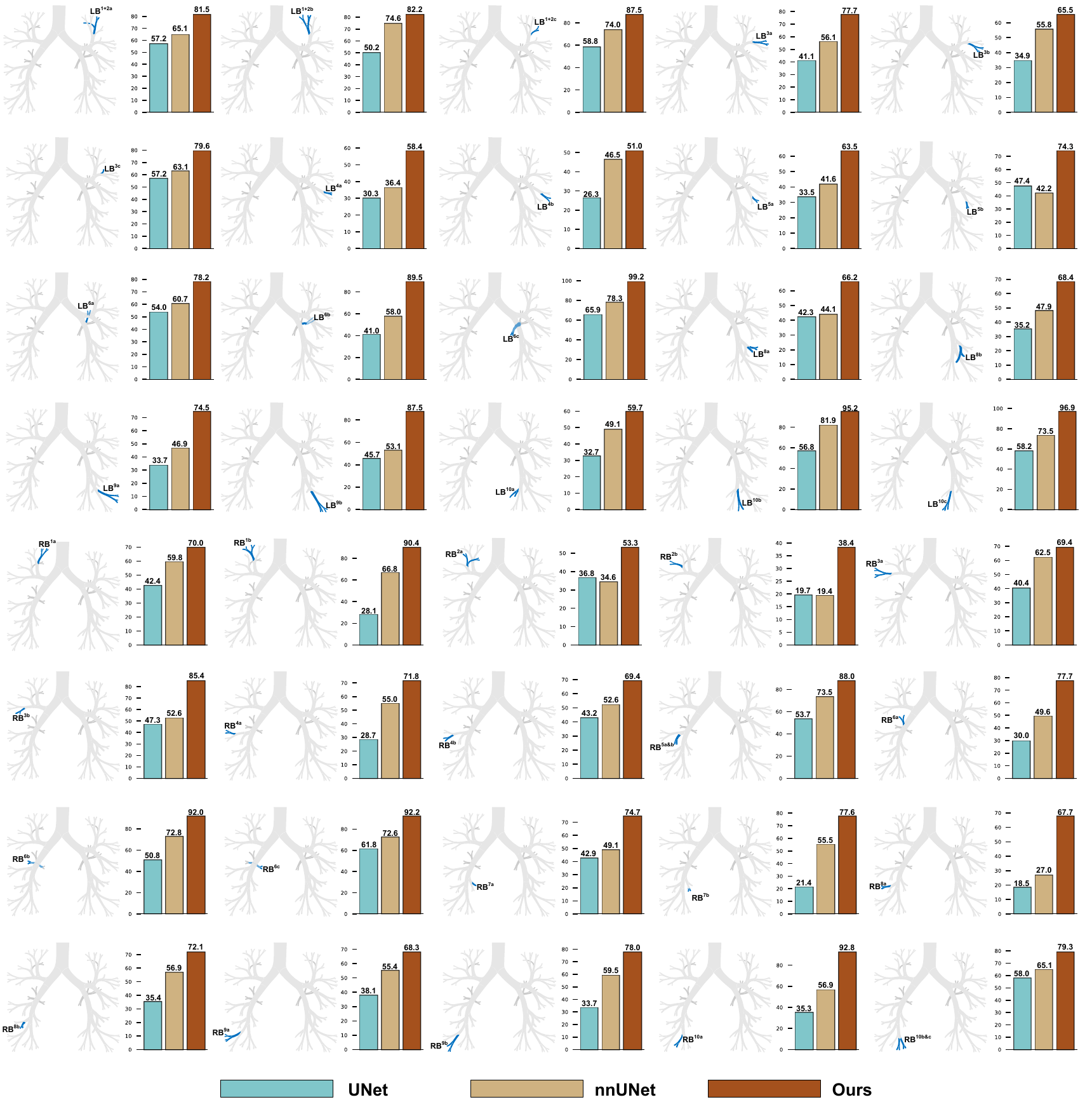}
\caption{
Visualization of the detailed evaluation of the AirMorph on subsegmental 
bronchi. TLD is reported. AirMorph is compared with UNet and nnUNet.
}\label{fig::finegrained_evaluation_seg}
\addcontentsline{toc}{section}{Fig. S1: Visualization of the detailed evaluation}
\end{figure}


\begin{figure}[h]
\centering
\includegraphics[width=1.0\linewidth]{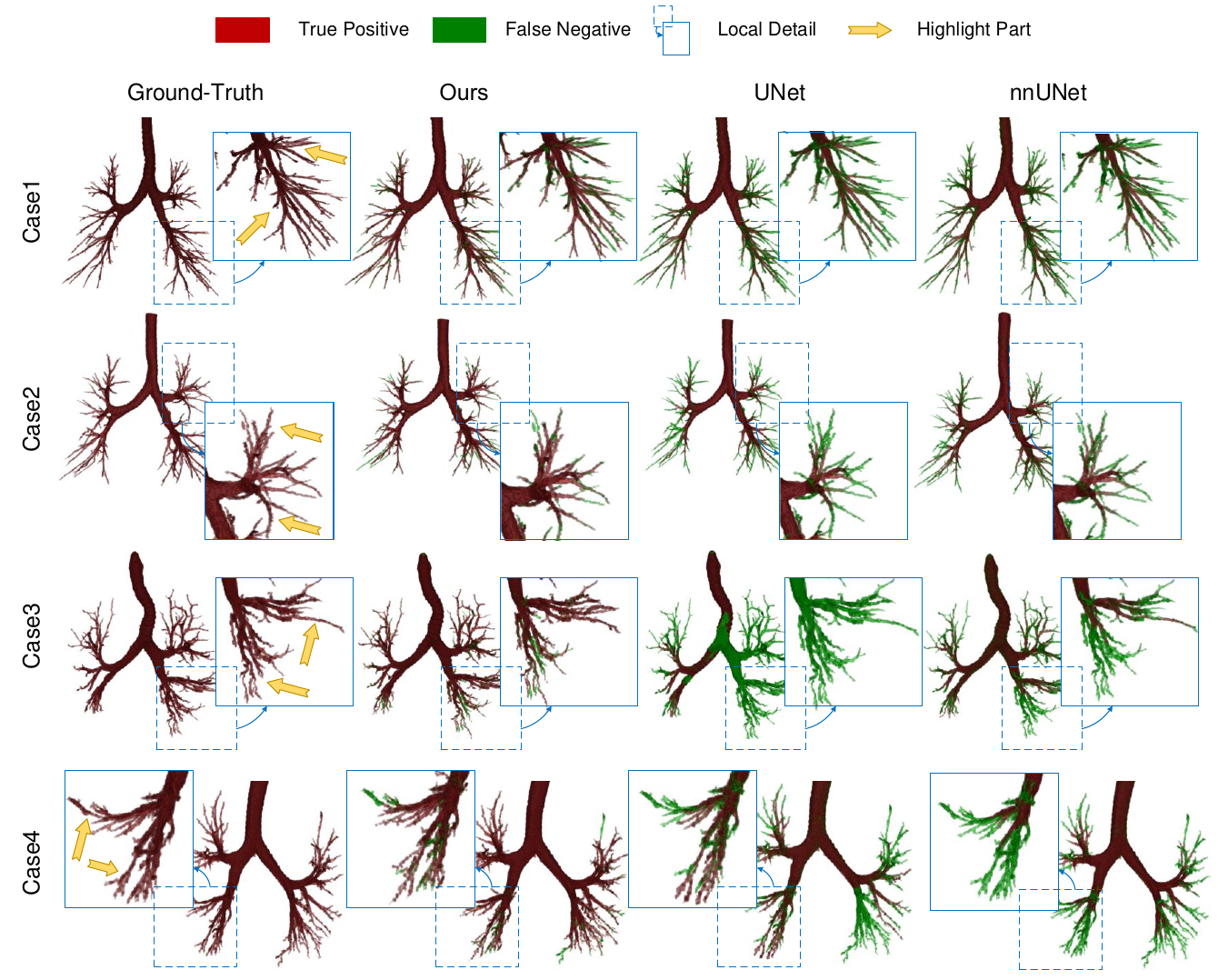}
\caption{Visualization of the quantitative comparison of the binary airway modeling. Case 1 and Case 2 are from the 
ATM'22 Dataset, Case 3 and Case 4 are from AIIB'23 Dataset.}\label{fig::comparison_airway_seg_quantative}
\addcontentsline{toc}{section}{Fig. S2: Visualization of the quantitative comparison of the binary airway modeling}
\end{figure}

\begin{figure}[h]
\centering
\includegraphics[width=1.0\linewidth]{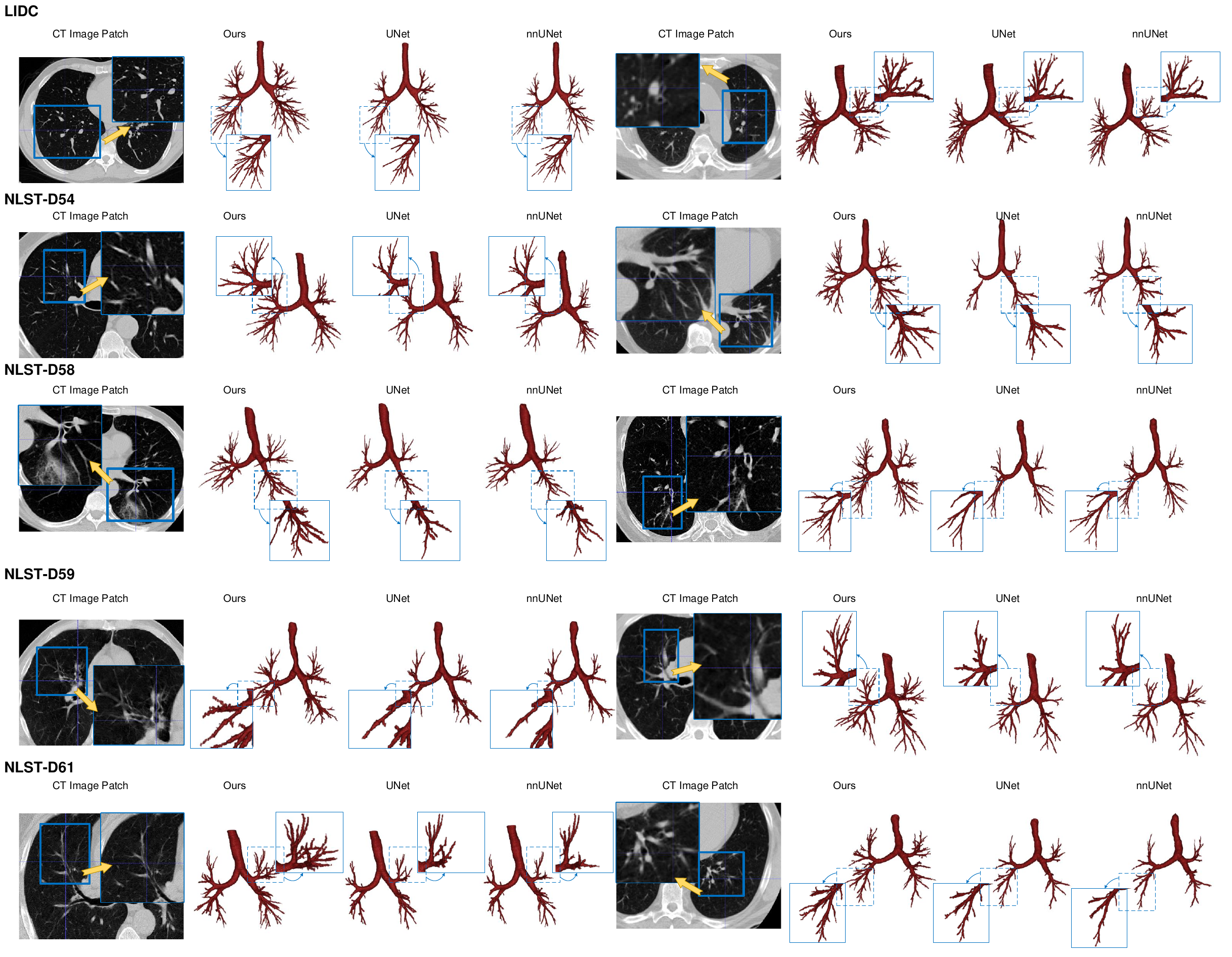}
\caption{Visualization of the qualitative comparison of the binary airway modeling.}\label{fig::comparison_airway_seg_qualitative}
\addcontentsline{toc}{section}{Fig. S3: Visualization of the qualitative comparison of the binary airway modeling}
\end{figure}


\begin{figure}[h]
\centering
\includegraphics[width=1.0\linewidth]{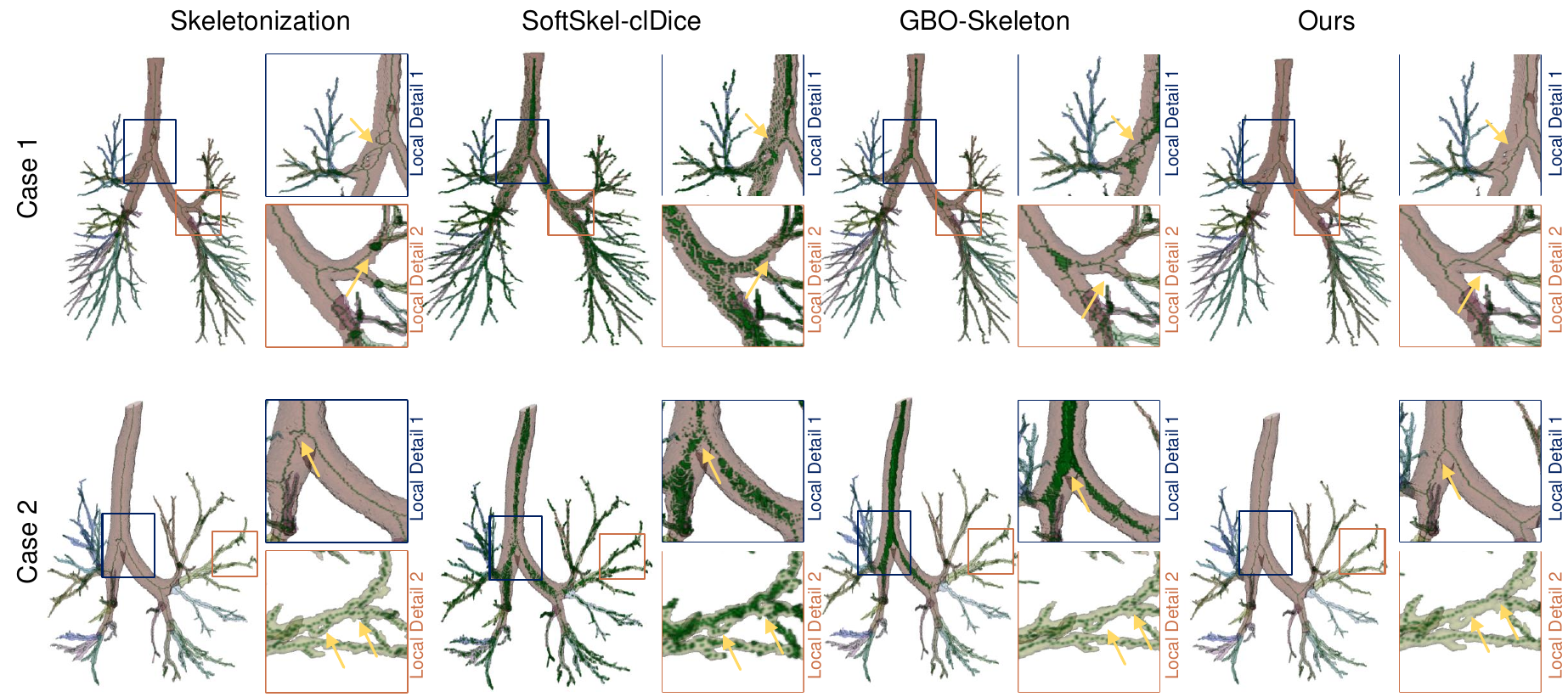}
\caption{Comparison of graph construction strategies for feature extraction in airway anatomical labeling. Case1 and Case2 are selected from ATM'22 and AIIB'23, respectively.}
\label{fig::comparison_graph_building}
\addcontentsline{toc}{section}{Fig. S4: Comparison of graph construction.}
\end{figure}

\begin{figure}[h]
\centering
\includegraphics[width=1.0\linewidth]{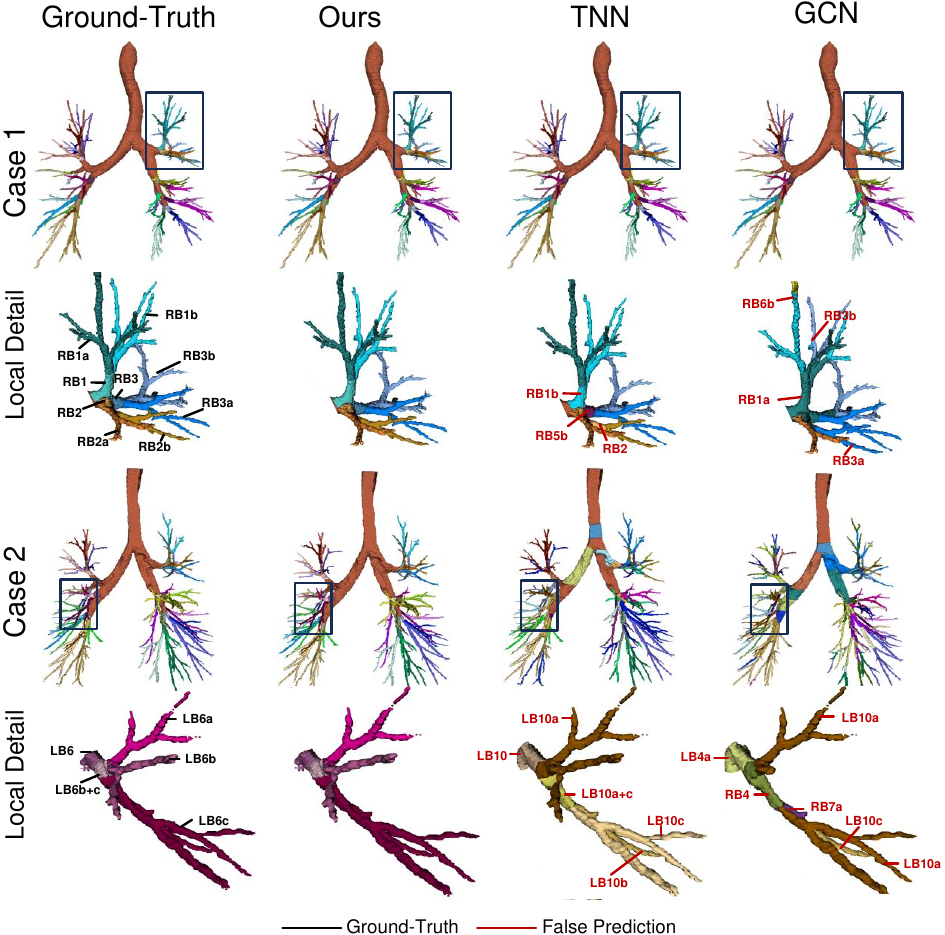}
\caption{Visualization of the quantitative comparison of the airway anatomical labeling. Case1 and Case2 are selected from ATM'22 and AIIB'23, respectively.}\label{fig::comparison_airway_cls_quantative}
\addcontentsline{toc}{section}{Fig. S5: Visualization of the quantitative comparison of the airway anatomical labeling.}
\end{figure}

\begin{figure}[h]
\centering
\includegraphics[width=1.0\linewidth]{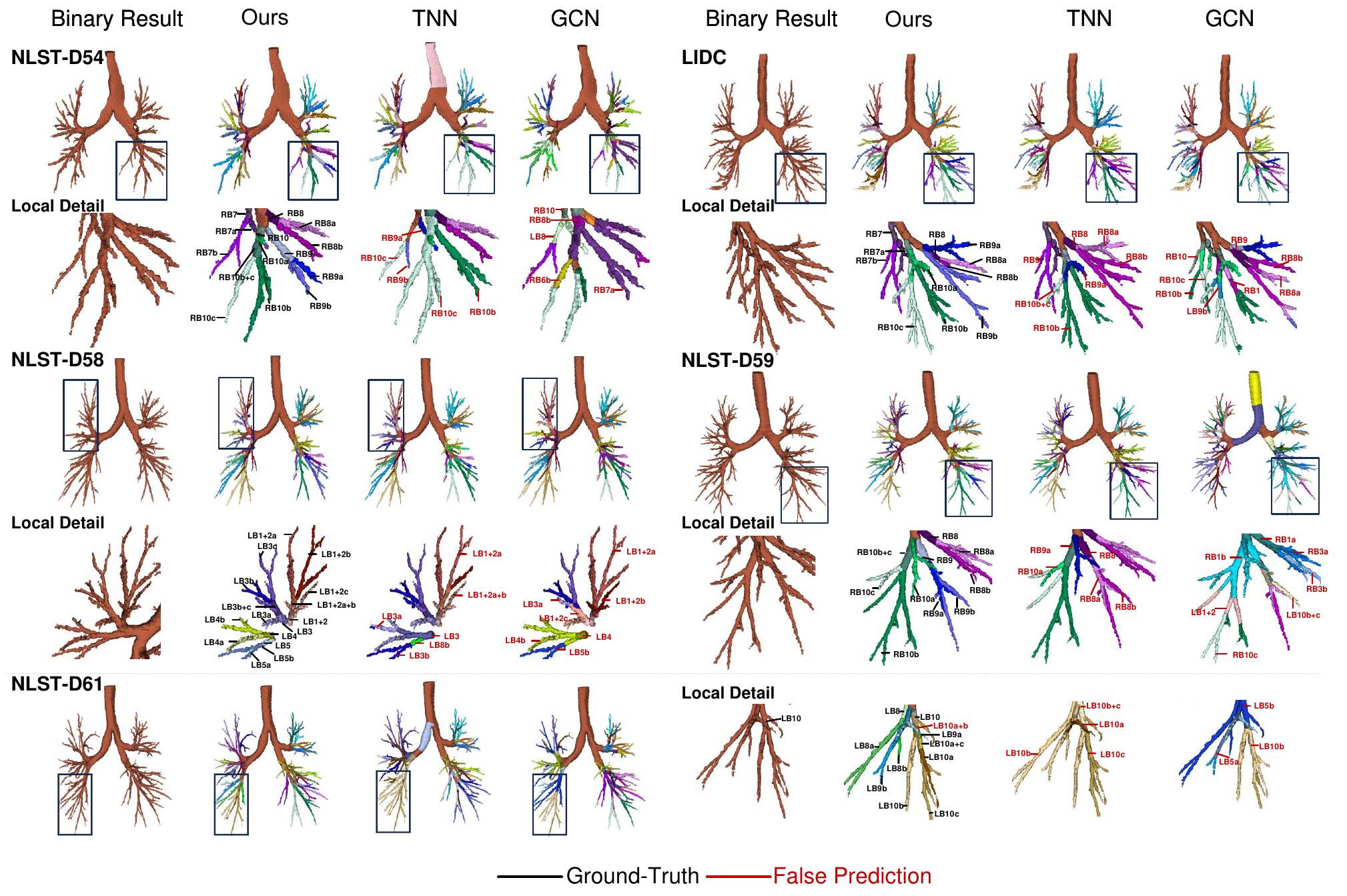}
\caption{Visualization of the qualitative comparison of the airway anatomical labeling.}\label{fig::comparison_airway_cls_qualitative}
\addcontentsline{toc}{section}{Fig. S6: Visualization of the qualitative comparison of the airway anatomical labeling.}
\end{figure}

\begin{figure}[h]
\centering
\includegraphics[width=0.9\linewidth]{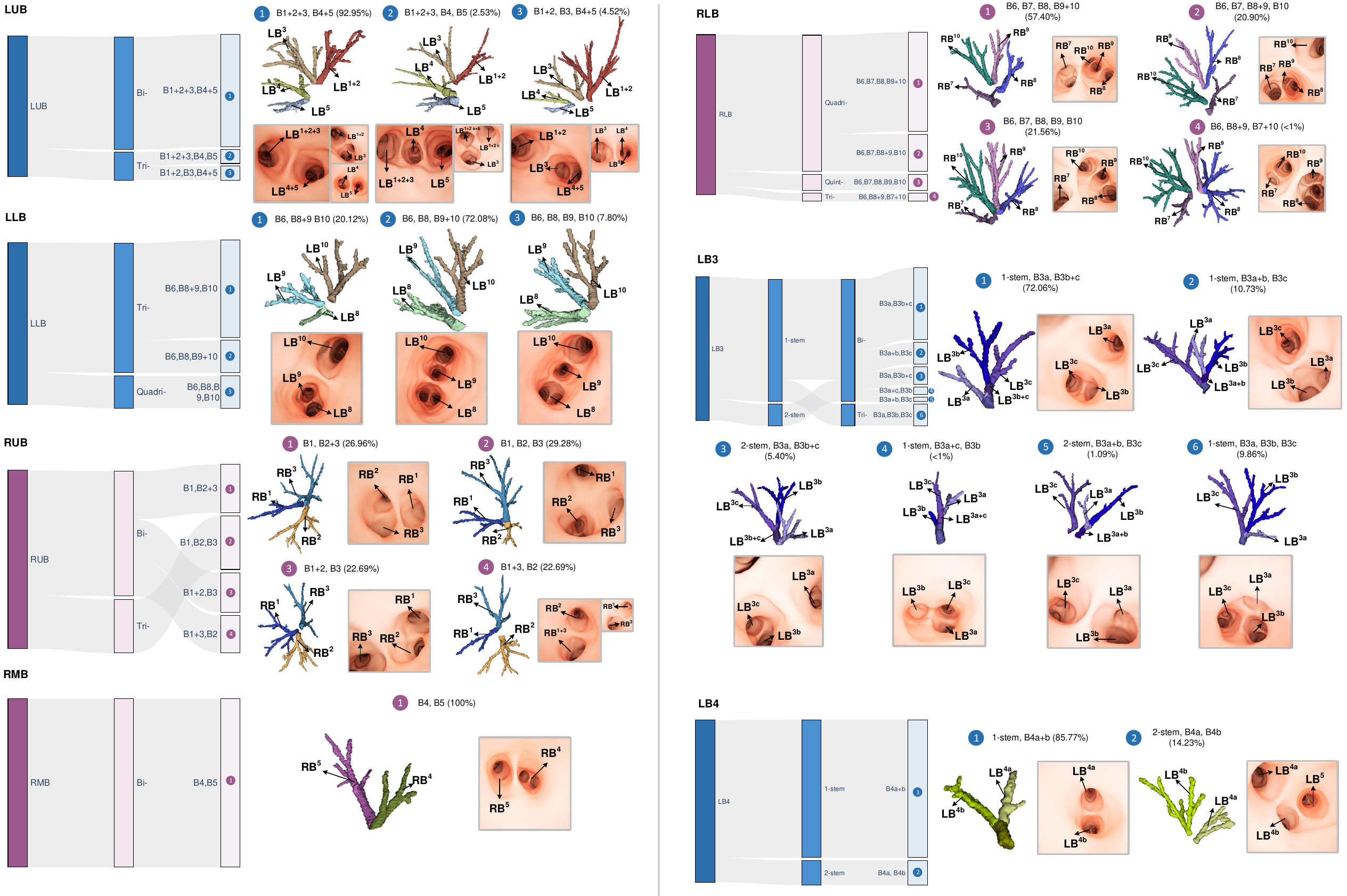}
\caption{Subfigure of the full branching patterns with endoscopic views ($LUB$, $LLB$, $RUB$, $RLB$, $LB^{3}$, $LB^{4}$).}\label{fig::branching_pattern_with_endo_view_supplementary_part1}
\addcontentsline{toc}{section}{Fig. S7: branching patterns with endoscopic views ($LUB$, $LLB$, $RUB$, $RLB$, $LB^{3}$, $LB^{4}$).}
\end{figure}

\begin{figure}[h]
\centering
\includegraphics[width=0.9\linewidth]{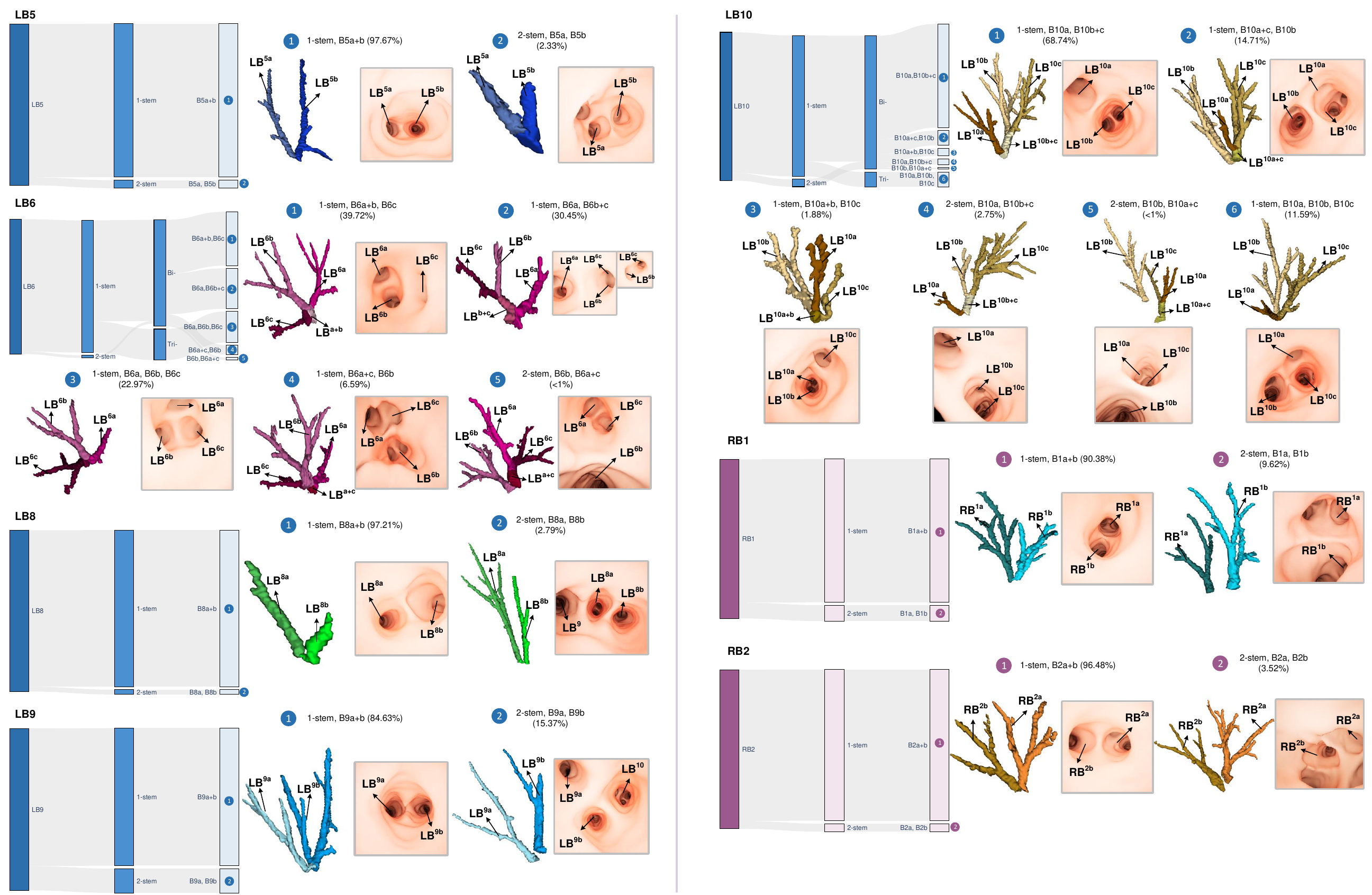}
\caption{Subfigure of the full branching patterns with endoscopic views ($LB^{5}$, $LB^{6}$, $LB^{8}$, $LB^{9}$, $LB^{10}$, $RB^{1}$, $RB^{2}$).}\label{fig::branching_pattern_with_endo_view_supplementary_part2}
\addcontentsline{toc}{section}{Fig. S8: branching patterns with endoscopic views ($LB^{5}$, $LB^{6}$, $LB^{8}$, $LB^{9}$, $LB^{10}$, $RB^{1}$, $RB^{2}$).}
\end{figure}

\begin{figure}[h]
\centering
\includegraphics[width=0.9\linewidth]{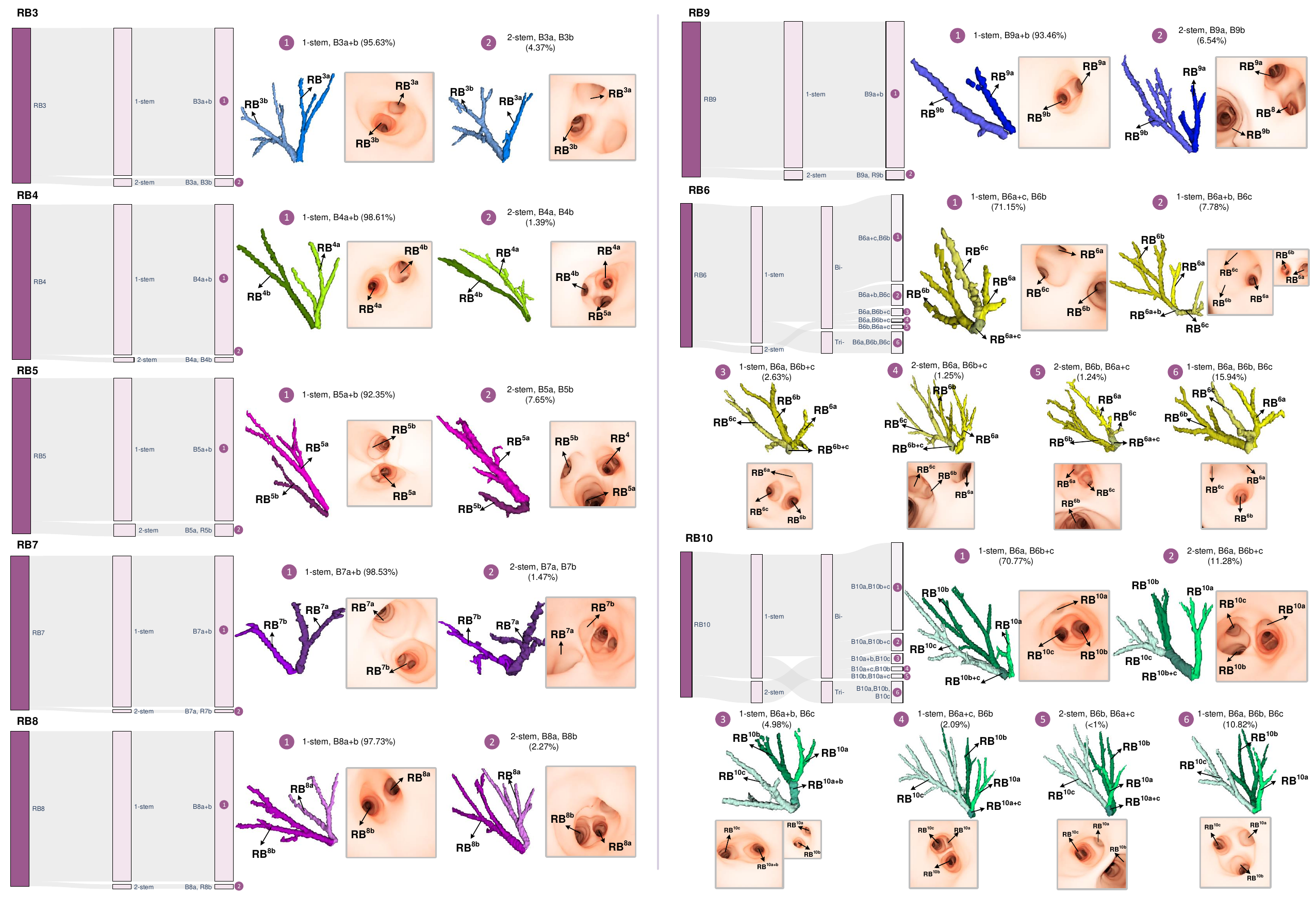}
\caption{Subfigure of the full branching patterns with endoscopic views ($RB^{3}$, $RB^{4}$, $RB^{5}$, $RB^{6}$, $RB^{7}$, $RB^{8}$, $RB^{9}$, $RB^{10}$).}\label{fig::branching_pattern_with_endo_view_supplementary_part3}
\addcontentsline{toc}{section}{Fig. S9: branching patterns with endoscopic views ($RB^{3}$, $RB^{4}$, $RB^{5}$, $RB^{6}$, $RB^{7}$, $RB^{8}$, $RB^{9}$, $RB^{10}$).}
\end{figure}

\begin{algorithm}[ht]
\caption{Intra-Segment Pattern Analysis}
\label{alog::intra_seg_pattern_analysis}
\begin{algorithmic}[1]
\Statex \textbf{Input:}
\Statex \quad \texttt{segment\_labels}: Array of segment labels for each node
\Statex \quad \texttt{generation}: Array of generation (depth) values for each node
\Statex \quad \texttt{descendant\_mask}: Binary matrix ($N \times N$), where \texttt{descendant\_mask[i,j] = 1} if node $i$ is an ancestor of node $j$
\Statex \quad \texttt{lca\_matrix}: Matrix ($N \times N$), where \texttt{lca\_matrix[i,j]} is the lowest common ancestor (LCA) of nodes $i$ and $j$
\Statex \textbf{Output:}
\Statex \quad \texttt{Pattern\_seg\_intra}: A dictionary mapping each lobar block (from \texttt{LOBAR\_SEGMENTS}) to clusters of segments based on the cotunk condition
\State \texttt{num\_segment\_classes} $\gets 18$
\For{$i = 0$ to \texttt{num\_segment\_classes} $-1$}
    \State Identify nodes in segment $i$ $\rightarrow$ \texttt{segment\_indices[$i$]}
    \If{\texttt{segment\_indices[$i$]} is empty}
        \State Mark segment $i$ as invalid
        \State \textbf{continue}
    \EndIf
    \State Select representative node for segment $i$ with minimum generation $\rightarrow$ \texttt{rep[$i$]}
\EndFor
\For{each pair of segments $(i, j)$ with $i \neq j$}
    \If{either segment $i$ or $j$ is invalid}
        \State \textbf{continue}
    \EndIf
    \State $lca_{ij} \gets$ \texttt{lca\_matrix[rep[$i$], rep[$j$]]}
    \State Extract descendant labels under $lca_{ij}$ via \texttt{descendant\_mask}
    \If{$lca_{ij}$ = main bronchus \textbf{and} all descendant labels $\in \{i, j, \text{main bronchus}\}$}
        \State \textbf{Union} segments $i$ and $j$ using union-find
    \EndIf
\EndFor
\For{each lobe in \texttt{LOBAR\_SEGMENTS} (key = lobe, value = list of segment indices)}
    \If{any segment in the lobe is invalid}
        \State \textbf{skip} this lobe
    \Else
        \State Initialize union-find structure over segments within this lobe
        \For{each pair $(i, j)$ in the lobe}
            \If{segments $i$ and $j$ satisfy the cotunk condition}
                \State \textbf{Union} $i$ and $j$
            \EndIf
        \EndFor
        \State Extract clusters from the union-find structure
        \State \texttt{Pattern\_seg\_intra[lobe]} $\gets$ clusters
    \EndIf
\EndFor
\State \Return \texttt{Pattern\_seg\_intra}
\addcontentsline{toc}{section}{Algorithm S1. Intra-Segment Pattern Analysis.}
\end{algorithmic}
\end{algorithm}


\begin{algorithm}[htbp]
\caption{Intra-SubSegment Pattern Analysis}
\label{alog::intra_subseg_pattern_analysis}
\begin{algorithmic}[1]
\Statex \textbf{Input:}
\Statex \quad \texttt{segment\_labels}: Array of segment labels for each node
\Statex \quad \texttt{annotations}: Array of subsegment annotation values (0--6), where 0 represents root subsegment, 1–3 represent basic subsegments (a, b, c), and 4–6 represent co-tunk subsegments (a+b, b+c, a+c)
\Statex \textbf{Output:}
\Statex \quad \texttt{Pattern\_sub\_intra}: Dictionary mapping each valid segment index (0–17) to a tuple \texttt{(stem\_number, cotunk\_type)}
\Statex \quad \emph{Note: \texttt{cotunk\_type} will later be interpreted as monofurcation, bifurcation, or trifurcation}
\State \texttt{num\_segment\_classes} $\gets 18$
\State Initialize \texttt{RESULT} as an array of shape $(18 \times 3)$: [valid flag, stem number, cotunk type]

\For{$i = 0$ to \texttt{num\_segment\_classes} $-1$}
    \State Create a mask for nodes in segment $i$
    \State Determine existence flags for each subsegment (a, b, c, a+b, b+c, a+c)
    \If{segment $i$ inherently allows only two basic subsegments (a and b)}
        \If{both a and b are detected}
            \State \texttt{RESULT[i, valid]} $\gets$ TRUE
            \State Set \texttt{RESULT[i, stem\_number]} based on optional root presence
            \State \texttt{RESULT[i, cotunk\_type]} $\gets 0$ \Comment{Interpret as monofurcation}
        \EndIf
    \Else
        \If{a, b, c are detected and at most one co-tunk subsegment is present}
            \State \texttt{RESULT[i, valid]} $\gets$ TRUE
            \State Determine \texttt{cotunk\_type} based on detected co-tunk subsegment
            \Comment{e.g., a+b → bifurcation; none → trifurcation}
            \State Set \texttt{RESULT[i, stem\_number]} based on root presence and co-tunk configuration
        \EndIf
    \EndIf
\EndFor

\State Initialize \texttt{Pattern\_sub\_intra} as empty dictionary
\For{each $i$ such that \texttt{RESULT[i, valid]} is TRUE}
    \State Record \texttt{(stem\_number, cotunk\_type)} into \texttt{Pattern\_sub\_intra[i]}
\EndFor

\State \Return \texttt{Pattern\_sub\_intra}
\addcontentsline{toc}{section}{Algorithm S2. Intra-SubSegment Pattern Analysis.}
\end{algorithmic}
\end{algorithm}

\begin{figure}[htbp]
\centering
\resizebox{0.72\textwidth}{!}{ 
\begin{minipage}{\textwidth}
\begin{algorithm}[H]
\caption{Inter-Subsegmental Pattern Analysis}
\label{alog::inter_subseg_pattern_analysis}
\begin{algorithmic}[1]
\Statex \textbf{Input:}
\Statex \quad \texttt{segment\_labels}: Array of segment labels for each node
\Statex \quad \texttt{subsegment\_labels}: Array of subsegment labels for each node
\Statex \quad \texttt{generation}: Array of generation values for each node
\Statex \quad \texttt{descendant\_mask}: Binary matrix ($N \times N$), where \texttt{descendant\_mask[i,j] = 1} if node $i$ is an ancestor of node $j$
\Statex \quad \texttt{lca\_matrix}: Matrix ($N \times N$), where \texttt{lca\_matrix[i,j]} is the lowest common ancestor (LCA) of nodes $i$ and $j$
\Statex \textbf{Output:}
\Statex \quad \texttt{Pattern\_sub\_inter}: Dictionary mapping each inter-subsegment block to its clustered pattern result
\State Group nodes by subsegment label $\rightarrow$ \texttt{indices\_by\_sub}
\State Group nodes by segment label $\rightarrow$ \texttt{indices\_by\_seg}
\State Initialize \texttt{SUB\_COTUNK\_INTER} as a zero matrix

\For{each subsegment label $i$}
    \If{no nodes exist for $i$}
        \State \textbf{continue}
    \EndIf
    \State Select representative node for $i$ with minimum generation

    \For{$j = 1$ to $i - 1$}
        \If{no nodes exist for $j$}
            \State \textbf{continue}
        \EndIf
        \If{subsegments $i$ and $j$ belong to the same segment}
            \State Check predefined co-tunk rules
            \If{rule is satisfied}
                \State Mark connection between $i$ and $j$ in \texttt{SUB\_COTUNK\_INTER}
            \EndIf
        \EndIf
        \State Select representative node for subsegment $j$
        \State $lca_{ij} \gets$ \texttt{lca\_matrix[rep$_i$, rep$_j$]}
        \If{$lca_{ij}$ is main bronchus \textbf{and} all descendants $\in \{i, j, \text{main bronchus}\}$}
            \State Mark connection between $i$ and $j$
        \EndIf
    \EndFor

    \For{each segment $k$}
        \If{$i \in$ segment $k$}
            \State \textbf{continue}
        \EndIf
        \If{nodes exist for segment $k$}
            \State Select representative node for segment $k$
            \State $lca_{ik} \gets$ \texttt{lca\_matrix[rep$_i$, rep$_k$]}
            \If{all descendant nodes have \texttt{segment\_label} = $k$ \textbf{or} \texttt{subsegment\_label} $\in \{i, \text{main bronchus}\}$}
                \For{each subsegment $s$ in segment $k$}
                    \State Mark connection between $i$ and $s$ in \texttt{SUB\_COTUNK\_INTER}
                \EndFor
            \EndIf
        \EndIf
    \EndFor
\EndFor

\For{each block in \texttt{BLOCKS\_INTER\_SUB}}
    \If{any corresponding segment is invalid based on intra-subsegment analysis}
        \State \textbf{continue}
    \Else
        \State Apply union-find clustering to subsegments in the block
        \State Apply uniform clustering to standardize the clusters
        \State Record result in \texttt{Pattern\_sub\_inter[block]}
    \EndIf
\EndFor

\State \Return \texttt{Pattern\_sub\_inter}
\end{algorithmic}
\end{algorithm}
\end{minipage}}
\addcontentsline{toc}{section}{Algorithm S3. Inter-Subsegmental Pattern Analysis.}
\end{figure}

\begin{figure}[htbp]
\centering
\resizebox{0.9\textwidth}{!}{
\begin{minipage}{1.0\textwidth}
\begin{algorithm}[H]
\caption{Computation of Stenosis ($\mathcal{S}$)}
\label{alg::stenosis}
\begin{algorithmic}[1]
\Require Binary airway mask $V_{\text{airway}}$, spacing $s$, skeleton $S$, parsed skeleton $S_{\text{parse}}$, segment and lobe labels $Y_{\text{seg}}, Y_{\text{lob}}$
\Ensure Segmental and lobar stenosis scores
\State $D \gets \text{EDT}(V_{\text{airway}}, \text{spacing}=s)$ \Comment{Euclidean distance transform}
\State $R \gets D \cdot S$ \Comment{Radius field along skeleton}
\State $L \gets \text{Unique labels from } S_{\text{parse}}$ where $l > 0$
\ForAll{$l \in L$}
    \State $r_{\min}^{(l)} \gets \min(R[S_{\text{parse}} = l])$
    \State $r_{\text{mean}}^{(l)} \gets \text{mean}(R[S_{\text{parse}} = l])$
\EndFor
\For{$j = 0$ to $N_{\text{seg}} - 1$}
    \If{$\exists l$ such that $Y_{\text{seg}}[l] = j$}
        \State $\text{Stenosis}_{\text{seg}}[j] \gets \text{mean} \left(1 - \frac{r_{\min}^{(l)}}{r_{\text{mean}}^{(l)}} \right)$
    \Else
        \State $\text{Stenosis}_{\text{seg}}[j] \gets -1$
    \EndIf
\EndFor
\For{$k = 0$ to $N_{\text{lob}} - 1$}
    \If{$\exists l$ such that $Y_{\text{lob}}[l] = k + 1$}
        \State $\text{Stenosis}_{\text{lob}}[k] \gets \text{mean} \left(1 - \frac{r_{\min}^{(l)}}{r_{\text{mean}}^{(l)}} \right)$
    \Else
        \State $\text{Stenosis}_{\text{lob}}[k] \gets -1$
    \EndIf
\EndFor
\State \Return $\text{Stenosis}_{\text{seg}}$, $\text{Stenosis}_{\text{lob}}$
\end{algorithmic}
\end{algorithm}
\end{minipage}
}
\addcontentsline{toc}{section}{Algorithm S4. Computation of Stenosis ($\mathcal{S}$).}
\end{figure}

\begin{figure}[htbp]
\centering
\resizebox{0.9\textwidth}{!}{
\begin{minipage}{1.0\textwidth}
\begin{algorithm}[H]
\caption{Computation of Ectasia ($\mathcal{E}$)}
\label{alg::ectasia}
\begin{algorithmic}[1]
\Require Same inputs as Algorithm~\ref{alg::stenosis}
\Ensure Segmental and lobar ectasia scores
\ForAll{$l \in L$}
\State $r_{\max}^{(l)} \gets \max(R[S_{\text{parse}} = l])$
\State $r_{\text{mean}}^{(l)} \gets \text{mean}(R[S_{\text{parse}} = l])$
\EndFor
\For{$j = 0$ to $N_{\text{seg}} - 1$}
\If{$\exists l$ such that $Y_{\text{seg}}[l] = j$}
    \State $\text{Ectasia}_{\text{seg}}[j] \gets \text{mean} \left( \frac{r_{\max}^{(l)}}{r_{\text{mean}}^{(l)}} \right)$
\Else
    \State $\text{Ectasia}_{\text{seg}}[j] \gets -1$
\EndIf
\EndFor
\For{$k = 0$ to $N_{\text{lob}} - 1$}
\If{$\exists l$ such that $Y_{\text{lob}}[l] = k + 1$}
    \State $\text{Ectasia}_{\text{lob}}[k] \gets \text{mean} \left( \frac{r_{\max}^{(l)}}{r_{\text{mean}}^{(l)}} \right)$
\Else
    \State $\text{Ectasia}_{\text{lob}}[k] \gets -1$
\EndIf
\EndFor
\State \Return $\text{Ectasia}_{\text{seg}}$, $\text{Ectasia}_{\text{lob}}$
\end{algorithmic}
\end{algorithm}
\end{minipage}
}
\addcontentsline{toc}{section}{Algorithm S5. Computation of Ectasia ($\mathcal{E}$).}
\end{figure}

\begin{figure}[htbp]
\centering
\resizebox{0.9\textwidth}{!}{
\begin{minipage}{1.0\textwidth}
\begin{algorithm}[H]
\caption{Computation of Tortuosity ($\mathcal{T}$)}
\label{alg::tortuosity}
\begin{algorithmic}[1]
\Require Voxel set $\Omega_l$ for a branch, voxel spacing $s$
\Ensure Tortuosity angle $\alpha$
\State Convert $\Omega_l$ to physical space using $s$
\State Use PCA to find principal axis and endpoints $S$, $E$
\State Identify $P = \arg\max_{x_i \in \Omega_l} \; \text{dist}(x_i, \overline{SE})$
\State Compute $\alpha = \angle(\vec{PS}, \vec{PE})$
\State \Return $\alpha$
\end{algorithmic}
\end{algorithm}
\end{minipage}
}
\addcontentsline{toc}{section}{Algorithm S6.Computation of Tortuosity ($\mathcal{T}$).}
\end{figure}

\begin{figure}[htbp]
\centering
\resizebox{0.9\textwidth}{!}{
\begin{minipage}{1.0\textwidth}
\begin{algorithm}[H]
\caption{Computation of Length ($\mathcal{L}$) }
\label{alg::geodesic_length}
\begin{algorithmic}[1]
\Require Tree edge list, generation array, label array $Y$, nodewise lengths $\ell$, target class $K$
\Ensure Mean geodesic length for class $K$
\State Identify all leaf nodes with label $K$
\State Compute their lowest common ancestor (LCA)
\If{LCA $\in$ class $K$}
    \State Accumulate all $\text{Path(LCA} \rightarrow \text{Leaf)}$ lengths
\Else
    \ForAll{leaf nodes}
        \State Trace from LCA down to nearest node with label $K$
        \State Compute path length from this node to leaf
    \EndFor
\EndIf
\State \Return $\mathcal{L}_K$
\end{algorithmic}
\end{algorithm}
\end{minipage}
}
\addcontentsline{toc}{section}{Algorithm S7. Computation of Length ($\mathcal{L}$).}
\end{figure}

\begin{figure}[htbp]
\centering
\resizebox{0.9\textwidth}{!}{
\begin{minipage}{1.0\textwidth}
\begin{algorithm}[H]
\caption{Computation of Divergence ($\mathcal{D}$)}
\label{alg::divergence_angle}
\begin{algorithmic}[1]
\Require Tree edge list, generation array, label array $Y$, node coordinates $\mathbf{x}$, class label $K$
\Ensure Divergence angle $\theta$ in radians

\State Identify leaf nodes labeled as $K$
\State Compute their lowest common ancestor (LCA)
\State Let apex $\gets$ coordinate of LCA
\ForAll{leaf nodes $i$}
    \State $\mathbf{v}_i \gets$ normalized vector from apex to leaf $i$
\EndFor
\State Optimize unit vector $\mathbf{u}$ to maximize $\min_i \langle \mathbf{u}, \mathbf{v}_i \rangle$
\State Compute: $\theta = 2 \cdot \arccos\left( \min_i \langle \mathbf{u}, \mathbf{v}_i \rangle \right)$
\State \Return $\theta$
\end{algorithmic}
\end{algorithm}
\end{minipage}
}
\addcontentsline{toc}{section}{Algorithm S8. Computation of Divergence ($\mathcal{D}$).}
\end{figure}

\begin{figure}[htbp]
\centering
\resizebox{0.9\textwidth}{!}{
\begin{minipage}{1.0\textwidth}
\begin{algorithm}[H]
\caption{Computation of Complexity ($\mathcal{C}$) }
\label{alg::branch_complexity}
\begin{algorithmic}[1]
\Require 3D skeleton volume $S$, class label mask $Y$, target class $K$
\Ensure Branch Complexity $\mathcal{C}$
\State Extract binary skeleton mask $S_K = (S > 0) \land (Y = K)$
\State Crop $S_K$ to its minimal bounding box and pad to fixed size
\For{each box size $s \in \{2^1, \dots, 2^k\}$}
    \State Count number $N(s)$ of non-empty boxes of size $s$
\EndFor
\State Perform linear regression on $(\log(1/s), \log N(s))$
\State \Return slope $\mathcal{C}$
\end{algorithmic}
\end{algorithm}
\end{minipage}
}
\addcontentsline{toc}{section}{Algorithm S9. Computation of Complexity ($\mathcal{C}$).}
\end{figure}

\begin{figure}[h]
\centering
\includegraphics[width=1.0\linewidth]{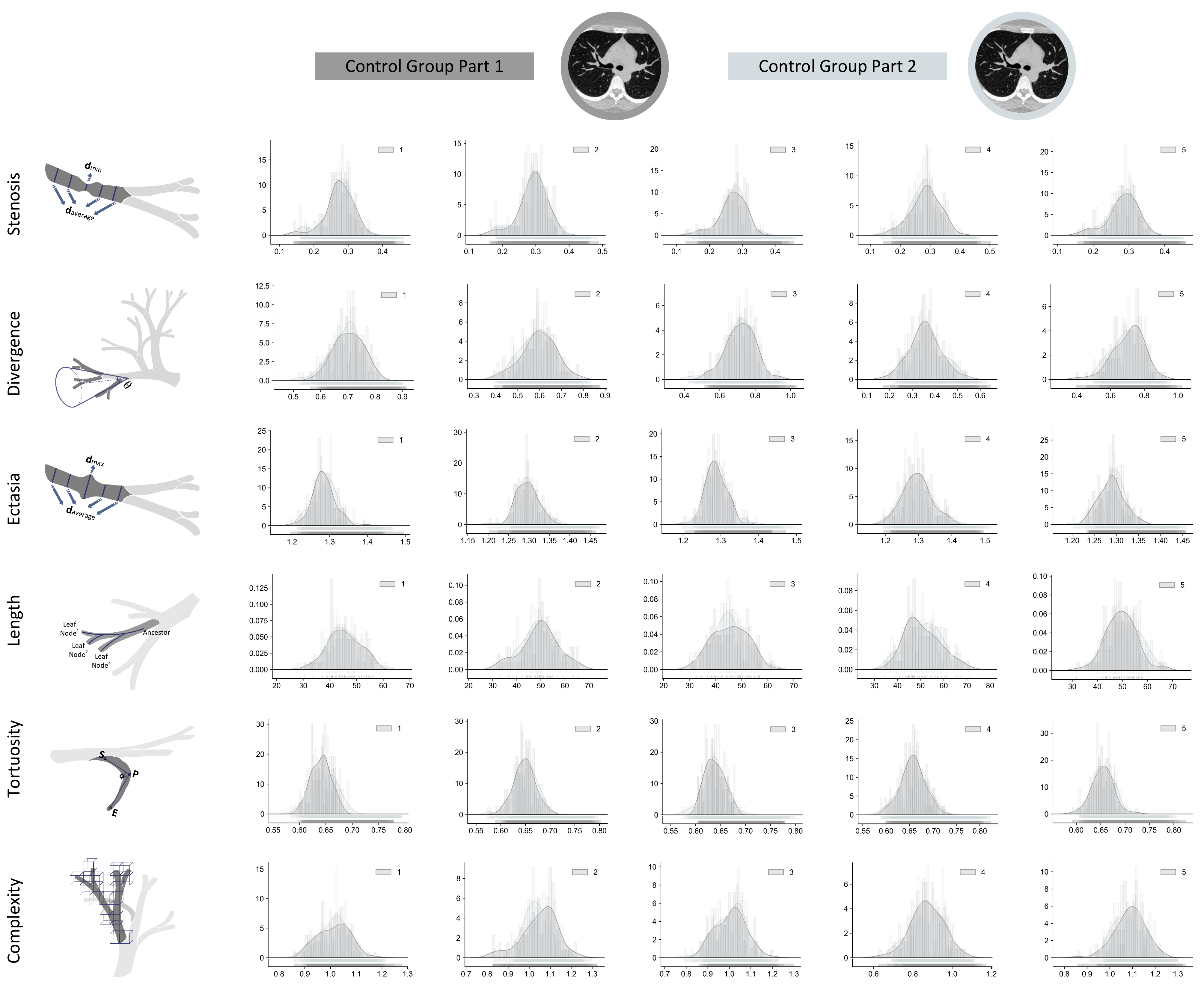}
\caption{Ablation study of the AirwaySignature within the healthy controls.}\label{fig::AirMorph_morpho_distribution_within_controlgroup}
\addcontentsline{toc}{section}{Fig. S10: Ablation study of the AirwaySignature within the healthy controls.}
\end{figure}

\begin{figure}[h]
\centering
\includegraphics[width=1.0\linewidth]{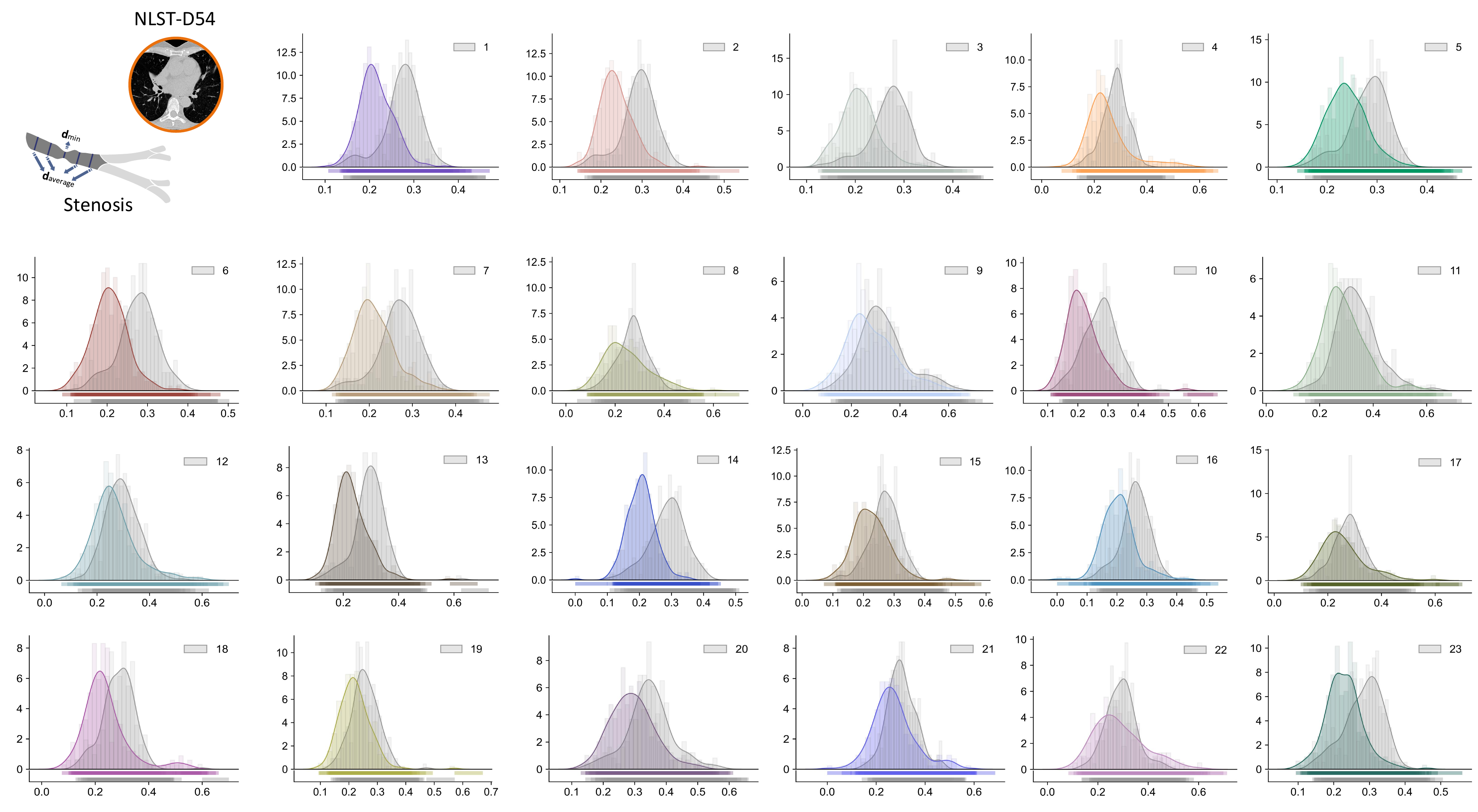}
\caption{Example of the Stenosis signature on all branches of the NLST-D54}\label{fig::AirMorph_morpho_distribution_of_D54Stenosis}
\addcontentsline{toc}{section}{Fig. S11: Example of the Stenosis signature.}
\end{figure}

\begin{figure}[ht]
\centering
\includegraphics[width=0.7\linewidth]{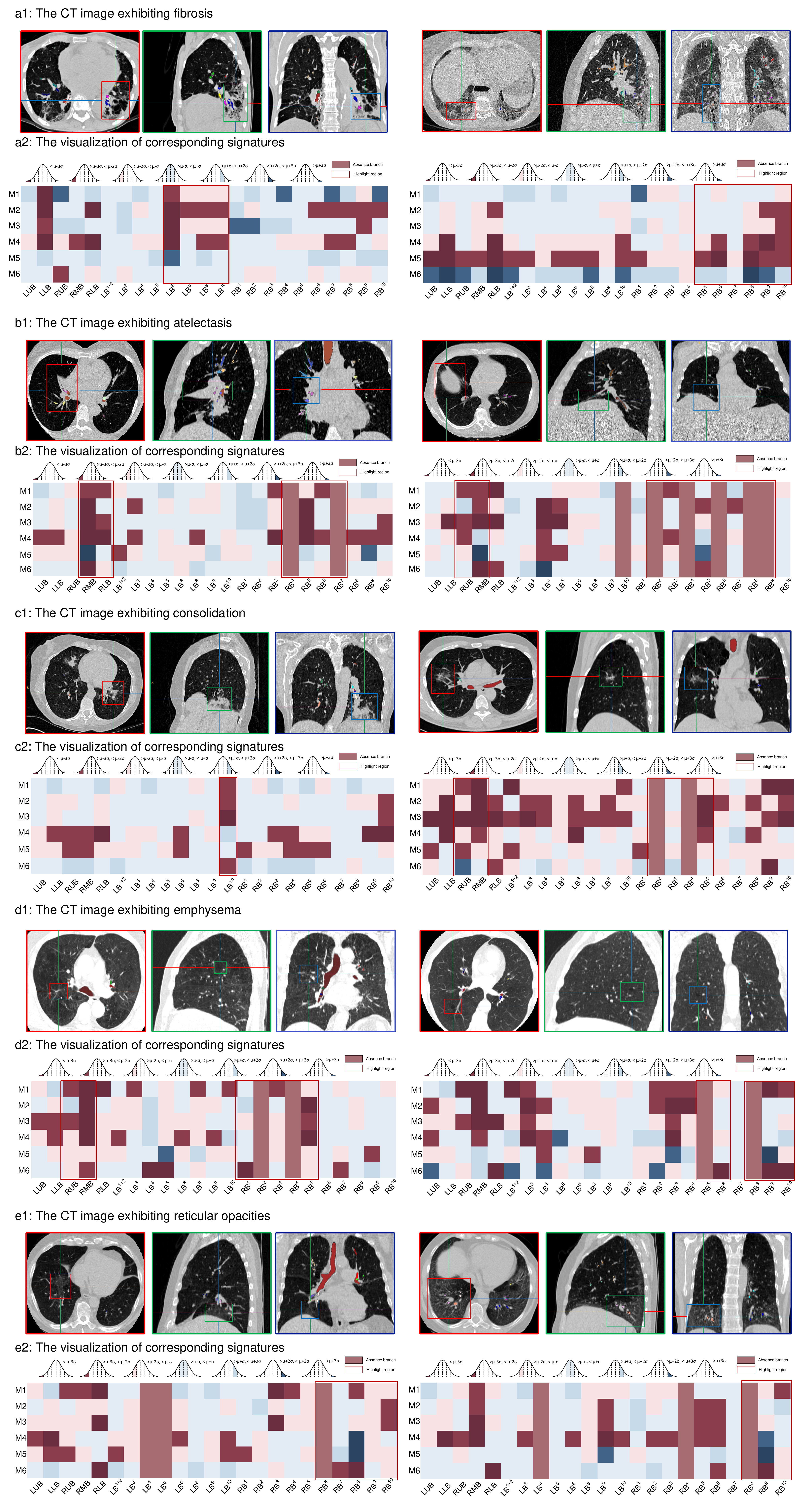}
\caption{AirwaySignature across all five lung diseases.}\label{fig::AirMorph_morpho_heatmapwithimge_all_five_diseases}
\addcontentsline{toc}{section}{Fig. S12: AirwaySignature across all five lung diseases.}
\end{figure}

\begin{figure}[ht]
\centering
\includegraphics[width=1.0\linewidth]{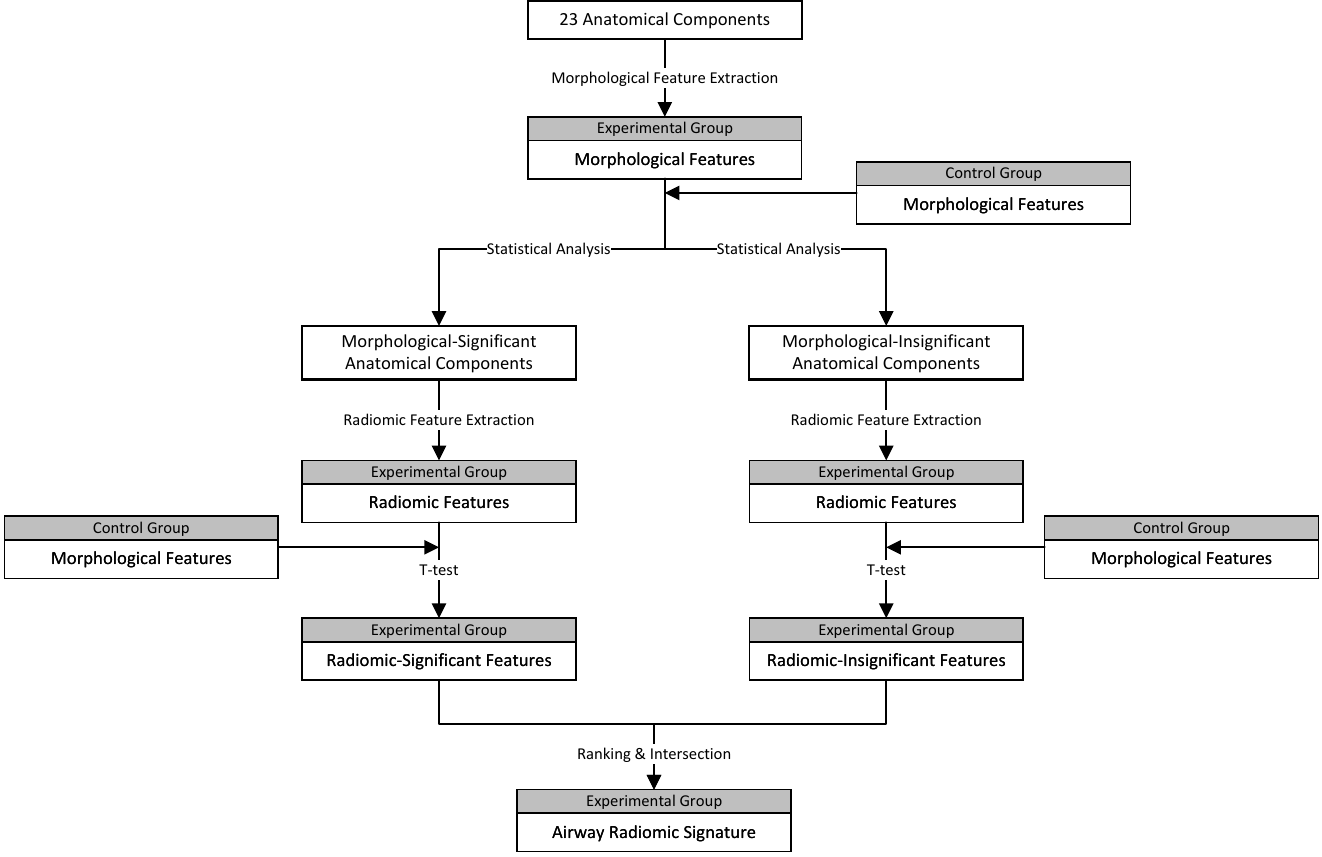}
\caption{
Pipeline for Radiomic signature selection.
}\label{fig::radiomic_feature_selection}
\addcontentsline{toc}{section}{Fig. S13: Pipeline for Radiomic signature selection.}
\end{figure}

\begin{figure}[ht]
\centering
\includegraphics[width=0.8\linewidth]{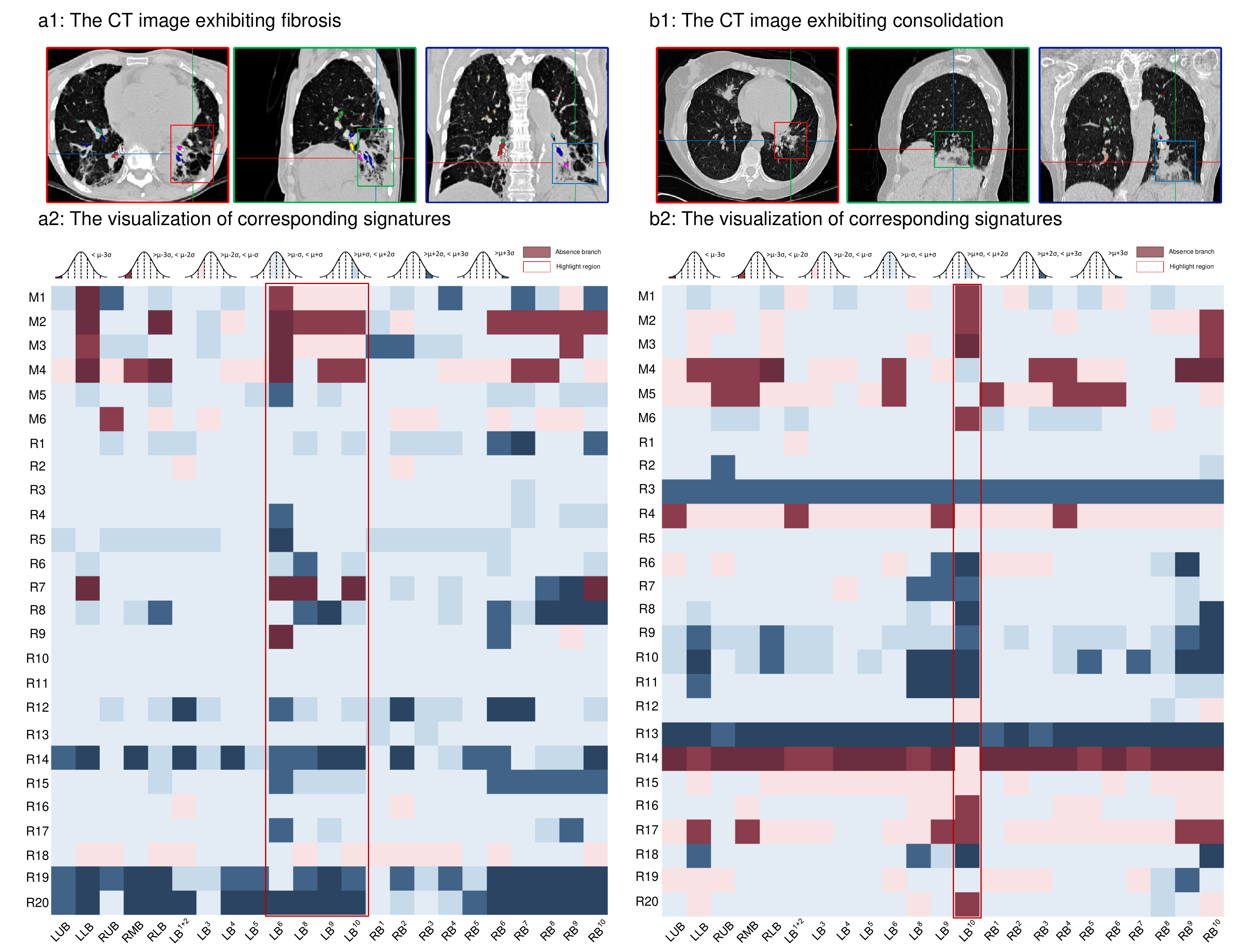}
\caption
{Representative cases with the AirwaySignature, along with the radiomics features.}\label{fig::AirMorph_full_features_with_radiomics}
\addcontentsline{toc}{section}{Fig. S14: Representative cases with the AirwaySignature and radiomics features.}
\end{figure}
\makeatletter
\def\hlinew#1{%
	\noalign{\ifnum0=`}\fi\hrule \@height #1 \futurelet
	\reserved@a\@xhline}
\makeatother
\begin{table*}[ht]
\renewcommand\arraystretch{1.3}
\caption{Quantative results on generalization performance with regard to the binary airway modeling on the cases with severe pulmonary fibrosis. }
\label{tab::AirMorph_seg_generalization_on_pulmonary_fibrosis}
\centering 
\resizebox{0.9\textwidth}{!}{
\begin{tabular}{>{\centering\arraybackslash}p{2cm}>{\centering\arraybackslash}p{2cm}>{\centering\arraybackslash}p{2cm}>{\centering\arraybackslash}p{2cm}>{\centering\arraybackslash}p{2cm}>{\centering\arraybackslash}p{2cm}}
\hlinew{1pt}
\rowcolor{tablecolor1}
\multicolumn{6}{c}{\textbf{Airway Atlas: Binary Airway Modeling}}       \\ \hlinew{1pt}
\rowcolor{tablecolor2}
\multicolumn{6}{l}{\textit{Lobar}}                                      \\ \hlinew{0.8pt}
\textbf{Method}      & \textbf{TLD}    & \textbf{BND}    & \textbf{DSC}   & \textbf{clDice} & \textbf{Sensitivity}    \\
UNet        & 46.43 & 49.36 & 62.39 & 58.76  & 54.89 \\
nnUNet      & 48.06 & 49.97 & 61.97 & 59.67  & 54.65 \\
AirMorph & \textbf{78.49} &\textbf{81.49} & \textbf{74.78} & \textbf{76.02}  & \textbf{81.91} \\ \hlinew{1pt}
\rowcolor{tablecolor2}
\multicolumn{6}{l}{\textit{Segmental}}                          \\ \hlinew{0.8pt}
\textbf{Method}      & \textbf{TLD}    & \textbf{BND}    & \textbf{DSC}   & \textbf{clDice} & \textbf{Sensitivity}    \\
UNet        & 42.81 & 45.45 & 52.88 & 50.49  & 49.02  \\ 
nnUNet      & 45.00    & 46.78 & 53.63 & 52.15  & 49.89 \\ 
AirMorph & \textbf{73.48} & \textbf{76.80}  & \textbf{68.30}  & \textbf{69.69}  & \textbf{75.99} \\ \hlinew{1pt}
\rowcolor{tablecolor2}
\multicolumn{6}{l}{\textit{SubSegmental}}                         \\ \hlinew{0.8pt}
\textbf{Method}      & \textbf{TLD}    & \textbf{BND}    & \textbf{DSC}   & \textbf{clDice} & \textbf{Sensitivity}    \\ 
UNet        & 40.23 & 43.73 & 42.43 & 43.7   & 41.13  \\
nnUNet      & 43.03 & 46.29 & 44.15 & 46.41  & 42.56 \\
AirMorph & \textbf{66.46} & \textbf{70.88} & \textbf{57.63} & \textbf{61.94}  & \textbf{66.09} \\\hlinew{1pt}
\end{tabular}}
\addcontentsline{toc}{section}{Table S1: Quantative results on generalization performance with regard to the binary airway modeling on the cases with severe pulmonary fibrosis.}
\end{table*}

\makeatletter
\def\hlinew#1{%
	\noalign{\ifnum0=`}\fi\hrule \@height #1 \futurelet
	\reserved@a\@xhline}
\makeatother
\begin{table*}[ht]
\renewcommand\arraystretch{1.3}
\caption{Quantative results on generalization performance with regard to the anatomical airway labeling on the cases with severe pulmonary fibrosis.}
\label{tab::AirMorph_cls_generalization_on_pulmonary_fibrosis}
\centering 
\resizebox{0.9\textwidth}{!}{
\begin{tabular}{>{\centering\arraybackslash}p{2cm}>{\centering\arraybackslash}p{2cm}>{\centering\arraybackslash}p{2cm}>{\centering\arraybackslash}p{2cm}>{\centering\arraybackslash}p{2cm}>{\centering\arraybackslash}p{2cm}}
\hlinew{1pt}
\rowcolor{tablecolor1}
\multicolumn{6}{c}{\textbf{Airway Atlas: Airway Anatomical Labeling}}   \\\hlinew{1pt}
\rowcolor{tablecolor2}
\multicolumn{6}{l}{\textit{Lobar}}                                      \\\hlinew{0.8pt}
\textbf{Method}     & \textbf{TreeCons}    & \textbf{TopoDist}    & \textbf{Accuracy}   & \textbf{Precision}     & \textbf{Sensitivity}        \\
GCN         &   91.60    &  0.6058     &  90.13  &    90.52    &  88.55    \\
TNN         &   \textbf{99.35}   &  0.3069     &  95.18  &   93.93     &   94.04       \\
AirMorph &   98.73   &  \textbf{0.2382}     & \textbf{97.20}   &   \textbf{97.04}     &  \textbf{96.23}       \\\hlinew{1pt}
\rowcolor{tablecolor2}
\multicolumn{6}{l}{\textit{Segmental}}                                  \\\hlinew{0.8pt}
\textbf{Method}     & \textbf{TreeCons}    & \textbf{TopoDist}    & \textbf{Accuracy}   & \textbf{Precision}     & \textbf{Sensitivity}          \\
GCN         &  55.33      &   1.9646    &  66.93    &  60.25      &  65.69         \\
TNN         &   81.34     &  1.1834     &  80.08    &  69.35      &    77.05     \\
AirMorph &   \textbf{97.26}     &  \textbf{ 0.6198  }  & \textbf{ 91.46 }  & \textbf{  88.69    } &  \textbf{ 89.71   }     \\\hlinew{1pt}
\rowcolor{tablecolor2}
\multicolumn{6}{l}{\textit{SubSegmental}}                               \\\hlinew{0.8pt}
\textbf{Method}     & \textbf{TreeCons}    & \textbf{TopoDist}    & \textbf{Accuracy}   & \textbf{Precision}     & \textbf{Sensitivity}          \\
GCN         &  51.49     &  3.8864     &  46.38   &  33.85      &   39.22         \\
TNN         & 86.32     &   2.3539    &    65.61    &  53.39      &   59.48    \\
AirMorph & \textbf{91.37}     & \textbf{ 1.7423}     &  \textbf{83.52  }    &  \textbf{ 80.59  }   &   \textbf{81.62 }  \\ \hlinew{1pt}
\end{tabular}}
\addcontentsline{toc}{section}{Table S2: Quantative results on generalization performance with regard to the anatomical airway labeling on the cases with severe pulmonary fibrosis.}
\end{table*}

\makeatletter
\def\hlinew#1{%
	\noalign{\ifnum0=`}\fi\hrule \@height #1 \futurelet
	\reserved@a\@xhline}
\makeatother
\begin{table*}[ht]
\renewcommand\arraystretch{1.3}
\caption{Quantative results on graph building results.}
\label{tab::AirMorph_graph_building}
\centering 
\resizebox{\textwidth}{!}{
\begin{tabular}{>{\centering\arraybackslash}p{2cm}>{\centering\arraybackslash}p{2.5cm}>{\centering\arraybackslash}p{2.5cm}>{\centering\arraybackslash}p{2.5cm}>{\centering\arraybackslash}p{2cm}}
\hlinew{1pt}
\rowcolor{tablecolor1}
Metrics   & Skeletonization & SoftSkel-clDice & GBO-Skeleton  & Ours        \\ \hlinew{1pt}
$\beta_{0}$ (Euler)  & 1.00 ± 0.00      & 1242.8 ± 407.5      & 1.00 ± 0.00   & \textbf{1.00 ± 0.00} \\
$\beta_{1}$ (Euler)  & 16.55 ± 13.93   & 12.86 ± 7.23        & 16.55 ± 13.93 & \textbf{0.00 ± 0.00 }\\ \hlinew{1pt}
\end{tabular}
}
\addcontentsline{toc}{section}{Table S3: Quantative results on graph building results.}
\end{table*}

\makeatletter
\def\hlinew#1{%
	\noalign{\ifnum0=`}\fi\hrule \@height #1 \futurelet
	\reserved@a\@xhline}
\makeatother
\begin{table*}[ht]
\renewcommand\arraystretch{1.4}
\centering
\caption{The comparative results of the binary airway modeling on the multi-site test cohorts.}
\label{tab::AirMorph_qualitative_seg}
\resizebox{\textwidth}{!}{
\begin{tabular}{cccccc}
\hlinew{1pt}
\rowcolor{tablecolor1}
\multicolumn{6}{l}{\textbf{Reconstructed Tree Length}}                                                                                                     \\\hlinew{1pt}
            & LIDC-IDRI                  & NLST-D54                  & NLST-D58                  & NLST-D59                   & NLST-D61                   \\\hlinew{1pt}
UNet        & 1698.91 ± 842.67           & 1808.02 ± 636.40          & 1744.98 ± 597.87          & 2322.29 ± 817.41           & 2267.16 ± 790.48           \\
nnUNet      & 1968.21 ± 939.96           & 1969.59 ± 628.83          & 1960.64 ± 613.45          & 2539.81 ± 873.58           & 2512.42 ± 803.66           \\
AirMorph & \textbf{2876.13 ± 1355.83} & \textbf{3078.86 ± 943.34} & \textbf{2890.10 ± 939.45} & \textbf{3668.33 ± 1294.32} & \textbf{3643.27 ± 1205.12} \\
\hlinew{1pt}
\rowcolor{tablecolor1}
\multicolumn{6}{l}{\textbf{Reconstructed Branch Number}}                                                                                                   \\\hlinew{1pt}
            & LIDC-IDRI                  & NLST-D54                  & NLST-D58                  & NLST-D59                   & NLST-D61                   \\\hlinew{1pt}
UNet        & 88.61 ± 52.70              & 92.93 ± 36.47             & 87.82 ± 34.98             & 134.43 ± 55.66             & 137.23 ± 56.40             \\
nnUNet      & 97.98 ± 54.83              & 97.47 ± 34.75             & 95.28 ± 34.12             & 131.90 ± 48.86             & 133.04 ± 46.12             \\
AirMorph & \textbf{157.92 ± 84.45}    & \textbf{170.93 ± 54.34}   & \textbf{154.83 ± 52.48}   & \textbf{205.42 ± 78.45}    & \textbf{209.21 ± 74.86}   \\\hlinew{1pt}
\end{tabular}
}
\addcontentsline{toc}{section}{Table S4: The comparative results of the binary airway modeling on the multi-site test cohorts.}
\end{table*}

\makeatletter
\def\hlinew#1{%
	\noalign{\ifnum0=`}\fi\hrule \@height #1 \futurelet
	\reserved@a\@xhline}
\makeatother
\begin{table*}[ht]
\renewcommand\arraystretch{1.4}
\centering
\caption{The comparative results of the airway anatomical labeling on the multi-site test cohorts.}
\label{tab::AirMorph_qualitative_cls}
\resizebox{\textwidth}{!}{
\begin{tabular}{cccccc}
\hlinew{1pt}
\rowcolor{tablecolor1}
\multicolumn{6}{l}{\textbf{TreeCons}}                                                                                                                                   \\\hlinew{1pt}
            & \multicolumn{1}{c}{LIDC-IDRI} & \multicolumn{1}{c}{NLST-D54} & \multicolumn{1}{c}{NLST-D58} & \multicolumn{1}{c}{NLST-D59} & \multicolumn{1}{c}{NLST-D61} \\\hlinew{1pt}
GCN         & 36.86 ± 11.07                 & 40.63 ± 7.89                 & 39.22 ± 8.55                 & 44.13 ± 8.38                 & 44.56 ± 8.03                 \\
TNN         & 58.74 ± 12.52                 & 62.63 ± 8.14                 & 60.16 ± 8.47                 & 66.00 ± 8.92                 & 66.58 ± 8.15                 \\
AirMorph & \textbf{89.57 ± 12.52}        & \textbf{93.90 ± 4.89}        & \textbf{93.72 ± 5.07}        & \textbf{95.36 ± 3.54}        & \textbf{95.28 ± 4.22}        \\
\hlinew{1pt}
\rowcolor{tablecolor1}
\multicolumn{6}{l}{\textbf{TopoDist}}                                                                                                                                   \\\hlinew{1pt}
            & \multicolumn{1}{c}{LIDC-IDRI} & \multicolumn{1}{c}{NLST-D54} & \multicolumn{1}{c}{NLST-D58} & \multicolumn{1}{c}{NLST-D59} & \multicolumn{1}{c}{NLST-D61} \\\hlinew{1pt}
GCN         & 1.95 ± 0.37                   & 1.92 ± 0.28                  & 1.86 ± 0.24                  & 1.94 ± 0.28                  & 1.96 ± 0.31                  \\
nnUNet      & 1.42 ± 0.24                   & 1.40 ± 0.17                  & 1.40 ± 0.16                  & 1.35 ± 0.17                  & 1.35 ± 0.18                  \\
AirMorph & \textbf{1.17 ± 0.27}          & \textbf{1.11 ± 0.07}         & \textbf{1.11 ± 0.08}         & \textbf{1.10 ± 0.07}         & \textbf{1.11 ± 0.11}         \\ \hlinew{1pt}
\end{tabular}
}
\addcontentsline{toc}{section}{Table S5: The comparative results of the airway anatomical labeling on the multi-site test cohorts.}
\end{table*}

\makeatletter
\def\hlinew#1{%
	\noalign{\ifnum0=`}\fi\hrule \@height #1 \futurelet
	\reserved@a\@xhline}
\makeatother
\begin{table*}[ht]
\renewcommand\arraystretch{1.3}
\caption{Ablation study of the graph building on ATM'22 internal testset.}
\label{tab::AirMorph_graph_building_ablation_study_on_ATM22}
\centering 
\resizebox{\textwidth}{!}{
\begin{tabular}{>{\centering\arraybackslash}p{2cm}>{\raggedright\arraybackslash}p{3.0cm}>{\centering\arraybackslash}p{1.6cm}>{\centering\arraybackslash}p{1.6cm}>{\centering\arraybackslash}p{1.6cm}>{\centering\arraybackslash}p{1.6cm}>{\centering\arraybackslash}p{1.6cm}}
\hlinew{1pt}
\rowcolor{tablecolor1}
\multicolumn{7}{l}{\textit{Lobar}}                                      \\ \hlinew{0.8pt}
\textbf{Method}   & \textbf{GraphBuild}     & \textbf{TreeCons}    & \textbf{TopoDist}    & \textbf{Accuracy}   & \textbf{Precision}     & \textbf{Sensitivity}  \\ \hlinew{0.8pt}
\multirow{4}{*}{GCN}    & Skeletonization    & 95.71  & 0.29717 & 94.11 & 94.25 & 91.23  \\
                        & MPC-Skel (Ours)   & 98.16  & 0.12249 & 96.73 & 96.64 & 95.77 \\
                        & \,\,Ablation: HRLP  & 87.24  & 2.11652 & 74.60 & 78.70 & 70.23  \\
                        & \,\,Ablation: LRHP  & 98.88  & 0.0971  & 96.41 & 96.48 & 95.51 \\ \hlinew{1pt}
\multirow{4}{*}{TNN}    & Skeletonization    & 97.75  & 0.62867 & 91.95 & 91.82 & 90.61 \\
                        & MPC-Skel (Ours)   & 100.00 & 0.03492 & 98.00 & 98.02 & 96.82 \\
                        & \,\,Ablation: HRLP  & 94.79  & 2.25155 & 76.59 & 73.50 & 73.13  \\
                        & \,\,Ablation: LRHP  & 98.72  & 0.48659 & 91.29 & 89.54 & 90.16 \\ \hlinew{1pt}
\multirow{4}{*}{AirMorph}    & Skeletonization    & 98.45  & 0.09023 & 97.72 & 97.92 & 96.29  \\
                                & MPC-Skel (Ours)   & 99.99  & 0.02603 & 99.06 & 99.20 & 98.52  \\
                                & \,\,Ablation: HRLP  & 89.26  & 1.48449 & 81.57 & 82.19 & 79.39 \\
                                & \,\,Ablation: LRHP  & 99.87  & 0.03092 & 98.65 & 98.81 & 98.31  \\ \hlinew{1pt}
\rowcolor{tablecolor1}
\multicolumn{7}{l}{\textit{Segmental}}                                      \\ \hlinew{0.8pt}
\textbf{Method}   & \textbf{GraphBuild} & \textbf{TreeCons}  & \textbf{TopoDist}  & \textbf{Accuracy} & \textbf{Precision}& \textbf{Sensitivity} \\\hlinew{0.8pt}
\multirow{4}{*}{GCN}    & Skeletonization     & 63.93             & 1.63591 & 75.29 & 68.91 & 74.31         \\
                        & MPC-Skel (Ours)  & 67.38 & 0.90323 & 83.09 & 75.92 & 81.51  \\
                        & \,\,Ablation: HRLP  & 49.40 & 4.79132 & 46.80 & 42.73 & 49.21 \\
                        & \,\,Ablation: LRHP   & 63.45 & 0.87471 & 82.15 & 73.58 & 80.85  \\ \hlinew{1pt}
\multirow{4}{*}{TNN}    & Skeletonization     & 83.56 & 1.34551 & 80.56 & 70.94 & 78.18    \\
                        & MPC-Skel (Ours)   & 83.46 & 0.41358 & 89.16 & 79.59 & 87.18   \\
                        & \,\,Ablation: HRLP & 76.86 & 4.46636 & 53.86 & 43.88 & 52.72  \\
                        & \,\,Ablation: LRHP   & 71.56 & 1.70571 & 71.67 & 58.54 & 67.41 \\ \hlinew{1pt}
\multirow{4}{*}{AirMorph}    & Skeletonization    & 96.40 & 0.46364 & 92.71 & 91.92 & 92.88  \\
                                & MPC-Skel (Ours)    & 99.47 & 0.10241 & 97.42 & 96.84 & 97.12  \\
                                & \,\,Ablation: HRLP   & 82.16 & 4.55093 & 60.15 & 55.81 & 60.44  \\
                                & \,\,Ablation: LRHP   & 99.08 & 0.14258 & 96.48 & 94.77 & 95.41 \\ \hlinew{1pt}
\rowcolor{tablecolor1}
\multicolumn{7}{l}{\textit{SubSegmental}}                                      \\ \hlinew{0.8pt}
\textbf{Method}   & \textbf{GraphBuild} & \textbf{TreeCons}  & \textbf{TopoDist}  & \textbf{Accuracy} & \textbf{Precision}& \textbf{Sensitivity}  \\\hlinew{0.8pt}
\multirow{4}{*}{GCN}    & Skeletonization      & 57.73 & 3.27129 & 55.85 & 45.94 & 49.05 \\
                        & MPC-Skel (Ours)  & 65.93 & 2.27723 & 63.89 & 52.83 & 57.31 \\
                        & \,\,Ablation: HRLP  & 35.12 & 7.25278 & 25.52 & 14.69 & 17.67  \\
                        & \,\,Ablation: LRHP   & 67.93 & 3.15233 & 62.88 & 51.53 & 56.25  \\ \hlinew{1pt}
\multirow{4}{*}{TNN}    & Skeletonization     & 79.54 & 4.14947 & 52.20 & 39.50 & 45.46  \\
                        & MPC-Skel (Ours)   & 86.32 & 2.35388 & 65.61 & 53.39 & 59.48 \\
                        & \,\,Ablation: HRLP  & 69.46 & 7.01118 & 36.15 & 23.69 & 28.67  \\
                        & \,\,Ablation: LRHP   & 81.86 & 3.90828 & 57.38 & 43.84 & 48.62  \\ \hlinew{1pt}
\multirow{4}{*}{AirMorph}    & Skeletonization     & 90.76 & 1.06177 & 87.65 & 85.71 & 86.40  \\
                                & MPC-Skel (Ours)    & 98.02 & 0.48075 & 94.12 & 93.12 & 93.33  \\
                                & \,\,Ablation: HRLP   & 62.04 & 4.71749 & 53.18 & 43.34 & 45.88  \\
                                & \,\,Ablation: LRHP   & 97.44 & 1.73777 & 90.69 & 89.82 & 89.76  \\ \hlinew{1pt}
\end{tabular}}
\addcontentsline{toc}{section}{Table S6: Ablation study of the graph building on ATM'22 internal testset.}
\end{table*}

\makeatletter
\def\hlinew#1{%
	\noalign{\ifnum0=`}\fi\hrule \@height #1 \futurelet
	\reserved@a\@xhline}
\makeatother
\begin{table*}[ht]
\renewcommand\arraystretch{1.3}
\caption{Ablation study of the graph building on AIIB'23 internal testset.}
\label{tab::AirMorph_graph_building_ablation_study_on_AIIB23}
\centering 
\resizebox{\textwidth}{!}{
\begin{tabular}{>{\centering\arraybackslash}p{2cm}>{\raggedright\arraybackslash}p{3.0cm}>{\centering\arraybackslash}p{1.6cm}>{\centering\arraybackslash}p{1.6cm}>{\centering\arraybackslash}p{1.6cm}>{\centering\arraybackslash}p{1.6cm}>{\centering\arraybackslash}p{1.6cm}}
\hlinew{1pt}
\rowcolor{tablecolor1}
\multicolumn{7}{l}{\textit{Lobar}}                                      \\ \hlinew{0.8pt}
\textbf{Method}   & \textbf{GraphBuild} & \textbf{TreeCons}  & \textbf{TopoDist}  & \textbf{Accuracy} & \textbf{Precision}& \textbf{Sensitivity}  \\ \hlinew{0.8pt}
\multirow{4}{*}{GCN}    & Skeletonization    & 90.43 & 1.1117  & 85.49 & 86.82 & 81.82  \\
                        & MPC-Skel (Ours)   & 91.60 & 0.60576 & 90.13 & 90.52 & 88.55  \\
                        & \,\,Ablation: HRLP  & 87.45 & 2.16909 & 76.57 & 78.59 & 71.37 \\
                        & \,\,Ablation: LRHP   & 89.51 & 0.75638 & 87.41 & 87.84 & 87.02 \\ \hlinew{1pt}
\multirow{4}{*}{TNN}    & Skeletonization    & 96.41 & 1.06921 & 86.56 & 84.99 & 84.90  \\
                        & MPC-Skel (Ours)   & 99.35 & 0.30692 & 95.18 & 93.93 & 94.04  \\
                        & \,\,Ablation: HRLP  & 94.79 & 2.25155 & 76.59 & 73.50 & 73.13  \\
                        & \,\,Ablation: LRHP  & 98.72 & 0.48659 & 91.29 & 89.54 & 90.16  \\ \hlinew{1pt}
\multirow{4}{*}{AirMorph}    & Skeletonization    & 95.71 & 0.53392 & 93.96 & 94.06 & 92.18  \\
                                & MPC-Skel (Ours)  & 98.73 & 0.23820 & 97.20 & 97.04 & 96.23  \\
                                & \,\,Ablation: HRLP  & 89.13 & 1.37233 & 84.26 & 84.82 & 82.45  \\
                                & \,\,Ablation: LRHP  & 97.17 & 0.29594 & 95.62 & 95.36 & 94.99  \\ \hlinew{1pt}
\rowcolor{tablecolor1}
\multicolumn{7}{l}{\textit{Segmental}}                                      \\ \hlinew{0.8pt}
\textbf{Method}   & \textbf{GraphBuild} & \textbf{TreeCons}  & \textbf{TopoDist}  & \textbf{Accuracy} & \textbf{Precision}& \textbf{Sensitivity} \\\hlinew{0.8pt}
\multirow{4}{*}{GCN}    & Skeletonization     & 55.75 & 3.10851 & 59.43 & 54.45 & 58.90  \\
                        & MPC-Skel (Ours)  & 55.33 & 1.96458 & 66.93 & 60.25 & 65.69  \\
                        & \,\,Ablation: HRLP  & 49.62 & 5.35382 & 44.81 & 40.76 & 46.58  \\
                        & \,\,Ablation: LRHP   & 49.35 & 2.15344 & 65.08 & 55.61 & 64.40  \\ \hlinew{1pt}
\multirow{4}{*}{TNN}    & Skeletonization      & 81.04 & 2.60559 & 66.46 & 56.10 & 63.51 \\
                        & MPC-Skel (Ours)   & 81.34 & 1.18344 & 80.08 & 69.35 & 77.05 \\
                        & \,\,Ablation: HRLP  & 76.86 & 4.46636 & 53.86 & 43.88 & 52.72 \\
                        & \,\,Ablation: LRHP  & 71.56 & 1.70571 & 71.67 & 58.54 & 67.41  \\\hlinew{1pt}
\multirow{4}{*}{AirMorph}    & Skeletonization   & 90.52 & 1.61192 & 84.50 & 81.87 & 83.21  \\
                                & MPC-Skel (Ours)   & 97.26 & 0.61977 & 91.46 & 88.69 & 89.71  \\
                                & \,\,Ablation: HRLP   & 78.29 & 4.48236 & 61.79 & 58.44 & 62.38 \\
                                & \,\,Ablation: LRHP   & 96.71 & 0.90482 & 87.45 & 81.36 & 83.75  \\ \hlinew{1pt}
\rowcolor{tablecolor1}
\multicolumn{7}{l}{\textit{SubSegmental}}                                      \\ \hlinew{0.8pt}
\textbf{Method}   & \textbf{GraphBuild} & \textbf{TreeCons}  & \textbf{TopoDist}  & \textbf{Accuracy} & \textbf{Precision}& \textbf{Sensitivity} \\\hlinew{0.8pt}
\multirow{4}{*}{GCN}    & Skeletonization       & 44.09 & 5.20294 & 39.24 & 27.44 & 30.78 \\
                        & MPC-Skel (Ours) & 51.49 & 3.88637 & 46.38 & 33.85 & 39.22  \\
                        & \,\,Ablation: HRLP   & 35.62 & 7.80358 & 25.69 & 15.69 & 18.27  \\
                        & \,\,Ablation: LRHP   & 53.49 & 5.17098 & 44.27 & 30.43 & 36.20 \\ \hlinew{1pt}
\multirow{4}{*}{TNN}    & Skeletonization      & 69.46 & 7.01118 & 36.15 & 23.69 & 28.67 \\
                        & MPC-Skel (Ours)    & 79.54 & 4.14947 & 52.20 & 39.50 & 45.46  \\
                        & \,\,Ablation: HRLP   & 69.46 & 7.01118 & 36.15 & 23.69 & 28.67 \\
                        & \,\,Ablation: LRHP  & 81.86 & 3.90828 & 57.38 & 43.84 & 48.62  \\ \hlinew{1pt}
\multirow{4}{*}{AirMorph}    & Skeletonization    & 78.38 & 2.75415 & 75.57 & 71.48 & 72.81  \\
                                & MPC-Skel (Ours)   & 91.37 & 1.74228 & 83.52 & 80.59 & 81.62  \\
                                & \,\,Ablation: HRLP   & 57.39 & 5.35355 & 53.86 & 46.56 & 48.53  \\
                                & \,\,Ablation: LRHP    & 89.28 & 4.00417 & 77.41 & 72.95 & 73.22 \\ \hlinew{1pt}
\end{tabular}}
\addcontentsline{toc}{section}{Table S7: Ablation study of the graph building on AIIB'23 internal testset.}
\end{table*}

\makeatletter
\def\hlinew#1{%
\noalign{\ifnum0=`}\fi\hrule \@height #1 \futurelet
\reserved@a\@xhline}
\makeatother
\begin{table*}[ht]
\renewcommand\arraystretch{1.4}
\caption
{
Statistical analysis of diverse airway branching patterns of ATM'22 dataset.
}
\centering
\label{tab::statis_branch_pattern_ATM22}
\resizebox{1.0\textwidth}{!}{
 &                                                                                           &                                                                    &                                                                                      \\ \hline
\end{tabular}}
\addcontentsline{toc}{section}{Table S8: Statistical analysis of diverse airway branching patterns of ATM'22 dataset.}
\end{table*}

\makeatletter
\def\hlinew#1{%
\noalign{\ifnum0=`}\fi\hrule \@height #1 \futurelet
\reserved@a\@xhline}
\makeatother
\begin{table*}[ht]
\renewcommand\arraystretch{1.4}
\caption
{
Statistical analysis of diverse airway branching patterns of AIIB'23 dataset.
}
\centering
\label{tab::statis_branch_pattern_AIIB23}
\resizebox{1.0\textwidth}{!}{
 &                                                                                           &                                                                    &                                                                                      \\ \hline
\end{tabular}}
\addcontentsline{toc}{section}{Table S9: Statistical analysis of diverse airway branching patterns of AIIB'23 dataset.}
\end{table*}

\makeatletter
\def\hlinew#1{%
\noalign{\ifnum0=`}\fi\hrule \@height #1 \futurelet
\reserved@a\@xhline}
\makeatother
\begin{table*}[ht]
\renewcommand\arraystretch{1.4}
\caption
{
Statistical analysis of diverse airway branching patterns of LIDC-IDRI dataset.
}
\centering
\label{tab::statis_branch_pattern_LIDC}
\resizebox{1.0\textwidth}{!}{
 &                                                                                           &                                                                    &                                                                                      \\ \hline
\end{tabular}}
\addcontentsline{toc}{section}{Table S10: Statistical analysis of diverse airway branching patterns of LIDC-IDRI dataset.}
\end{table*}

\makeatletter
\def\hlinew#1{%
\noalign{\ifnum0=`}\fi\hrule \@height #1 \futurelet
\reserved@a\@xhline}
\makeatother
\begin{table*}[ht]
\renewcommand\arraystretch{1.4}
\caption
{
Statistical analysis of diverse airway branching patterns of NLST-D54 dataset.
}
\centering
\label{tab::statis_branch_pattern_NLST-D54}
\resizebox{1.0\textwidth}{!}{
 &                                                                                           &                                                                    &                                                                                      \\ \hline
\end{tabular}}
\addcontentsline{toc}{section}{Table S11: Statistical analysis of diverse airway branching patterns of NLST-D54 dataset.}
\end{table*}

\makeatletter
\def\hlinew#1{%
\noalign{\ifnum0=`}\fi\hrule \@height #1 \futurelet
\reserved@a\@xhline}
\makeatother
\begin{table*}[ht]
\renewcommand\arraystretch{1.4}
\caption
{
Statistical analysis of diverse airway branching patterns of NLST-D58 dataset.
}
\centering
\label{tab::statis_branch_pattern_NLST-D58}
\resizebox{1.0\textwidth}{!}{
 &                                                                                           &                                                                    &                                                                                      \\ \hline
\end{tabular}}
\addcontentsline{toc}{section}{Table S12: Statistical analysis of diverse airway branching patterns of NLST-D58 dataset.}
\end{table*}

\makeatletter
\def\hlinew#1{%
\noalign{\ifnum0=`}\fi\hrule \@height #1 \futurelet
\reserved@a\@xhline}
\makeatother
\begin{table*}[ht]
\renewcommand\arraystretch{1.4}
\caption
{
Statistical analysis of diverse airway branching patterns of NLST-D59 dataset.
}
\centering
\label{tab::statis_branch_pattern_NLST-D59}
\resizebox{1.0\textwidth}{!}{
 &                                                                                           &                                                                    &                                                                                      \\ \hline
\end{tabular}}
\addcontentsline{toc}{section}{Table S13: Statistical analysis of diverse airway branching patterns of NLST-D59 dataset.}
\end{table*}

\makeatletter
\def\hlinew#1{%
\noalign{\ifnum0=`}\fi\hrule \@height #1 \futurelet
\reserved@a\@xhline}
\makeatother
\begin{table*}[ht]
\renewcommand\arraystretch{1.4}
\caption
{
Statistical analysis of diverse airway branching patterns of NLST-D61 dataset.
}
\centering
\label{tab::statis_branch_pattern_NLST-D61}
\resizebox{1.0\textwidth}{!}{
}
\addcontentsline{toc}{section}{Table S14: Statistical analysis of diverse airway branching patterns of NLST-D61 dataset.}
\end{table*}
\begin{sidewaystable*}
\makeatletter
\def\hlinew#1{%
\noalign{\ifnum0=`}\fi\hrule \@height #1 \futurelet
\reserved@a\@xhline}
\makeatother
\centering
\renewcommand\arraystretch{1.3}
\caption{The quantative morphological features of the AIIB23 group (experimental group).}
\label{tab::AIIB23_Morpho_Quantative_Result}
\resizebox{\textheight}{!}{
\begin{tabular}{@{}ccccccc!{\vrule width 1.5pt}cccccc@{}}
\hlinew{1pt}
\rowcolor{tablecolor1}
& \multicolumn{6}{c}{Health Control Group} & \multicolumn{6}{c}{AIIB23 Group} \\
Branch Name & Stenosis & Ectasia & Tortuosity & Divergence & Length & Complexity & Stenosis(p-value) & Ectasia(p-value) & Tortuosity(p-value) & Divergence(p-value) & Length(p-value) & Complexity(p-value) \\
LUB & 0.27 ± 0.04 & 1.29 ± 0.03 & 0.64 ± 0.02 & 0.70 ± 0.05 & 45.33 ± 5.87 & 1.01 ± 0.06 & 0.27 ± 0.16(\textless{}0.01) & 1.28 ± 0.25(\textless{}0.01) & 0.60 ± 0.18(0.87) & 0.63 ± 0.19(\textless{}0.01) & 47.55 ± 9.98(0.13) & 1.01 ± 0.13(\textless{}0.01) \\
LDB & 0.29 ± 0.04 & 1.30 ± 0.03 & 0.65 ± 0.02 & 0.59 ± 0.07 & 49.70 ± 7.19 & 1.05 ± 0.08 & 0.30 ± 0.08(0.57) & 1.32 ± 0.05(0.9) & 0.62 ± 0.03(0.77) & 0.59 ± 0.10(\textless{}0.01) & 49.56 ± 9.42(0.68) & 1.05 ± 0.10(\textless{}0.01) \\
RUB & 0.27 ± 0.04 & 1.29 ± 0.03 & 0.64 ± 0.02 & 0.71 ± 0.08 & 45.26 ± 6.67 & 1.00 ± 0.07 & 0.27 ± 0.07(\textless{}0.01) & 1.30 ± 0.05(\textless{}0.01) & 0.61 ± 0.02(0.02) & 0.67 ± 0.09(\textless{}0.01) & 50.82 ± 8.00(0.63) & 1.03 ± 0.09(0.04) \\
RMB & 0.28 ± 0.05 & 1.29 ± 0.04 & 0.66 ± 0.03 & 0.36 ± 0.07 & 50.91 ± 7.58 & 0.87 ± 0.08 & 0.28 ± 0.07(\textless{}0.01) & 1.30 ± 0.06(0.04) & 0.63 ± 0.04(0.94) & 0.33 ± 0.08(\textless{}0.01) & 53.31 ± 9.11(0.81) & 0.87 ± 0.09(0.19) \\
RDB & 0.28 ± 0.04 & 1.29 ± 0.03 & 0.66 ± 0.02 & 0.70 ± 0.09 & 50.10 ± 6.24 & 1.09 ± 0.06 & 0.29 ± 0.08(\textless{}0.01) & 1.32 ± 0.06(\textless{}0.01) & 0.62 ± 0.03(\textless{}0.01) & 0.64 ± 0.12(\textless{}0.01) & 46.29 ± 7.93(0.23) & 1.05 ± 0.10(\textless{}0.01) \\
LB1+2 & 0.28 ± 0.05 & 1.28 ± 0.04 & 0.64 ± 0.03 & 0.41 ± 0.08 & 47.31 ± 8.13 & 0.79 ± 0.09 & 0.27 ± 0.17(0.93) & 1.27 ± 0.25(0.03) & 0.60 ± 0.18(0.91) & 0.39 ± 0.18(\textless{}0.01) & 49.77 ± 12.39(0.23) & 0.80 ± 0.13(0.03) \\
\end{tabular}}
\addcontentsline{toc}{section}{Table S15: The quantative morphological features of the pulmonary fibrosis group.}
\end{sidewaystable*}

\begin{sidewaystable*}
\makeatletter
\def\hlinew#1{%
\noalign{\ifnum0=`}\fi\hrule \@height #1 \futurelet
\reserved@a\@xhline}
\makeatother
\centering
\renewcommand\arraystretch{1.3}
\caption{The quantative morphological features of the NLST-D54 group (experimental group).}
\label{tab::NLSTD54_Morpho_Quantative_Result}
\resizebox{\textheight}{!}{
\begin{tabular}{@{}ccccccc!{\vrule width 1.5pt}cccccc@{}}
\hlinew{1pt}
\rowcolor{tablecolor1}
& \multicolumn{6}{c}{Health Control Group} & \multicolumn{6}{c}{NLST-D54 Group} \\
BranchName & Stenosis & Ectasia & Tortuosity & Divergence & Length & Complexity & Stenosis(p-value) & Ectasia(p-value) & Tortuosity(p-value) & Divergence(p-value) & Length(p-value) & Complexity(p-value) \\
LUB & 0.27 ± 0.04 & 1.29 ± 0.03 & 0.64 ± 0.02 & 0.70 ± 0.05 & 45.33 ± 5.87 & 1.01 ± 0.06 & 0.22 ± 0.04(\textless{}0.01) & 1.28 ± 0.04(0.03) & 0.62 ± 0.02(\textless{}0.01) & 0.68 ± 0.07(\textless{}0.01) & 44.00 ± 7.13(\textless{}0.01) & 0.93 ± 0.08(0.08) \\
LDB & 0.29 ± 0.04 & 1.30 ± 0.03 & 0.65 ± 0.02 & 0.59 ± 0.07 & 49.70 ± 7.19 & 1.05 ± 0.08 & 0.24 ± 0.04(\textless{}0.01) & 1.30 ± 0.04(\textless{}0.01) & 0.62 ± 0.03(\textless{}0.01) & 0.61 ± 0.08(\textless{}0.01) & 44.23 ± 7.40(\textless{}0.01) & 0.95 ± 0.10(0.94) \\
RUB & 0.27 ± 0.04 & 1.29 ± 0.03 & 0.64 ± 0.02 & 0.71 ± 0.08 & 45.26 ± 6.67 & 1.00 ± 0.07 & 0.21 ± 0.04(\textless{}0.01) & 1.27 ± 0.04(0.5) & 0.62 ± 0.02(\textless{}0.01) & 0.64 ± 0.09(\textless{}0.01) & 44.83 ± 6.56(\textless{}0.01) & 0.93 ± 0.07(\textless{}0.01) \\
RMB & 0.28 ± 0.05 & 1.29 ± 0.04 & 0.66 ± 0.03 & 0.36 ± 0.07 & 50.91 ± 7.58 & 0.87 ± 0.08 & 0.25 ± 0.08(\textless{}0.01) & 1.32 ± 0.11(\textless{}0.01) & 0.61 ± 0.05(\textless{}0.01) & 0.23 ± 0.08(\textless{}0.01) & 37.70 ± 11.41(\textless{}0.01) & 0.66 ± 0.20(\textless{}0.01) \\
RDB & 0.28 ± 0.04 & 1.29 ± 0.03 & 0.66 ± 0.02 & 0.70 ± 0.09 & 50.10 ± 6.24 & 1.09 ± 0.06 & 0.24 ± 0.04(\textless{}0.01) & 1.31 ± 0.04(\textless{}0.01) & 0.63 ± 0.02(\textless{}0.01) & 0.66 ± 0.11(\textless{}0.01) & 43.91 ± 6.67(\textless{}0.01) & 0.96 ± 0.08(\textless{}0.01) \\
LB1+2 & 0.28 ± 0.05 & 1.28 ± 0.04 & 0.64 ± 0.03 & 0.41 ± 0.08 & 47.31 ± 8.13 & 0.79 ± 0.09 & 0.21 ± 0.04(\textless{}0.01) & 1.27 ± 0.07(\textless{}0.01) & 0.64 ± 0.04(\textless{}0.01) & 0.32 ± 0.09(0.52) & 50.29 ± 9.00(\textless{}0.01) & 0.73 ± 0.15(0.12) \\
LB3 & 0.27 ± 0.05 & 1.29 ± 0.05 & 0.64 ± 0.03 & 0.49 ± 0.09 & 47.16 ± 6.15 & 0.79 ± 0.08 & 0.21 ± 0.05(\textless{}0.01) & 1.28 ± 0.07(\textless{}0.01) & 0.61 ± 0.04(\textless{}0.01) & 0.39 ± 0.12(\textless{}0.01) & 43.16 ± 8.72(\textless{}0.01) & 0.69 ± 0.17(0.49) \\
LB4 & 0.27 ± 0.06 & 1.30 ± 0.07 & 0.65 ± 0.04 & 0.37 ± 0.11 & 38.14 ± 7.30 & 0.63 ± 0.10 & 0.25 ± 0.09(\textless{}0.01) & 1.33 ± 0.12(\textless{}0.01) & 0.62 ± 0.07(\textless{}0.01) & 0.26 ± 0.15(\textless{}0.01) & 33.14 ± 11.82(\textless{}0.01) & 0.47 ± 0.24(\textless{}0.01) \\
LB5 & 0.33 ± 0.10 & 1.34 ± 0.13 & 0.67 ± 0.07 & 0.16 ± 0.11 & 43.40 ± 13.66 & 0.61 ± 0.12 & 0.28 ± 0.10(0.12) & 1.36 ± 0.16(\textless{}0.01) & 0.63 ± 0.08(\textless{}0.01) & 0.15 ± 0.12(\textless{}0.01) & 34.21 ± 13.46(\textless{}0.01) & 0.49 ± 0.19(0.07) \\
LB6 & 0.27 ± 0.06 & 1.32 ± 0.15 & 0.62 ± 0.03 & 0.43 ± 0.09 & 39.23 ± 6.42 & 0.71 ± 0.09 & 0.22 ± 0.06(0.59) & 1.32 ± 0.11(\textless{}0.01) & 0.59 ± 0.04(\textless{}0.01) & 0.43 ± 0.11(\textless{}0.01) & 36.37 ± 6.65(\textless{}0.01) & 0.65 ± 0.11(0.81) \\
LB8 & 0.33 ± 0.12 & 1.31 ± 0.18 & 0.65 ± 0.13 & 0.30 ± 0.15 & 49.27 ± 12.29 & 0.71 ± 0.12 & 0.29 ± 0.08(0.38) & 1.33 ± 0.10(\textless{}0.01) & 0.62 ± 0.05(\textless{}0.01) & 0.29 ± 0.13(\textless{}0.01) & 42.43 ± 11.44(\textless{}0.01) & 0.60 ± 0.17(0.26) \\
LB9 & 0.28 ± 0.14 & 1.29 ± 0.25 & 0.66 ± 0.17 & 0.27 ± 0.16 & 47.70 ± 12.69 & 0.69 ± 0.14 & 0.27 ± 0.09(\textless{}0.01) & 1.33 ± 0.11(\textless{}0.01) & 0.63 ± 0.06(\textless{}0.01) & 0.26 ± 0.12(\textless{}0.01) & 40.97 ± 11.88(\textless{}0.01) & 0.58 ± 0.20(0.16) \\
LB10 & 0.30 ± 0.06 & 1.28 ± 0.04 & 0.65 ± 0.03 & 0.36 ± 0.10 & 57.44 ± 9.83 & 0.82 ± 0.09 & 0.23 ± 0.06(\textless{}0.01) & 1.28 ± 0.06(\textless{}0.01) & 0.64 ± 0.05(\textless{}0.01) & 0.28 ± 0.11(\textless{}0.01) & 54.55 ± 9.63(\textless{}0.01) & 0.70 ± 0.17(0.87) \\
RB1 & 0.29 ± 0.06 & 1.29 ± 0.04 & 0.63 ± 0.03 & 0.37 ± 0.12 & 43.88 ± 7.74 & 0.75 ± 0.09 & 0.21 ± 0.04(0.3) & 1.27 ± 0.06(\textless{}0.01) & 0.64 ± 0.04(0.02) & 0.36 ± 0.11(0.15) & 47.45 ± 8.35(\textless{}0.01) & 0.73 ± 0.12(\textless{}0.01) \\
RB2 & 0.27 ± 0.05 & 1.29 ± 0.05 & 0.64 ± 0.03 & 0.42 ± 0.09 & 40.88 ± 5.78 & 0.70 ± 0.08 & 0.23 ± 0.06(\textless{}0.01) & 1.31 ± 0.12(\textless{}0.01) & 0.61 ± 0.05(\textless{}0.01) & 0.37 ± 0.11(\textless{}0.01) & 38.61 ± 7.94(\textless{}0.01) & 0.63 ± 0.17(0.08) \\
RB3 & 0.26 ± 0.05 & 1.29 ± 0.07 & 0.65 ± 0.03 & 0.46 ± 0.09 & 48.99 ± 9.16 & 0.78 ± 0.10 & 0.21 ± 0.06(\textless{}0.01) & 1.27 ± 0.07(\textless{}0.01) & 0.60 ± 0.04(\textless{}0.01) & 0.37 ± 0.10(\textless{}0.01) & 45.31 ± 9.62(\textless{}0.01) & 0.68 ± 0.17(\textless{}0.01) \\
RB4 & 0.28 ± 0.05 & 1.29 ± 0.06 & 0.66 ± 0.04 & 0.28 ± 0.07 & 47.43 ± 8.04 & 0.72 ± 0.09 & 0.26 ± 0.08(\textless{}0.01) & 1.34 ± 0.14(\textless{}0.01) & 0.61 ± 0.05(\textless{}0.01) & 0.21 ± 0.10(\textless{}0.01) & 35.27 ± 10.24(\textless{}0.01) & 0.47 ± 0.25(\textless{}0.01) \\
RB5 & 0.29 ± 0.06 & 1.31 ± 0.07 & 0.66 ± 0.04 & 0.23 ± 0.07 & 53.31 ± 9.37 & 0.75 ± 0.10 & 0.24 ± 0.08(\textless{}0.01) & 1.30 ± 0.09(\textless{}0.01) & 0.62 ± 0.05(\textless{}0.01) & 0.20 ± 0.10(\textless{}0.01) & 41.32 ± 12.81(\textless{}0.01) & 0.51 ± 0.27(0.21) \\
RB6 & 0.26 ± 0.05 & 1.31 ± 0.08 & 0.63 ± 0.03 & 0.54 ± 0.10 & 40.39 ± 5.28 & 0.73 ± 0.08 & 0.22 ± 0.05(\textless{}0.01) & 1.32 ± 0.08(\textless{}0.01) & 0.60 ± 0.04(\textless{}0.01) & 0.41 ± 0.10(\textless{}0.01) & 35.33 ± 6.50(\textless{}0.01) & 0.63 ± 0.13(0.25) \\
RB7 & 0.21 ± 0.41 & 1.10 ± 0.73 & 0.48 ± 0.50 & 0.13 ± 0.40 & 38.37 ± 15.49 & 0.59 ± 0.22 & 0.29 ± 0.07(0.02) & 1.36 ± 0.16(0.78) & 0.65 ± 0.05(0.08) & 0.24 ± 0.10(0.24) & 43.09 ± 9.82(\textless{}0.01) & 0.55 ± 0.20(0.12) \\
RB8 & 0.30 ± 0.06 & 1.30 ± 0.06 & 0.68 ± 0.04 & 0.29 ± 0.10 & 53.62 ± 9.52 & 0.75 ± 0.09 & 0.27 ± 0.09(\textless{}0.01) & 1.33 ± 0.13(\textless{}0.01) & 0.63 ± 0.06(\textless{}0.01) & 0.22 ± 0.10(\textless{}0.01) & 42.53 ± 10.98(\textless{}0.01) & 0.55 ± 0.23(\textless{}0.01) \\
RB9 & 0.29 ± 0.06 & 1.31 ± 0.09 & 0.68 ± 0.05 & 0.24 ± 0.07 & 48.76 ± 10.63 & 0.68 ± 0.11 & 0.29 ± 0.10(\textless{}0.01) & 1.39 ± 0.21(\textless{}0.01) & 0.65 ± 0.08(\textless{}0.01) & 0.18 ± 0.11(\textless{}0.01) & 40.35 ± 12.32(0.51) & 0.52 ± 0.22(\textless{}0.01) \\
RB10 & 0.28 ± 0.05 & 1.28 ± 0.04 & 0.66 ± 0.03 & 0.36 ± 0.08 & 58.25 ± 9.77 & 0.83 ± 0.09 & 0.23 ± 0.05(\textless{}0.01) & 1.30 ± 0.12(\textless{}0.01) & 0.65 ± 0.05(\textless{}0.01) & 0.29 ± 0.09(0.04) & 51.69 ± 9.81(\textless{}0.01) & 0.71 ± 0.18(\textless{}0.01) \\ 
\hlinew{1pt}
\end{tabular}}
\addcontentsline{toc}{section}{Table S16: The quantative morphological features of the pulmonary atelecta group.}
\end{sidewaystable*}

\begin{sidewaystable*}
\makeatletter
\def\hlinew#1{%
\noalign{\ifnum0=`}\fi\hrule \@height #1 \futurelet
\reserved@a\@xhline}
\makeatother
\centering
\renewcommand\arraystretch{1.3}
\caption{The quantative morphological features of the NLST-D58 group (experimental group).}
\label{tab::NLSTD58_Morpho_Quantative_Result}
\resizebox{\textheight}{!}{
\begin{tabular}{@{}ccccccc!{\vrule width 1.5pt}cccccc@{}}
\hlinew{1pt}
\rowcolor{tablecolor1}
& \multicolumn{6}{c}{Health Control Group} & \multicolumn{6}{c}{NLST-D58 Group} \\
BranchName & Stenosis & Ectasia & Tortuosity & Divergence & Length & Complexity & Stenosis(p-value) & Ectasia(p-value) & Tortuosity(p-value) & Divergence(p-value) & Length(p-value) & Complexity(p-value) \\
LUB & 0.27 ± 0.04 & 1.29 ± 0.03 & 0.64 ± 0.02 & 0.70 ± 0.05 & 45.33 ± 5.87 & 1.01 ± 0.06 & 0.22 ± 0.04(0.01) & 1.28 ± 0.04(0.01) & 0.62 ± 0.02(\textless{}0.01) & 0.68 ± 0.08(\textless{}0.01) & 43.70 ± 6.78(\textless{}0.01) & 0.92 ± 0.08(0.23) \\
LDB & 0.29 ± 0.04 & 1.30 ± 0.03 & 0.65 ± 0.02 & 0.59 ± 0.07 & 49.70 ± 7.19 & 1.05 ± 0.08 & 0.25 ± 0.04(0.13) & 1.30 ± 0.04(\textless{}0.01) & 0.62 ± 0.03(\textless{}0.01) & 0.60 ± 0.11(\textless{}0.01) & 45.44 ± 7.40(\textless{}0.01) & 0.94 ± 0.09(0.09) \\
RUB & 0.27 ± 0.04 & 1.29 ± 0.03 & 0.64 ± 0.02 & 0.71 ± 0.08 & 45.26 ± 6.67 & 1.00 ± 0.07 & 0.22 ± 0.03(\textless{}0.01) & 1.28 ± 0.04(0.02) & 0.61 ± 0.02(\textless{}0.01) & 0.61 ± 0.10(\textless{}0.01) & 43.60 ± 6.69(\textless{}0.01) & 0.91 ± 0.08(0.03) \\
RMB & 0.28 ± 0.05 & 1.29 ± 0.04 & 0.66 ± 0.03 & 0.36 ± 0.07 & 50.91 ± 7.58 & 0.87 ± 0.08 & 0.26 ± 0.07(\textless{}0.01) & 1.32 ± 0.09(\textless{}0.01) & 0.61 ± 0.05(\textless{}0.01) & 0.22 ± 0.08(\textless{}0.01) & 35.65 ± 11.57(\textless{}0.01) & 0.60 ± 0.24(\textless{}0.01) \\
RDB & 0.28 ± 0.04 & 1.29 ± 0.03 & 0.66 ± 0.02 & 0.70 ± 0.09 & 50.10 ± 6.24 & 1.09 ± 0.06 & 0.25 ± 0.04(\textless{}0.01) & 1.32 ± 0.04(\textless{}0.01) & 0.63 ± 0.03(\textless{}0.01) & 0.66 ± 0.11(\textless{}0.01) & 43.71 ± 7.62(\textless{}0.01) & 0.94 ± 0.09(\textless{}0.01) \\
LB1+2 & 0.28 ± 0.05 & 1.28 ± 0.04 & 0.64 ± 0.03 & 0.41 ± 0.08 & 47.31 ± 8.13 & 0.79 ± 0.09 & 0.22 ± 0.05(\textless{}0.01) & 1.27 ± 0.06(\textless{}0.01) & 0.64 ± 0.04(\textless{}0.01) & 0.30 ± 0.09(0.18) & 51.09 ± 10.33(\textless{}0.01) & 0.73 ± 0.11(0.27) \\
LB3 & 0.27 ± 0.05 & 1.29 ± 0.05 & 0.64 ± 0.03 & 0.49 ± 0.09 & 47.16 ± 6.15 & 0.79 ± 0.08 & 0.22 ± 0.05(\textless{}0.01) & 1.28 ± 0.07(\textless{}0.01) & 0.61 ± 0.04(\textless{}0.01) & 0.38 ± 0.12(\textless{}0.01) & 41.99 ± 7.24(\textless{}0.01) & 0.66 ± 0.18(0.23) \\
LB4 & 0.27 ± 0.06 & 1.30 ± 0.07 & 0.65 ± 0.04 & 0.37 ± 0.11 & 38.14 ± 7.30 & 0.63 ± 0.10 & 0.26 ± 0.09(\textless{}0.01) & 1.34 ± 0.12(\textless{}0.01) & 0.61 ± 0.07(\textless{}0.01) & 0.25 ± 0.14(\textless{}0.01) & 29.80 ± 10.77(0.12) & 0.44 ± 0.23(\textless{}0.01) \\
LB5 & 0.33 ± 0.10 & 1.34 ± 0.13 & 0.67 ± 0.07 & 0.16 ± 0.11 & 43.40 ± 13.66 & 0.61 ± 0.12 & 0.28 ± 0.10(0.02) & 1.36 ± 0.16(\textless{}0.01) & 0.64 ± 0.08(\textless{}0.01) & 0.14 ± 0.10(\textless{}0.01) & 36.38 ± 19.09(\textless{}0.01) & 0.48 ± 0.21(0.24) \\
LB6 & 0.27 ± 0.06 & 1.32 ± 0.15 & 0.62 ± 0.03 & 0.43 ± 0.09 & 39.23 ± 6.42 & 0.71 ± 0.09 & 0.23 ± 0.06(0.3) & 1.34 ± 0.13(\textless{}0.01) & 0.60 ± 0.04(\textless{}0.01) & 0.44 ± 0.12(\textless{}0.01) & 37.07 ± 7.73(\textless{}0.01) & 0.62 ± 0.15(0.21) \\
LB8 & 0.33 ± 0.12 & 1.31 ± 0.18 & 0.65 ± 0.13 & 0.30 ± 0.15 & 49.27 ± 12.29 & 0.71 ± 0.12 & 0.31 ± 0.08(\textless{}0.01) & 1.34 ± 0.11(\textless{}0.01) & 0.63 ± 0.05(\textless{}0.01) & 0.27 ± 0.12(\textless{}0.01) & 44.16 ± 10.79(\textless{}0.01) & 0.63 ± 0.12(0.04) \\
LB9 & 0.28 ± 0.14 & 1.29 ± 0.25 & 0.66 ± 0.17 & 0.27 ± 0.16 & 47.70 ± 12.69 & 0.69 ± 0.14 & 0.28 ± 0.09(0.02) & 1.34 ± 0.11(\textless{}0.01) & 0.63 ± 0.07(\textless{}0.01) & 0.25 ± 0.12(\textless{}0.01) & 41.37 ± 13.15(0.19) & 0.56 ± 0.21(0.05) \\
LB10 & 0.30 ± 0.06 & 1.28 ± 0.04 & 0.65 ± 0.03 & 0.36 ± 0.10 & 57.44 ± 9.83 & 0.82 ± 0.09 & 0.24 ± 0.06(\textless{}0.01) & 1.28 ± 0.05(0.06) & 0.65 ± 0.04(\textless{}0.01) & 0.28 ± 0.11(0.03) & 55.53 ± 9.32(\textless{}0.01) & 0.70 ± 0.18(0.76) \\
RB1 & 0.29 ± 0.06 & 1.29 ± 0.04 & 0.63 ± 0.03 & 0.37 ± 0.12 & 43.88 ± 7.74 & 0.75 ± 0.09 & 0.22 ± 0.05(0.01) & 1.27 ± 0.05(\textless{}0.01) & 0.64 ± 0.04(\textless{}0.01) & 0.35 ± 0.10(0.23) & 47.12 ± 8.51(\textless{}0.01) & 0.72 ± 0.10(0.01) \\
RB2 & 0.27 ± 0.05 & 1.29 ± 0.05 & 0.64 ± 0.03 & 0.42 ± 0.09 & 40.88 ± 5.78 & 0.70 ± 0.08 & 0.24 ± 0.06(\textless{}0.01) & 1.31 ± 0.09(0.02) & 0.61 ± 0.05(\textless{}0.01) & 0.37 ± 0.12(\textless{}0.01) & 39.06 ± 7.80(\textless{}0.01) & 0.59 ± 0.22(0.03) \\
RB3 & 0.26 ± 0.05 & 1.29 ± 0.07 & 0.65 ± 0.03 & 0.46 ± 0.09 & 48.99 ± 9.16 & 0.78 ± 0.10 & 0.22 ± 0.06(\textless{}0.01) & 1.29 ± 0.08(\textless{}0.01) & 0.59 ± 0.04(\textless{}0.01) & 0.36 ± 0.12(\textless{}0.01) & 42.43 ± 10.10(\textless{}0.01) & 0.63 ± 0.20(0.75) \\
RB4 & 0.28 ± 0.05 & 1.29 ± 0.06 & 0.66 ± 0.04 & 0.28 ± 0.07 & 47.43 ± 8.04 & 0.72 ± 0.09 & 0.27 ± 0.09(\textless{}0.01) & 1.32 ± 0.12(\textless{}0.01) & 0.61 ± 0.06(\textless{}0.01) & 0.22 ± 0.10(\textless{}0.01) & 34.23 ± 10.87(0.27) & 0.41 ± 0.28(\textless{}0.01) \\
RB5 & 0.29 ± 0.06 & 1.31 ± 0.07 & 0.66 ± 0.04 & 0.23 ± 0.07 & 53.31 ± 9.37 & 0.75 ± 0.10 & 0.26 ± 0.07(\textless{}0.01) & 1.33 ± 0.13(\textless{}0.01) & 0.62 ± 0.06(\textless{}0.01) & 0.20 ± 0.10(\textless{}0.01) & 39.05 ± 12.15(\textless{}0.01) & 0.45 ± 0.29(0.08) \\
RB6 & 0.26 ± 0.05 & 1.31 ± 0.08 & 0.63 ± 0.03 & 0.54 ± 0.10 & 40.39 ± 5.28 & 0.73 ± 0.08 & 0.23 ± 0.06(\textless{}0.01) & 1.34 ± 0.10(\textless{}0.01) & 0.59 ± 0.04(\textless{}0.01) & 0.39 ± 0.11(\textless{}0.01) & 34.58 ± 7.24(\textless{}0.01) & 0.61 ± 0.14(\textless{}0.01) \\
RB7 & 0.21 ± 0.41 & 1.10 ± 0.73 & 0.48 ± 0.50 & 0.13 ± 0.40 & 38.37 ± 15.49 & 0.59 ± 0.22 & 0.30 ± 0.08(\textless{}0.01) & 1.35 ± 0.17(0.49) & 0.65 ± 0.06(\textless{}0.01) & 0.22 ± 0.10(0.08) & 43.58 ± 10.69(\textless{}0.01) & 0.53 ± 0.21(0.46) \\
RB8 & 0.30 ± 0.06 & 1.30 ± 0.06 & 0.68 ± 0.04 & 0.29 ± 0.10 & 53.62 ± 9.52 & 0.75 ± 0.09 & 0.29 ± 0.08(\textless{}0.01) & 1.34 ± 0.13(\textless{}0.01) & 0.63 ± 0.07(\textless{}0.01) & 0.23 ± 0.12(\textless{}0.01) & 42.70 ± 11.00(0.02) & 0.53 ± 0.24(\textless{}0.01) \\
RB9 & 0.29 ± 0.06 & 1.31 ± 0.09 & 0.68 ± 0.05 & 0.24 ± 0.07 & 48.76 ± 10.63 & 0.68 ± 0.11 & 0.31 ± 0.11(\textless{}0.01) & 1.40 ± 0.19(\textless{}0.01) & 0.65 ± 0.09(\textless{}0.01) & 0.15 ± 0.10(\textless{}0.01) & 40.92 ± 14.45(0.1) & 0.51 ± 0.22(\textless{}0.01) \\
RB10 & 0.28 ± 0.05 & 1.28 ± 0.04 & 0.66 ± 0.03 & 0.36 ± 0.08 & 58.25 ± 9.77 & 0.83 ± 0.09 & 0.25 ± 0.06(\textless{}0.01) & 1.31 ± 0.08(\textless{}0.01) & 0.65 ± 0.05(\textless{}0.01) & 0.27 ± 0.09(0.09) & 51.74 ± 10.97(\textless{}0.01) & 0.70 ± 0.17(\textless{}0.01) \\
\hlinew{1pt}
\end{tabular}}
\addcontentsline{toc}{section}{Table S17: The quantative morphological features of the pulmonary consolidation group.}
\end{sidewaystable*}

\begin{sidewaystable*}
\makeatletter
\def\hlinew#1{%
\noalign{\ifnum0=`}\fi\hrule \@height #1 \futurelet
\reserved@a\@xhline}
\makeatother
\centering
\renewcommand\arraystretch{1.3}
\caption{The quantative morphological features of the NLST-D59 group (experimental group).}
\label{tab::NLSTD59_Morpho_Quantative_Result}
\resizebox{\textheight}{!}{
\begin{tabular}{@{}ccccccc!{\vrule width 1.5pt}cccccc@{}}
\hlinew{1pt}
\rowcolor{tablecolor1}
& \multicolumn{6}{c}{Health Control Group} & \multicolumn{6}{c}{NLST-D59 Group} \\
BranchName & Stenosis & Ectasia & Tortuosity & Divergence & Length & Complexity & Stenosis(p-value) & Ectasia(p-value) & Tortuosity(p-value) & Divergence(p-value) & Length(p-value) & Complexity(p-value) \\
LUB & 0.27 ± 0.04 & 1.29 ± 0.03 & 0.64 ± 0.02 & 0.70 ± 0.05 & 45.33 ± 5.87 & 1.01 ± 0.06 & 0.21 ± 0.03(0.02) & 1.27 ± 0.03(0.09) & 0.63 ± 0.02(\textless{}0.01) & 0.71 ± 0.07(\textless{}0.01) & 46.20 ± 7.34(\textless{}0.01) & 0.97 ± 0.09(\textless{}0.01) \\
LDB & 0.29 ± 0.04 & 1.30 ± 0.03 & 0.65 ± 0.02 & 0.59 ± 0.07 & 49.70 ± 7.19 & 1.05 ± 0.08 & 0.23 ± 0.04(\textless{}0.01) & 1.29 ± 0.04(\textless{}0.01) & 0.64 ± 0.03(\textless{}0.01) & 0.63 ± 0.08(\textless{}0.01) & 47.87 ± 8.18(\textless{}0.01) & 1.00 ± 0.10(\textless{}0.01) \\
RUB & 0.27 ± 0.04 & 1.29 ± 0.03 & 0.64 ± 0.02 & 0.71 ± 0.08 & 45.26 ± 6.67 & 1.00 ± 0.07 & 0.20 ± 0.03(\textless{}0.01) & 1.27 ± 0.04(0.03) & 0.62 ± 0.02(\textless{}0.01) & 0.66 ± 0.10(\textless{}0.01) & 46.46 ± 7.58(\textless{}0.01) & 0.97 ± 0.10(\textless{}0.01) \\
RMB & 0.28 ± 0.05 & 1.29 ± 0.04 & 0.66 ± 0.03 & 0.36 ± 0.07 & 50.91 ± 7.58 & 0.87 ± 0.08 & 0.24 ± 0.07(\textless{}0.01) & 1.30 ± 0.08(\textless{}0.01) & 0.63 ± 0.04(\textless{}0.01) & 0.27 ± 0.08(\textless{}0.01) & 41.76 ± 10.99(\textless{}0.01) & 0.73 ± 0.20(0.98) \\
RDB & 0.28 ± 0.04 & 1.29 ± 0.03 & 0.66 ± 0.02 & 0.70 ± 0.09 & 50.10 ± 6.24 & 1.09 ± 0.06 & 0.23 ± 0.04(\textless{}0.01) & 1.29 ± 0.04(\textless{}0.01) & 0.64 ± 0.03(\textless{}0.01) & 0.68 ± 0.11(\textless{}0.01) & 46.46 ± 7.66(\textless{}0.01) & 1.01 ± 0.10(0.52) \\
LB1+2 & 0.28 ± 0.05 & 1.28 ± 0.04 & 0.64 ± 0.03 & 0.41 ± 0.08 & 47.31 ± 8.13 & 0.79 ± 0.09 & 0.20 ± 0.04(\textless{}0.01) & 1.26 ± 0.05(\textless{}0.01) & 0.65 ± 0.04(\textless{}0.01) & 0.33 ± 0.09(0.01) & 52.64 ± 9.30(\textless{}0.01) & 0.76 ± 0.11(\textless{}0.01) \\
LB3 & 0.27 ± 0.05 & 1.29 ± 0.05 & 0.64 ± 0.03 & 0.49 ± 0.09 & 47.16 ± 6.15 & 0.79 ± 0.08 & 0.20 ± 0.04(\textless{}0.01) & 1.27 ± 0.06(\textless{}0.01) & 0.62 ± 0.04(\textless{}0.01) & 0.44 ± 0.12(\textless{}0.01) & 45.55 ± 8.94(\textless{}0.01) & 0.74 ± 0.16(\textless{}0.01) \\
LB4 & 0.27 ± 0.06 & 1.30 ± 0.07 & 0.65 ± 0.04 & 0.37 ± 0.11 & 38.14 ± 7.30 & 0.63 ± 0.10 & 0.24 ± 0.08(\textless{}0.01) & 1.31 ± 0.11(\textless{}0.01) & 0.62 ± 0.06(\textless{}0.01) & 0.29 ± 0.14(\textless{}0.01) & 35.40 ± 10.36(\textless{}0.01) & 0.56 ± 0.19(0.2) \\
LB5 & 0.33 ± 0.10 & 1.34 ± 0.13 & 0.67 ± 0.07 & 0.16 ± 0.11 & 43.40 ± 13.66 & 0.61 ± 0.12 & 0.25 ± 0.09(0.97) & 1.33 ± 0.15(\textless{}0.01) & 0.65 ± 0.07(\textless{}0.01) & 0.16 ± 0.10(\textless{}0.01) & 39.69 ± 12.72(\textless{}0.01) & 0.57 ± 0.16(0.3) \\
LB6 & 0.27 ± 0.06 & 1.32 ± 0.15 & 0.62 ± 0.03 & 0.43 ± 0.09 & 39.23 ± 6.42 & 0.71 ± 0.09 & 0.21 ± 0.05(\textless{}0.01) & 1.31 ± 0.09(0.83) & 0.61 ± 0.04(0.01) & 0.46 ± 0.10(\textless{}0.01) & 39.12 ± 7.47(\textless{}0.01) & 0.69 ± 0.12(0.14) \\
LB8 & 0.33 ± 0.12 & 1.31 ± 0.18 & 0.65 ± 0.13 & 0.30 ± 0.15 & 49.27 ± 12.29 & 0.71 ± 0.12 & 0.28 ± 0.07(0.79) & 1.32 ± 0.10(0.15) & 0.64 ± 0.05(\textless{}0.01) & 0.30 ± 0.12(\textless{}0.01) & 48.16 ± 11.62(\textless{}0.01) & 0.68 ± 0.15(0.55) \\
LB9 & 0.28 ± 0.14 & 1.29 ± 0.25 & 0.66 ± 0.17 & 0.27 ± 0.16 & 47.70 ± 12.69 & 0.69 ± 0.14 & 0.25 ± 0.07(0.17) & 1.31 ± 0.10(\textless{}0.01) & 0.64 ± 0.05(\textless{}0.01) & 0.29 ± 0.12(\textless{}0.01) & 44.94 ± 11.90(\textless{}0.01) & 0.65 ± 0.18(0.55) \\
LB10 & 0.30 ± 0.06 & 1.28 ± 0.04 & 0.65 ± 0.03 & 0.36 ± 0.10 & 57.44 ± 9.83 & 0.82 ± 0.09 & 0.22 ± 0.05(\textless{}0.01) & 1.27 ± 0.05(0.49) & 0.65 ± 0.04(\textless{}0.01) & 0.31 ± 0.10(0.45) & 56.89 ± 10.27(\textless{}0.01) & 0.76 ± 0.14(\textless{}0.01) \\
RB1 & 0.29 ± 0.06 & 1.29 ± 0.04 & 0.63 ± 0.03 & 0.37 ± 0.12 & 43.88 ± 7.74 & 0.75 ± 0.09 & 0.20 ± 0.04(0.53) & 1.27 ± 0.05(\textless{}0.01) & 0.64 ± 0.04(0.57) & 0.37 ± 0.12(\textless{}0.01) & 48.73 ± 8.68(\textless{}0.01) & 0.76 ± 0.11(\textless{}0.01) \\
RB2 & 0.27 ± 0.05 & 1.29 ± 0.05 & 0.64 ± 0.03 & 0.42 ± 0.09 & 40.88 ± 5.78 & 0.70 ± 0.08 & 0.21 ± 0.05(\textless{}0.01) & 1.29 ± 0.08(0.95) & 0.62 ± 0.04(\textless{}0.01) & 0.38 ± 0.11(\textless{}0.01) & 40.84 ± 8.41(\textless{}0.01) & 0.66 ± 0.17(0.3) \\
RB3 & 0.26 ± 0.05 & 1.29 ± 0.07 & 0.65 ± 0.03 & 0.46 ± 0.09 & 48.99 ± 9.16 & 0.78 ± 0.10 & 0.20 ± 0.05(\textless{}0.01) & 1.27 ± 0.05(0.02) & 0.61 ± 0.04(\textless{}0.01) & 0.41 ± 0.10(\textless{}0.01) & 47.16 ± 9.87(\textless{}0.01) & 0.73 ± 0.18(\textless{}0.01) \\
RB4 & 0.28 ± 0.05 & 1.29 ± 0.06 & 0.66 ± 0.04 & 0.28 ± 0.07 & 47.43 ± 8.04 & 0.72 ± 0.09 & 0.24 ± 0.07(\textless{}0.01) & 1.30 ± 0.11(\textless{}0.01) & 0.64 ± 0.05(\textless{}0.01) & 0.23 ± 0.09(\textless{}0.01) & 40.18 ± 10.28(\textless{}0.01) & 0.56 ± 0.24(0.03) \\
RB5 & 0.29 ± 0.06 & 1.31 ± 0.07 & 0.66 ± 0.04 & 0.23 ± 0.07 & 53.31 ± 9.37 & 0.75 ± 0.10 & 0.23 ± 0.07(\textless{}0.01) & 1.29 ± 0.08(\textless{}0.01) & 0.63 ± 0.05(\textless{}0.01) & 0.20 ± 0.08(\textless{}0.01) & 45.00 ± 11.71(\textless{}0.01) & 0.59 ± 0.25(\textless{}0.01) \\
RB6 & 0.26 ± 0.05 & 1.31 ± 0.08 & 0.63 ± 0.03 & 0.54 ± 0.10 & 40.39 ± 5.28 & 0.73 ± 0.08 & 0.21 ± 0.05(\textless{}0.01) & 1.31 ± 0.08(\textless{}0.01) & 0.60 ± 0.04(\textless{}0.01) & 0.46 ± 0.12(\textless{}0.01) & 38.16 ± 7.14(\textless{}0.01) & 0.68 ± 0.14(0.41) \\
RB7 & 0.21 ± 0.41 & 1.10 ± 0.73 & 0.48 ± 0.50 & 0.13 ± 0.40 & 38.37 ± 15.49 & 0.59 ± 0.22 & 0.27 ± 0.07(\textless{}0.01) & 1.36 ± 0.21(\textless{}0.01) & 0.66 ± 0.05(0.87) & 0.23 ± 0.10(\textless{}0.01) & 44.83 ± 10.43(\textless{}0.01) & 0.58 ± 0.19(0.09) \\
RB8 & 0.30 ± 0.06 & 1.30 ± 0.06 & 0.68 ± 0.04 & 0.29 ± 0.10 & 53.62 ± 9.52 & 0.75 ± 0.09 & 0.26 ± 0.07(\textless{}0.01) & 1.31 ± 0.10(\textless{}0.01) & 0.65 ± 0.05(\textless{}0.01) & 0.24 ± 0.10(\textless{}0.01) & 47.31 ± 11.39(\textless{}0.01) & 0.63 ± 0.21(0.06) \\
RB9 & 0.29 ± 0.06 & 1.31 ± 0.09 & 0.68 ± 0.05 & 0.24 ± 0.07 & 48.76 ± 10.63 & 0.68 ± 0.11 & 0.27 ± 0.09(\textless{}0.01) & 1.35 ± 0.17(\textless{}0.01) & 0.66 ± 0.07(\textless{}0.01) & 0.20 ± 0.10(\textless{}0.01) & 42.85 ± 12.76(\textless{}0.01) & 0.58 ± 0.20(\textless{}0.01) \\
RB10 & 0.28 ± 0.05 & 1.28 ± 0.04 & 0.66 ± 0.03 & 0.36 ± 0.08 & 58.25 ± 9.77 & 0.83 ± 0.09 & 0.22 ± 0.05(\textless{}0.01) & 1.28 ± 0.06(\textless{}0.01) & 0.65 ± 0.04(\textless{}0.01) & 0.31 ± 0.09(0.26) & 53.92 ± 10.26(\textless{}0.01) & 0.76 ± 0.15(0.74) \\
\hlinew{1pt}
\end{tabular}}
\addcontentsline{toc}{section}{Table S18: The quantative morphological features of the pulmonary emphysema group.}
\end{sidewaystable*}

\begin{sidewaystable*}
\makeatletter
\def\hlinew#1{%
\noalign{\ifnum0=`}\fi\hrule \@height #1 \futurelet
\reserved@a\@xhline}
\makeatother
\centering
\renewcommand\arraystretch{1.3}
\caption{The quantative morphological features of the NLST-D61 group (experimental group).}
\label{tab::NLSTD61_Morpho_Quantative_Result}
\resizebox{\textheight}{!}{
    \begin{tabular}{@{}ccccccc!{\vrule width 1.5pt}cccccc@{}}
\hlinew{1pt}
\rowcolor{tablecolor1}
& \multicolumn{6}{c}{Health Control Group} & \multicolumn{6}{c}{NLST-D61 Group} \\
BranchName & Stenosis & Ectasia & Tortuosity & Divergence & Length & Complexity & Stenosis(p-value) & Ectasia(p-value) & Tortuosity(p-value) & Divergence(p-value) & Length(p-value) & Complexity(p-value) \\
LUB & 0.27 ± 0.04 & 1.29 ± 0.03 & 0.64 ± 0.02 & 0.70 ± 0.05 & 45.33 ± 5.87 & 1.01 ± 0.06 & 0.21 ± 0.03(0.28) & 1.27 ± 0.03(0.93) & 0.63 ± 0.02(\textless{}0.01) & 0.70 ± 0.07(\textless{}0.01) & 45.28 ± 6.96(\textless{}0.01) & 0.97 ± 0.08(\textless{}0.01) \\
LDB & 0.29 ± 0.04 & 1.30 ± 0.03 & 0.65 ± 0.02 & 0.59 ± 0.07 & 49.70 ± 7.19 & 1.05 ± 0.08 & 0.23 ± 0.04(\textless{}0.01) & 1.28 ± 0.04(\textless{}0.01) & 0.63 ± 0.03(\textless{}0.01) & 0.63 ± 0.09(\textless{}0.01) & 47.38 ± 8.10(\textless{}0.01) & 1.00 ± 0.09(\textless{}0.01) \\
RUB & 0.27 ± 0.04 & 1.29 ± 0.03 & 0.64 ± 0.02 & 0.71 ± 0.08 & 45.26 ± 6.67 & 1.00 ± 0.07 & 0.20 ± 0.03(\textless{}0.01) & 1.27 ± 0.04(0.2) & 0.62 ± 0.02(\textless{}0.01) & 0.66 ± 0.10(\textless{}0.01) & 46.00 ± 7.20(\textless{}0.01) & 0.97 ± 0.10(\textless{}0.01) \\
RMB & 0.28 ± 0.05 & 1.29 ± 0.04 & 0.66 ± 0.03 & 0.36 ± 0.07 & 50.91 ± 7.58 & 0.87 ± 0.08 & 0.24 ± 0.07(\textless{}0.01) & 1.29 ± 0.07(\textless{}0.01) & 0.63 ± 0.04(\textless{}0.01) & 0.27 ± 0.09(\textless{}0.01) & 42.01 ± 10.95(\textless{}0.01) & 0.74 ± 0.19(0.63) \\
RDB & 0.28 ± 0.04 & 1.29 ± 0.03 & 0.66 ± 0.02 & 0.70 ± 0.09 & 50.10 ± 6.24 & 1.09 ± 0.06 & 0.23 ± 0.04(0.02) & 1.29 ± 0.04(\textless{}0.01) & 0.64 ± 0.02(\textless{}0.01) & 0.68 ± 0.12(\textless{}0.01) & 45.88 ± 7.53(\textless{}0.01) & 1.01 ± 0.09(0.62) \\
LB1+2 & 0.28 ± 0.05 & 1.28 ± 0.04 & 0.64 ± 0.03 & 0.41 ± 0.08 & 47.31 ± 8.13 & 0.79 ± 0.09 & 0.20 ± 0.04(\textless{}0.01) & 1.26 ± 0.05(\textless{}0.01) & 0.64 ± 0.04(\textless{}0.01) & 0.33 ± 0.09(0.48) & 51.47 ± 9.34(\textless{}0.01) & 0.76 ± 0.11(\textless{}0.01) \\
LB3 & 0.27 ± 0.05 & 1.29 ± 0.05 & 0.64 ± 0.03 & 0.49 ± 0.09 & 47.16 ± 6.15 & 0.79 ± 0.08 & 0.20 ± 0.05(\textless{}0.01) & 1.27 ± 0.06(\textless{}0.01) & 0.62 ± 0.03(\textless{}0.01) & 0.44 ± 0.12(\textless{}0.01) & 44.49 ± 8.54(\textless{}0.01) & 0.74 ± 0.15(\textless{}0.01) \\
LB4 & 0.27 ± 0.06 & 1.30 ± 0.07 & 0.65 ± 0.04 & 0.37 ± 0.11 & 38.14 ± 7.30 & 0.63 ± 0.10 & 0.23 ± 0.08(\textless{}0.01) & 1.30 ± 0.10(\textless{}0.01) & 0.62 ± 0.06(\textless{}0.01) & 0.31 ± 0.14(\textless{}0.01) & 34.67 ± 9.82(\textless{}0.01) & 0.55 ± 0.21(0.39) \\
LB5 & 0.33 ± 0.10 & 1.34 ± 0.13 & 0.67 ± 0.07 & 0.16 ± 0.11 & 43.40 ± 13.66 & 0.61 ± 0.12 & 0.25 ± 0.09(0.83) & 1.32 ± 0.13(\textless{}0.01) & 0.64 ± 0.07(\textless{}0.01) & 0.17 ± 0.10(\textless{}0.01) & 38.28 ± 11.45(\textless{}0.01) & 0.56 ± 0.17(0.09) \\
LB6 & 0.27 ± 0.06 & 1.32 ± 0.15 & 0.62 ± 0.03 & 0.43 ± 0.09 & 39.23 ± 6.42 & 0.71 ± 0.09 & 0.20 ± 0.05(\textless{}0.01) & 1.30 ± 0.08(0.36) & 0.61 ± 0.04(0.01) & 0.46 ± 0.10(\textless{}0.01) & 38.71 ± 7.54(\textless{}0.01) & 0.68 ± 0.13(0.02) \\
LB8 & 0.33 ± 0.12 & 1.31 ± 0.18 & 0.65 ± 0.13 & 0.30 ± 0.15 & 49.27 ± 12.29 & 0.71 ± 0.12 & 0.28 ± 0.07(0.67) & 1.31 ± 0.10(\textless{}0.01) & 0.64 ± 0.05(\textless{}0.01) & 0.31 ± 0.12(\textless{}0.01) & 46.91 ± 11.30(\textless{}0.01) & 0.67 ± 0.16(0.19) \\
LB9 & 0.28 ± 0.14 & 1.29 ± 0.25 & 0.66 ± 0.17 & 0.27 ± 0.16 & 47.70 ± 12.69 & 0.69 ± 0.14 & 0.26 ± 0.08(0.5) & 1.32 ± 0.11(\textless{}0.01) & 0.65 ± 0.06(\textless{}0.01) & 0.29 ± 0.12(\textless{}0.01) & 44.19 ± 11.93(\textless{}0.01) & 0.65 ± 0.17(0.82) \\
LB10 & 0.30 ± 0.06 & 1.28 ± 0.04 & 0.65 ± 0.03 & 0.36 ± 0.10 & 57.44 ± 9.83 & 0.82 ± 0.09 & 0.22 ± 0.05(\textless{}0.01) & 1.27 ± 0.05(0.41) & 0.65 ± 0.04(\textless{}0.01) & 0.31 ± 0.10(0.01) & 56.74 ± 10.57(\textless{}0.01) & 0.76 ± 0.13(\textless{}0.01) \\
RB1 & 0.29 ± 0.06 & 1.29 ± 0.04 & 0.63 ± 0.03 & 0.37 ± 0.12 & 43.88 ± 7.74 & 0.75 ± 0.09 & 0.20 ± 0.04(0.49) & 1.27 ± 0.05(\textless{}0.01) & 0.64 ± 0.03(0.86) & 0.37 ± 0.11(0.02) & 48.30 ± 8.46(\textless{}0.01) & 0.75 ± 0.11(\textless{}0.01) \\
RB2 & 0.27 ± 0.05 & 1.29 ± 0.05 & 0.64 ± 0.03 & 0.42 ± 0.09 & 40.88 ± 5.78 & 0.70 ± 0.08 & 0.21 ± 0.05(\textless{}0.01) & 1.29 ± 0.09(0.42) & 0.62 ± 0.04(\textless{}0.01) & 0.38 ± 0.11(\textless{}0.01) & 40.43 ± 7.84(\textless{}0.01) & 0.66 ± 0.16(0.69) \\
RB3 & 0.26 ± 0.05 & 1.29 ± 0.07 & 0.65 ± 0.03 & 0.46 ± 0.09 & 48.99 ± 9.16 & 0.78 ± 0.10 & 0.20 ± 0.05(\textless{}0.01) & 1.26 ± 0.06(\textless{}0.01) & 0.61 ± 0.03(\textless{}0.01) & 0.42 ± 0.11(\textless{}0.01) & 46.45 ± 9.00(\textless{}0.01) & 0.73 ± 0.17(\textless{}0.01) \\
RB4 & 0.28 ± 0.05 & 1.29 ± 0.06 & 0.66 ± 0.04 & 0.28 ± 0.07 & 47.43 ± 8.04 & 0.72 ± 0.09 & 0.25 ± 0.08(\textless{}0.01) & 1.30 ± 0.10(\textless{}0.01) & 0.63 ± 0.05(\textless{}0.01) & 0.24 ± 0.09(\textless{}0.01) & 39.64 ± 10.07(\textless{}0.01) & 0.57 ± 0.23(0.18) \\
RB5 & 0.29 ± 0.06 & 1.31 ± 0.07 & 0.66 ± 0.04 & 0.23 ± 0.07 & 53.31 ± 9.37 & 0.75 ± 0.10 & 0.23 ± 0.07(\textless{}0.01) & 1.29 ± 0.08(\textless{}0.01) & 0.63 ± 0.05(\textless{}0.01) & 0.21 ± 0.09(\textless{}0.01) & 45.85 ± 11.40(\textless{}0.01) & 0.61 ± 0.24(\textless{}0.01) \\
RB6 & 0.26 ± 0.05 & 1.31 ± 0.08 & 0.63 ± 0.03 & 0.54 ± 0.10 & 40.39 ± 5.28 & 0.73 ± 0.08 & 0.21 ± 0.05(\textless{}0.01) & 1.31 ± 0.09(\textless{}0.01) & 0.60 ± 0.03(\textless{}0.01) & 0.46 ± 0.12(\textless{}0.01) & 37.72 ± 6.95(\textless{}0.01) & 0.68 ± 0.14(0.36) \\
RB7 & 0.21 ± 0.41 & 1.10 ± 0.73 & 0.48 ± 0.50 & 0.13 ± 0.40 & 38.37 ± 15.49 & 0.59 ± 0.22 & 0.28 ± 0.07(\textless{}0.01) & 1.34 ± 0.16(0.03) & 0.66 ± 0.05(0.84) & 0.23 ± 0.09(\textless{}0.01) & 44.59 ± 10.45(\textless{}0.01) & 0.58 ± 0.19(0.77) \\
RB8 & 0.30 ± 0.06 & 1.30 ± 0.06 & 0.68 ± 0.04 & 0.29 ± 0.10 & 53.62 ± 9.52 & 0.75 ± 0.09 & 0.26 ± 0.08(\textless{}0.01) & 1.31 ± 0.10(\textless{}0.01) & 0.65 ± 0.06(\textless{}0.01) & 0.24 ± 0.11(\textless{}0.01) & 45.96 ± 11.38(\textless{}0.01) & 0.63 ± 0.20(0.02) \\
RB9 & 0.29 ± 0.06 & 1.31 ± 0.09 & 0.68 ± 0.05 & 0.24 ± 0.07 & 48.76 ± 10.63 & 0.68 ± 0.11 & 0.27 ± 0.09(\textless{}0.01) & 1.35 ± 0.17(\textless{}0.01) & 0.65 ± 0.07(\textless{}0.01) & 0.20 ± 0.10(\textless{}0.01) & 41.84 ± 12.47(\textless{}0.01) & 0.57 ± 0.20(\textless{}0.01) \\
RB10 & 0.28 ± 0.05 & 1.28 ± 0.04 & 0.66 ± 0.03 & 0.36 ± 0.08 & 58.25 ± 9.77 & 0.83 ± 0.09 & 0.22 ± 0.05(\textless{}0.01) & 1.28 ± 0.06(\textless{}0.01) & 0.65 ± 0.04(\textless{}0.01) & 0.31 ± 0.09(0.05) & 53.06 ± 9.47(\textless{}0.01) & 0.77 ± 0.11(0.49) \\
\hlinew{1pt}
\end{tabular}}
\addcontentsline{toc}{section}{Table S18: The quantative morphological features of the pulmonary reticular opacities group.}
\end{sidewaystable*}

\end{document}